\documentclass[twocolumn,pre,tightenlines,superscriptaddress,showpacs]{revtex4-2}

\usepackage{amsmath}
\usepackage{amssymb,amsfonts,latexsym}
\usepackage{bm}
\usepackage[mathcal]{euscript}
\usepackage{graphicx}
\usepackage{epsfig}
\usepackage{color}

\newcommand{\be}{\begin{equation}}
\newcommand{\ee}{\end{equation}}
\newcommand{\ve}{\varepsilon}
\newcommand{\la}{\langle}
\newcommand{\ra}{\rangle}
\newcommand{\mC}{\mathcal{C}}

\begin{document}
\title{Discontinuous phase transition from ferromagnetic to oscillating states in a nonequilibrium mean-field spin model}
\author{Laura Guislain}
\affiliation{Univ.~Grenoble Alpes, CNRS, LIPhy, 38000 Grenoble, France}
\author{Eric Bertin}
\affiliation{Univ.~Grenoble Alpes, CNRS, LIPhy, 38000 Grenoble, France}
\date{\today}
\begin{abstract}
We study a nonequilibrium ferromagnetic mean-field spin model exhibiting a phase with spontaneous temporal oscillations of the magnetization, on top of the usual paramagnetic and ferromagnetic phases.
This behavior is obtained by introducing dynamic field variables coupled to the spins through non-reciprocal couplings.
We determine a nonequilibrium generalization of the Landau free energy in terms of the large deviation function of the magnetization and of an appropriately defined smoothed stochastic time derivative of the magnetization.
While the transition between paramagnetic and oscillating phase is continuous, the transition between ferromagnetic and oscillating phases is found to be discontinuous, with a coexistence of both phases, one being stable and the other one metastable. Depending on parameter values, the ferromagnetic points may either be inside or outside the limit cycle, leading to different transition scenarios.
The stability of these steady states is determined from the large deviation function.
We also show that in the coexistence region, the entropy production has a pronounced maximum as a function of system size.
\end{abstract}

\maketitle

%%%%%%%%%%%%%%%%%%%%%%%%%%%%%%%%%%%%%%%%

\section{Introduction}

A number of systems driven far from equilibrium are known to exhibit spontaneous collective oscillations.
This is the case for instance for coupled oscillators, like the Kuramoto model with distributed frequencies \cite{acebron_kuramoto_2005,gupta_kuramoto_2014}, or in models of identical coupled noisy oscillators \cite{risler_universal_2004,risler_universal_2005}.
Interestingly, spontaneous oscillations have also been reported in systems composed of a large number of coupled units which individually do not oscillate in the absence of interaction.
Standard examples include different types of chemical oscillators \cite{nicolis_dissipative_1986}, and recent experimental and theoretical studies have also reported spontaneous oscillations in populations of biological cells \cite{Kamino_fold2017,Wang_emergence2019}, assemblies of active particles with non-reciprocal interactions \cite{saha_scalar_2020,you_nonreciprocity_2020}, biochemical clocks \cite{Cao_free_energy2015,nguyen_phase_2018,Aufinger_complex2022}, droplets in a fluid binary mixture \cite{devailly_phase_2015},
models of population dynamics \cite{andrae_entropy_2010,Duan_Hopf2019}, socio-economic models \cite{Gualdi2015,yi_symmetry_2015} or nonequilibrium spin systems \cite{collet_rhythmic_2016,de_martino_oscillations_2019,daipra_oscillatory_2020}.

At the deterministic level of description, valid in the infinite system size limit, the spontaneous emergence of oscillations can be described as a Hopf bifurcation \cite{crawford_introduction_1991} in the framework of dynamical system theory.
However, many situations of experimental relevance involve mesoscopic systems for which fluctuations cannot be neglected, as in the case of biochemical clocks for instance
\cite{Fei_design2018}. An important consequence of the presence of fluctuations is that the coherence time of oscillations becomes finite \cite{gaspard_correlation_2002,barato_cost_2016,barato_coherence_2017,oberreiter_universal_2022,remlein_coherence_2022}.
At a phenomenological level, the onset of oscillations in a fluctuating system may be described as a stochastic Hopf bifurcation \cite{Sagues2007,Xu_Langevin2020}.
Yet, a deeper understanding would require to cast this phenomenon in the general framework of nonequilibrium phase transitions, by explicitly connecting the collective level of description to the microscopic dynamics.
One may in particular interpret the onset of oscillations in a large system of interacting degrees of freedom by extending to far-from-equilibrium systems the thermodynamic framework of phase transition introduced at equilibrium. Along this line, the entropy production density has been shown to play the role of a generalized thermodynamic potential, with a discontinuous derivative at the onset of spontaneous oscillations \cite{crochik_entropy_2005,xiao_entropy_2008,xiao_stochastic_2009,andrae_entropy_2010,barato_entropy_2012,tome_entropy_2021,nguyen_phase_2018,noa_entropy_2019,martynec_entropy_2020,seara_irreversibility_2021}.
Another approach to phase transitions consists in introducing order parameters associated with spontaneously broken symmetries \cite{LeBellac}.
At a mean-field level of description, one may then introduce a Landau free energy and determine its expansion close to the phase transition.
While this approach has been originally designed for equilibrium systems, several recent works have extended it to different types of nonequilibrium situations in the context of spin models, to describe relaxation effects \cite{Meibohm_2022,Holtzman_2022}, or the driving by an oscillatory field or by multiple heat baths \cite{Aron_2020}.
Based on a large deviation theory approach, the spontaneous transition from a paramagnetic to an oscillating state has been recently described in a nonequilibrium Landau framework \cite{guislain_nonequil2023}.

In this paper, we extend the results of Ref.~\cite{guislain_nonequil2023} by considering within the same nonequilibrium Landau framework the transition from a ferromagnetic state to a state of spontaneous collective oscillations. We study a mean-field spin model where spins are coupled to dynamic fields in a non-reciprocal way, resulting in a breaking of detailed balance which allows for the onset of oscillations in some parameter range. The presence of spontaneous oscillations may be interpreted as an instance of a non-reciprocal phase transition \cite{Fruchart_non-reciprocal_2021,Martin_exact_2023}. Both spin and field variables are also subjected to ferromagnetic couplings, with different values. 
The phase transition is characterized by determining a large deviation function of the magnetization and of a stochastic variable playing the role of a smoothed time derivative of the magnetization.
The transition from the ferromagnetic state to the oscillating state is found to be discontinuous, with a coexistence of the two phases in the transition region. The large deviation function allows us to determine which phase is stable or metastable. We also characterize finite size effects in terms of entropy production.

The paper is organized as follows. The model is defined in Sec.~\ref{sec:model:description}, the method is presented in Sec.~\ref{sec:method} and the main results of Ref.~\cite{guislain_nonequil2023} on the continuous phase transition from paramagnetic to oscillating states are summarized and extended in Sec.~\ref{sec:cas0}.
Then, Sec.~\ref{sec:cas1} characterizes a first scenario of discontinuous phase transition from ferromagnetic to oscillating states, whereby the limit cycle surrounds the ferromagnetic points.
A second scenario, in which ferromagnetic points stand outside the limit cycle, is studied in Sec.~\ref{sec:cas2}.
Conclusions and perspectives are summarized in Sec.~\ref{sec:conclusion}.

%%%%%%%%%%%%%%%%%%%%%%%%%%%%%%%%%%%%%%%%%%%%%%%%%%%
%%%%%%%%%%%%%%%%%%%%%%%%%%%%%%%%%%%%%%%%%%%%%%%%%%%

\section{Model description}
\label{sec:model:description}

\subsection{Definition of the model}
We consider a generalization of the kinetic mean-field Ising model with ferromagnetic interactions introduced in \cite{guislain_nonequil2023}, and sharing some similarities with related models having two spin populations \cite{collet_macroscopic_2014,collet_rhythmic_2016}
or subjected to a feedback control \cite{de_martino_oscillations_2019,Sinelschikov_emergence_2023}.
The model involves $2N$ microscopic variables: $N$ spins $s_i=\pm1$ and $N$ fields $h_i=\pm 1$. 
We define the magnetization $m$ and average field $h$ as
\be
m = \frac{1}{N} \sum_{i=1}^N s_i\,, \qquad h = \frac{1}{N} \sum_{i=1}^N h_i\,.
\ee
The stochastic dynamics consists in randomly flipping a single spin $s_i=\pm 1$ with rate $w_s^{\pm}$, or a single field $h_i=\pm 1$ with rate $w_h^{\pm}$.
In a mean-field spirit, the flipping rates $w_s^{\pm}$ and $w_h^{\pm}$ are independent of $i$, and depend only on $m$ and $h$,
\be \label{eq:transition:rates}
w_s^{\pm}(m,h) = \frac{1}{1+e^{\beta \Delta E_s^{\pm}(m,h)}}, \quad w_h^{\pm} = \frac{1}{1+e^{\beta \Delta E_h^{\pm}(m,h)}},
\ee
with $\beta=T^{-1}$ the inverse temperature and $\Delta E_{s,h}^{\pm}(m,h)$ the variation of $E_{s,h}(m,h)$ when flipping a spin $s_i=\pm 1$ or a field $h_i=\pm 1$, where
\begin{align}
E_s(m,h) &= -N\left(\frac{J_1}{2} m^2+\frac{J_2}{2} h^2+mh\right),\\
E_h(m,h) &= E_s(m,h)+\mu Nhm\,.
\end{align}
When $\mu=0$, $E_h=E_s$ and the transition rates satisfy detailed balance with respect to the equilibrium probability distribution $P_{\rm eq}\propto e^{-\beta E_s}$.
Detailed balance is broken as soon as $\mu \ne 0$, and $\mu$ can thus be interpreted as a parameter controling the distance to equilibrium.
For fixed values of the interactions $J_1$ and $J_2$, the temperature $T$ and the distance to equilibrium $\mu$ are the two control parameters of the model. 

We denote as $\mC=(s_1,\dots,s_N,h_1,\dots,h_N)$ the microscopic configuration of the system. Flips of the variables $s_i$ and $h_i$ are encoded into formal transition rates $W(\mC'|\mC)$ from a configuration $\mC$ to a configuration $\mC'$.
The probability $P(\mC,t)$ of a configuration $\mC$ at time $t$ evolves according to the master equation
\be \label{eq:master:equation}
\partial_t P(\mC,t)=\sum_{\mC'(\ne\mC)}\big[W(\mC|\mC')P(\mC',t)-W(\mC'|\mC)P(\mC,t)\big].
\ee

%%%%%%%%%%%%%%%%%%%%%%%%%%%%%%%%%%%%%%%%%%%%%%%%%%%

\subsection{Phase diagram in the deterministic limit}
\label{sec:subsec:deter}

We first study the bifurcation diagram of the system obtained in the deterministic limit $N\to\infty$.
We compute the time derivatives $d_t\langle m\rangle$ and $d_t\langle h\rangle$ using the master equation Eq.~(\ref{eq:master:equation}), where $\langle x\rangle=\sum_{\mC}m(\mC)P(\mC)$ for any observable $x$. We assume that the law of large numbers holds in the limit $N\to\infty$ so that $\langle f(m, h)\rangle \to f(m, h)$ for any regular function $f$. Finally we obtain a set of deterministic equations on $m(t)$ and $h(t)$ (see Appendix \ref{appendix:deterministic:equations}):
\begin{align} \label{eq:deter:equations:m}
\frac{dm}{dt} &= -m+\tanh[\beta(J_1m+h)],\\
 \label{eq:deter:equations:h}
\frac{dh}{dt} &= -h+\tanh[\beta(J_2h+(1-\mu)m)].
\end{align}
We explore regimes where the magnetization $m(t)$ may exhibit oscillations. In dynamical systems theory, a limit cycle may generally be described in the plane defined by a variable and its time derivative, thus we introduce a new variable $\dot{m}=dm/dt$. The set of deterministic equations become 
\be \begin{aligned} \label{eq:deter:equations}
\frac{dm}{dt}=\dot{m}\,, \qquad 
\frac{d\dot{m}}{dt}=Y(m, \dot{m})
\end{aligned}\ee 
where $Y(m, \dot{m})$ has a lengthy expression, given in Appendix~\ref{appendix:values:coeff} [see Eq.~(\ref{eq:Ymdotm:appA})].
$Y(m, \dot{m})$ satisfies the symmetry $Y(-m,-\dot{m})=-Y(m, \dot{m})$.
To study the fixed points of Eq.~(\ref{eq:deter:equations}) and their stability, we decompose $Y(m, \dot{m})$ into a $\dot{m}$-independent contribution
\be \label{eq:def:Ym0}
Y(m, 0) = -V'(m)
\ee
[see Appendix~\ref{appendix:values:coeff}, Eq.~(\ref{eq:full:expr:Vm:appB}) for its explicit expression] and a $\dot{m}$-dependent contribution
\be \label{eq:def:gmmdot}
\dot{m} g(m, \dot{m}) = Y(m, \dot{m})-Y(m, 0)\,,
\ee
which defines the function $g(m,\dot{m})$. From Eq.~(\ref{eq:deter:equations}), the fixed points $(m,\dot{m})=(m_0,0)$ satisfy $Y(m_0, 0)=0$, and thus $V'(m_0)=0$ according to Eq.~(\ref{eq:def:Ym0}).
One finds in particular that $(m, \dot{m})=(0, 0)$ is always a fixed point of the system, because $V'(0)=0$ by symmetry.

Linearizing the dynamics around a fixed point $(m_0, 0)$, $m=m_0+\delta m$, $\dot{m}=\delta\dot{m}$, one has from Eq.~(\ref{eq:deter:equations})
\be
\frac{d}{dt} \begin{pmatrix} \delta m\\ \delta\dot{m}\end{pmatrix}
%= \begin{pmatrix}0 & 1\\ -V''(m) & g(m, 0)\end{pmatrix}
= \mathbf{M}
\begin{pmatrix} \delta m\\ \delta\dot{m}\end{pmatrix}
\ee
with
\be
\mathbf{M} =\begin{pmatrix}0 & 1\\ -V''(m_0) & g(m_0, 0)\end{pmatrix}.
\ee
The linear stability of the fixed point $(m_0,0)$ is determined by the sign of the eigenvalues %$\lambda_{\pm}$ 
of the matrix $\mathbf{M}$,
\be \label{eq:eigenvalues:M}
\lambda_{\pm}=\frac{1}{2} g(m_0, 0) \left(1\pm\sqrt{1-\frac{4V''(m_0)}{g(m_0,0)^2}}\right).
\ee
The fixed point $(m_0, 0)$ is stable if both $\lambda_{+}$ and $\lambda_{-}$ are negative (or have a negative real part), implying that $V''(m_0)>0$ and $g(m_0, 0)<0$.
We see in particular from Eq.~(\ref{eq:eigenvalues:M}) that %$g(m_0, 0)$ governs the stability of the fixed point when $V''(m_0)>0$, as 
the fixed point $(m_0, 0)$ becomes unstable when $g(m_0, 0)$ is positive.
We define the critical temperature $T_c=(J_1+J_2)/2$ and the dimensionless deviation from $T_c$,
\be
\ve = \frac{T_c-T}{T_c}\,.
\ee
Using expression (\ref{eq:Ymdotm:appA}) of $Y(m, \dot{m})$, we get for $m_0=0$ and small $\ve$ that $g(0, 0)=a_0\ve$, with $a_0=2T/T_c$.
%The fixed point $(0,0)$ is stable for $\ve<0$ and unstable for $\ve>0$.
We also have $V''(0)=(\mu-\mu_l(T))/T^2$, where we define $\mu_l(T)$ as 
\be
\label{eq:def:mul}
\mu_l(T)=1-(J_1-T)(J_2-T).
\ee
%One has $V''(0)=(\mu-\mu_l)/T^2$ and $g(0, 0)=2T_c/T((T_c-T)/T_c)$ with $T_c=(J_1+J_2)/2$ and $\mu_l=1-T^2(J_1/T-1)(J_2/T-1)$, 
Hence, the fixed point $(m, \dot{m})=(0,0)$ is linearly stable for $T>T_c$ [$\ve<0$, implying $g(0,0)<0$] provided that $\mu>\mu_l(T)$ [implying $V''(0)>0$], and unstable otherwise. We define $\mu_c=\mu_l(T_c)$. 

\begin{figure}[t]
    \centering
    \includegraphics{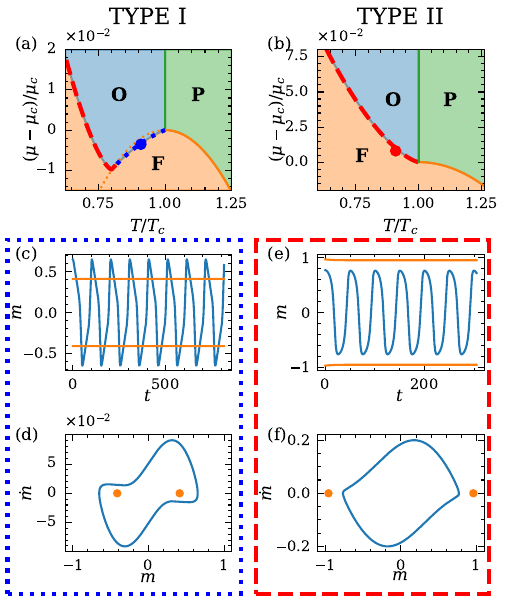}
    \caption{Phase diagram of the deterministic dynamics Eq.~(\ref{eq:deter:equations}) obtained numerically for (a) $J_1=0.6$, $J_2=0.4$ and (b) $J_1=J_2=0.5$. Three distinct behaviors are observed: a stable paramagnetic point (P), two stable ferromagnetic points (F), and an oscillating phase (O). On the thick dashed and dotted lines, the limit cycle and the ferromagnetic points coexist. On the blue dotted line, the ferromagnetic points are inside the limit cycle (Type-I coexistence) and on the red dashed line the ferromagnetic points are outside the limit cycle (Type-II coexistence). When the two lines meet, the bifurcation is heteroclinic. The orange line corresponds to $\mu_l(T)$ [Eq.~(\ref{eq:def:mul})]: in plain for $T>T_c$ where it represents the limit of stability of the paramagnetic points and dotted for $T<T_c$ as indication. (c)-(f) Examples of trajectories $m(t)$ and phase space ($m(t), \dot{m}(t))$: (c),(d) Type-I coexistence for $J_1=0.6$, $J_2=0.4$, $T/T_c=0.9$ and $(\mu-\mu_c)/\mu_c=-3.5\times 10^{-3}$ (blue dot in (a)); (e),(f) Type-II coexistence for $J_1=J_2=0.5$, $T/T_c=0.9$ and $(\mu-\mu_c)/\mu_c=8\times 10^{-3}$ (red dot in (b)). }
    \label{fig:phase:diagrams}
\end{figure}

Two examples of stability diagrams, obtained from the numerical evaluation of the fixed points %[minimum of $V(m)$ from Eq.~(\ref{eq:def:Ym0})]
and their stability [given by the sign of the eigenvalues of Eq.~(\ref{eq:eigenvalues:M})], are shown in Fig.~\ref{fig:phase:diagrams} for different values of $J_1$ and $J_2$. Trajectories and existence of limit cycles are obtained from the numerical integration of Eqs.~(\ref{eq:deter:equations:m}) and (\ref{eq:deter:equations:h}). A stable paramagnetic fixed point [denoted as P in Fig.~\ref{fig:phase:diagrams}(a,b)] is found at high enough temperature, while this point becomes unstable at low temperature.
For low values of $(T, \mu)$, two symmetric ferromagnetic stable fixed points (F) are observed.
At low $T$ and high $\mu$, an oscillating behavior (O) is observed. 
The transition lines between the three different behaviors meet at a tricritical point ($T_c$, $\mu_c$), see Fig.~\ref{fig:phase:diagrams}(a,b).

Depending on the value of $\mu$, the bifurcation from the paramagnetic point to a limit cycle at $T_c$ which occurs for $\mu>\mu_c$ can either be continuous (supercritical Hopf bifurcation) or discontinuous (subcritical Hopf bifurcation) \cite{guislain_nonequil2023}. It is generically continuous when the couplings $J_1$ and $J_2$ are positive.
The transition from the ferromagnetic points to a limit cycle is discontinuous [except for the particular values $J_2=\pm2+J_1$] and we observe small regions of the parameter space where the ferromagnetic points and the limit cycle coexist. In Fig.~\ref{fig:phase:diagrams}, they are represented by thick dotted and dashed lines.
Depending on the values of the parameters ($T$, $J_1$ and $J_2$), the ferromagnetic points can be either inside or outside the limit cycle, which leads to a topological classification of the transition into two differents types.
In the following, we call discontinuous transition of Type I the case when the ferromagnetic points are inside the limit cycle, and discontinuous transition of Type II the case when the ferromagnetic points are outside the limit cycle.
A discontinuous transition of Type I is typically found close to $T_c$ for $J_1>J_2$ (under additional assumptions that are specified below),
as illustrated for $J_1=0.6$ and $J_2=0.4$ by the dotted blue line in Fig.~\ref{fig:phase:diagrams}(a).
An example of trajectory $m(t)$ and phase space plot $(m(t), \dot{m}(t))$ is represented in Fig.~\ref{fig:phase:diagrams}(c),(d). 
A discontinuous transition of Type II is found for $J_1\leq J_2$ (under additional assumptions that are specified below) and $T<T_c$, see Fig.~\ref{fig:phase:diagrams}(b) for $J_1=J_2=0.5$. The corresponding trajectory $m(t)$ and its phase space representation $(m(t), \dot{m}(t))$ is plotted in Fig.~\ref{fig:phase:diagrams}(e),(f). The farther from $T_c$, the closer the ferromagnetic points and the limit cycle are.

Note that in the case $J_1=0.6$ and $J_2=0.4$, one observes that for lower temperatures, the ferromagnetic points sit outside the limit cycle [dashed red line in Fig.~\ref{fig:phase:diagrams}(a)], similarly to the behavior displayed in Fig.~\ref{fig:phase:diagrams}(e), (f). At the point where the dotted blue line meets the dashed red line, a limit cycle joining the two ferromagnetic points and with an infinite period appears when the ferromagnetic points loose stability, corresponding to a heteroclinic bifurcation.

Note also that for $J_2=\pm 2+J_1$, the transition is neither of type I or II, but is a continuous transition from the ferromagnetic points to the limit cycle corresponding to a heteroclinic bifurcation. We do not study this particular case in this paper, but a comment on the specificity of this case is made in Sec.~\ref{sec:cas2:discussion:heteroclinic}.

\subsection{Close to the tricritical point}
\label{sec:tricrit:deter}

\begin{figure}[t]
    \centering
    \includegraphics{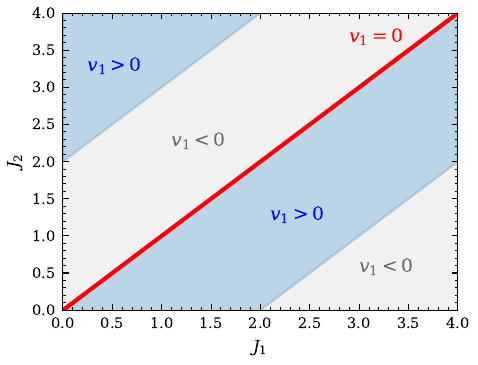}
    \caption{Sign of $v_1(T_c, \mu_c)$ from Eq.~(\ref{eq:def:v1}) in the plane ($J_1$, $J_2$).
    In regions with $v_1>0$, the discontinuous transition between ferromagnetic and oscillating phases in the $(T,\mu)$ phase diagram is of Type I: a limit cycle coexists with ferromagnetic points located inside the cycle (see Fig.~\ref{fig:phase:diagrams}(a) and Sec.~\ref{sec:cas1}).
    In the case $v_1=0$ with $J_1=J_2$ (red line), the discontinuous transition between ferromagnetic and oscillating phases in the $(T,\mu)$ phase diagram is of Type II: a limit cycle coexists with ferromagnetic points located outside the cycle (see Fig.~\ref{fig:phase:diagrams}(b) and Sec.~\ref{sec:cas2}).
    The case $v_1=0$ with $J_1\ne J_2$ is not discussed in details in this work.}
    \label{fig:sign:v1:J1:J2}
\end{figure}

In the following, we focus on the transition close to $T_c$, i.e., for small $\ve$ in order to use a perturbative theory. 
We consider that $m$ is small such that only the first orders of the series expansion of $V(m)$ are necessary. One finds at order $m^4$ for $V(m)$ and at order $m^2$ for $g(m, 0)$ that 
\begin{align}
    \label{eq:vpm:3}
    &V(m) =\frac{\mu-\mu_l(T)}{2T^2}m^2+\frac{v_1(T, \mu)}{4}m^4 + V_0,\\
    \label{eq:g(m,0):small:m}
    &g(m,0) = a_0\ve -a_1 m^2,
\end{align}
where $V_0$ is at this stage an arbitrary constant, and where $v_1(T, \mu)$, $a_0$ and $a_1$ are given in Appendix~\ref{appendix:values:coeff}. In particular one has $a_0, a_1>0$ and
\be \label{eq:def:v1}
v_1(T_c, \mu_c)=\frac{(J_1-J_2)[4-(J_1-J_2)^2]}{12(J_1+J_2)}\,.
\ee
The sign of $v_1(T_c, \mu_c)$, which plays a key role in the behavior of the model, thus depends on the relative values of $J_1$ and $J_2$ (see Fig.~\ref{fig:sign:v1:J1:J2}). 

\begin{figure}[t]
    \centering
    \includegraphics{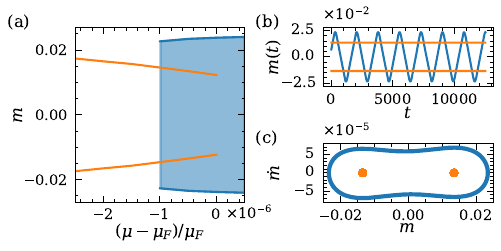}
    \caption{Type-I coexistence close to the tricritical point, corresponding to ferromagnetic points inside the limit cycle, for $\ve=(T_c-T)/T_c=10^{-4}$ and $J_1=0.6$, $J_2=0.4$. (a) Values of $m$ for the ferromagnetic points (orange lines) and for the limit cycle (blue shaded area) along the transition. At $\mu=\mu_F(T)$ [Eq.~(\ref{eq:def:muF})] ferromagnetic points become linearly unstable. (b) Examples of trajectories for $(\mu-\mu_F)/\mu_F=-5\times10^{-7}$. (c) Corresponding phase space representation in the plane ($m$, $\dot{m}$).}
    \label{fig:phase:diagrams_TC1}
\end{figure}

When $v_1>0$, ferromagnetic points exist for $\mu<\mu_l(T)$, and are given by
\be \label{eq:def:m0}
m_0^2 = \frac{\mu_l(T)-\mu}{T^2 v_1}\,,
\ee
i.e., nonzero solutions of the equation $V'(m_0)=0$.
According to Eq.~(\ref{eq:eigenvalues:M}), and given that $V''(m_0)>0$, ferromagnetic points are linearly stable when $g(m_0, 0)<0$, which corresponds to
\be \label{eq:def:muF}
\mu<\mu_F(T)\equiv\mu_l(T)- \frac{\ve a_0T^2 v_1}{a_1}\,.
\ee
%\textbf{[Are $\mu_l$ and $\mu_F$ distinguishable in Fig.1?]}
Numerically, one observes that before ferromagnetic points become linearly unstable upon increasing $\mu$, they coexist over a tiny parameter range with a limit cycle that surrounds them.
An example is given for $J_1=0.6$ and $J_1=0.4$ in Fig.~\ref{fig:phase:diagrams_TC1}, which displays the ferromagnetic points and the extension of the limit cycle as a function of $\mu$ at fixed $\ve$ [Fig.~\ref{fig:phase:diagrams_TC1}(a)], as well as the corresponding trajectories $m(t)$ [Fig.~\ref{fig:phase:diagrams_TC1}(b)] and the coexisting trajectories in the phase space $(m, \dot{m})$ [Fig.~\ref{fig:phase:diagrams_TC1}(c)]. Unlike for smaller values of $T$, we observe that close to the tricritical point, the ferromagnetic points and the limit cycle are well separated. We also observe in this regime that the two symmetries $m\,\mapsto\,-m$ and $\dot{m}\,\mapsto\,-\dot{m}$ are separately valid to a good approximation, while they were previously valid only under the simultaneous transformation $(m,\dot{m})\,\mapsto\,(-m,-\dot{m})$.

\begin{figure}[t]
    \centering
    \includegraphics{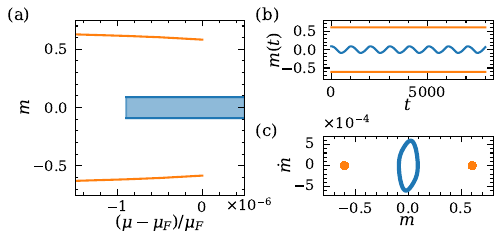}
    \caption{Type-II coexistence close to the tricritical point, corresponding to ferromagnetic points outside the limit cycle, for $\ve=10^{-3}$ and $J_1=J_2=0.5$. (a) Values of $m$ for the ferromagnetic points (orange lines) and for the limit cycle (blue shaded area) along the transition. At $\mu=\mu_F(T)$ [here, $(\mu_F-\mu_c)/\mu_c \approx 1.5\times10^{-5}$] the ferromagnetic points become unstable. (b) Examples of trajectories for $(\mu-\mu_F)/\mu_F=-5\times10^{-6}$. (c) Corresponding phase space representation in the plane ($m$, $\dot{m}$).}
    \label{fig:phase:diagrams_TC2}
\end{figure}

When $v_1\leq 0$, higher order terms in the expansion of $V(m)$ are necessary to obtain the ferromagnetic points and their stability. Numerically, we observe two scenarios: the first one is a heteroclinic bifurcation, the ferromagnetic points loose stability and a limit cycle with infinite period arises. This is the case in particular for $J_2=\pm 2+J_1$, for which $v_1(T_c, \mu_c)=0$. The second scenario is that before disappearing, the ferromagnetic points coexist with a small elliptic limit cycle. This is the case for $J_1=J_2=0.5$, see Fig.~\ref{fig:phase:diagrams_TC2} where an example of the evolution with $\mu$ (at fixed $\ve$) of the ferromagnetic fixed points and of the limit cycle are displayed, together with examples of trajectories $m(t)$. When $v_1(T_c, \mu_c)<0$, one finds numerically that the ferromagnetic phase and the paramagnetic phase coexist for $T\gtrsim T_c, \mu\gtrsim\mu_l(T)$,
so that $(T_c, \mu_c)$ is no longer a tricritical point, whereas for $v_1(T_c, \mu_c)=0$ which is verified for $J_1=J_2$ and for $J_2=\pm 2 +J_1$, one finds that the three transition lines meet at $(T_c, \mu_c)$ (see Fig.~\ref{fig:phase:diagrams} for $J_1=J_2=0.5$ for an example of bifurcation diagram). 
In the following, we focus on the case where $v_1(T_c, \mu_c)\geq 0$ where the three phases meet at the critical point $(T_c, \mu_c)$, in order to perform a perturbative analysis close to the tricritical point.


%%%%%%%%%%%%%%%%%%%%%%%%%%%%%%%%%%%%%%%%%%%%%%%%%%%%%%%%%%%%%%%%%%%%%%%%%%%
%%%%%%%%%%%%%%%%%%%%%%%%%%%%%%%%%%%%%%%%%%%%%%%%%%%%%%%%%%%%%%%%%%%%%%%%%%%

\section{Generalized Landau theory}
\label{sec:method}

\begin{figure}[t]
    \centering
    \includegraphics{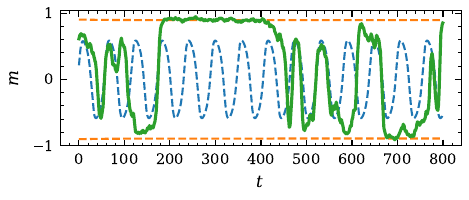}
    \caption{Trajectories $m(t)$: the dashed blue and orange lines correspond to deterministic trajectories and the green line to a trajectory for a finite system size $N=5000$ obtained from stochastic numerical simulations. Parameters: $J_1=J_2=0.5$, $\ve=5\times10^{-2}$ and $(\mu-\mu_c)/\mu_c=3.88\times 10^{-3}$. Jumps between noisy oscillatory states and ferromagnetic states are observed.}
    \label{fig:trajectory:m:t}
\end{figure}

The deterministic limit provides knowledge on the different stable fixed points or limit cycles that are present in the system.
However it lacks information on the behavior of the system at finite size $N$, such as knowledge on the macroscopic fluctuations around the stable points or cycles. But most importantly, in case of coexistence of solutions in the limit $N\to\infty$, the deterministic approach fails to predict which solution is the most stable one at finite but large size $N$.
In addition, for moderate size $N$, one observes jumps between noisy oscillatory states and ferromagnetic states, as illustrated in Fig.~\ref{fig:trajectory:m:t}. A statistical description of such a situation where the ferromagnetic points and the limit cycle are both linearly stable in the deterministic limit would thus be useful. 
We briefly recall in this section the nonequilibrium generalization of the Landau theory developed in \cite{guislain_nonequil2023}, which allows for a description of phase transitions to oscillating states. %in finite-size systems.

\subsection{Stochastic time derivative $\dot{m}$}

We first introduce a new variable $\dot{m}$ that plays the role of a smoothed time derivative of the magnetization for finite-size systems.
Following \cite{guislain_nonequil2023}, we formally define the stochastic derivative $\dot{m}(\mC)$ of the magnetization $m(\mC)$ as 
\be \label{eq:def:mdot:C}
\dot{m}(\mC)=\sum_{\mC'(\neq \mC)}\left[m\left(\mC'\right)-m\left(\mC\right)\right]W(\mC'|\mC),
\ee
such that on average $d\la m\ra /dt=\la \dot{m}\ra$.
Eq.~(\ref{eq:def:mdot:C}) thus associates with each microscopic configuration $\mC$ an observable $\dot{m}(\mC)$, which is a smoothed time derivative of $m$ because it is averaged over all possible transitions $\mC\to\mC'$, for a fixed configuration $\mC$. The advantage of this definition is that fluctuations of $\dot{m}$ are typically on the same scale as that of $m$, which is a key property for the large deviation approach described below. Taking instead the time derivative of $m\big(\mC(t)\big)$ would lead to diverging, white-noise-like fluctuations which are not appropriate to develop a generalized Landau theory.

Under the mean-field assumption, the formal transition rate $W(\mC'|\mC)$ can be reexpressed in terms of the flipping rates $w_s^{\pm}(m,h)$ to flip a spin $s_i=\pm 1$ defined in Eq.~(\ref{eq:transition:rates}).
When flipping a spin $s_i=\pm1$, the magnetization change is given by $m(\mC')-m(\mC)=\mp 2/N$. Since there are $N(1\pm m)/2$ possibilities to choose a spin $s_i=\pm1$, one finds:
\be \label{eq:def:mdot:v0}
\dot{m}=(1-m)w_s^{-}(m,h)- (1+m)w_s^{+}(m,h).
\ee
Using the expression (\ref{eq:transition:rates}) of the flipping rates $w_s^{\pm}(m,h)$, Eq.~(\ref{eq:def:mdot:v0}) becomes:
\be \label{eq:def:mdot}
\dot{m}= -m+\tanh[\beta(J_1 m+h)].
\ee
Note that the functional relation $\dot{m}(m, h)$ turns out to be identical to the functional relation (\ref{eq:deter:equations:m}) obtained in the deterministic limit $N\to \infty$.
However, Eq.~(\ref{eq:def:mdot}) is valid for any finite $N$, and the variables $m$ and $h$ are here stochastic variables.

\subsection{Large deviation function}
At finite size $N$, the dynamics of the system is determined by the master equation (\ref{eq:master:equation}). Instead of considering $P(\mC)$ which involves $2^{2N}$ configurations $\mC=\{s_1,\dots, s_N, h_1,\dots, h_N\}$, we consider $P_N(m, \dot{m})$ the joint stationary probability density of the global observables $m$ and $\dot{m}$. 
The variations of $m$ and $\dot{m}$ during a transition scale as $1/N$. We introduce $\mathbf{d}_k$ such that $(\Delta m, \Delta \dot{m})=\pm \mathbf{d}_k/N$ with $k=1$ when flipping a spin $s_i=\pm 1$ and $k=2$ when flipping a field $h_i=\pm1$, so that we have:
\begin{align} \textbf{d}_{1}&=\left(-2, 2 -2\beta J_1+2\beta J_1(m+\dot{m})^2\right),\\
 \textbf{d}_{2}&=-\left(0, -2\beta +2\beta (m+\dot{m})^2\right).\end{align}
We note $NW_k^{\pm}$ the coarse-grained transition rates from a configuration $(m, h)$ to $(m', h')$,
with $m'=m\mp 2/N$ and $h'=h$ if $k=1$, and $m'=m$ and $h'=h\mp 2/N$ if $k=2$. One has:
\be  \begin{aligned} \label{eq:transition:rate} 
W_1^{\pm} &= \frac{(1\pm m)/2}{1+\exp[\pm 2\beta (J_1 m+h)]},\\
W_2^{\pm} &= \frac{(1\pm h)/2}{1+\exp[\pm 2\beta (J_2h+(1-\mu)m)]}.
\end{aligned} \ee
The coarse-grained master equation governing the evolution of $P_N(m, \dot{m})$ reads \cite{guislain_nonequil2023}:
\be \begin{aligned}\label{eq:eq:p}
\partial_t& P_N(m, \dot{m})=N\sum_{k,\sigma} \bigg[ 
-W_k^{\sigma}(m, \dot{m})P_N(m, \dot{m})\\
&+W_k^{\sigma}\left((m, \dot{m})-\frac{\sigma\textbf{d}_{k}}{N}\right)P_N\left((m, \dot{m})-\frac{\sigma\textbf{d}_{k}}{N}\right)\bigg]
\end{aligned} \ee
where $k=1, 2$ and $\sigma=\pm$.
For large $N$, the stationary joint distribution $P_N(m, \dot{m})$ takes a large deviation form \cite{touchette_2009}
\be \label{eq:large:deviation:form}
P_N(m, \dot{m}) \underset{N\to\infty}{\sim} \exp[-N\phi(m, \dot{m})]\,,
\ee
a property justified by the theory of Markov jump processes with vanishing jump size \cite{knessl_1985}.
Beside providing information on the fluctuations at finite system size, the large deviation function (or rate function) $\phi(m, \dot{m})$ determines the macroscopic phase of the system.
Linearly stable solutions of the deterministic equations correspond to local minima of the large deviation function. When two or more linearly stable solutions are present in the deterministic equations, the global minima of $\phi$ gives the macroscopic phase of the system (i.e., the most stable one). 

Injecting the large deviation form (\ref{eq:large:deviation:form}) into Eq.~(\ref{eq:eq:p}) gives to order $(\nabla \phi)^2$,
\be \begin{aligned} \label{eq:phi:quadrat} 
&\dot{m}\partial_m\phi+Y(m, \dot{m})\partial_{\dot{m}}\phi  +D_{11}(\partial_m \phi)^2
\\
& +2D_{12}\partial_m\phi\partial_{\dot{m}}\phi+D_{22}(\partial_{\dot{m}}\phi)^2=0\,,\end{aligned} \ee 
with 
\be (\dot{m}, Y(m, \dot{m}))=\sum_{k,\sigma}-\sigma \mathbf{d}_kW_k^{\sigma}\,.\ee
The function $Y(m, \dot{m})$ is found to be the same function as the one introduced in the deterministic limit in Eq.~(\ref{eq:deter:equations}). We introduce $\mathbf{D}=\{D_{ij}(m, \dot{m})\}$ as
\be \label{eq:def:D} \mathbf{D}\equiv \sum_{k} \mathbf{d}_k\cdot \mathbf{d}_k^TW_{k}^{\sigma},\ee
whose explicit expression is given in Appendix~\ref{appendix:values:coeff}. 

We use the decomposition introduced in Eq.~(\ref{eq:def:gmmdot}),
\be
Y(m, \dot{m})=-V'(m)+\dot{m}g(m, \dot{m})\,
\ee 
and we focus, in this paper, on obtaining the large deviation function in regions where a fixed point ($m_0, 0$) looses stability in the deterministic limit, i.e., where $g(m_0, 0)$ changes sign, in order to use a perturbative framework in terms of the small parameter 
\be \label{eq:def:u0}
u_0 \equiv g(m_0, 0).
\ee
We assume $\nabla \phi = O(u_0)$ since quadratic terms in $\nabla \phi$ have to balance the contribution in $u_0 \dot{m} \partial_{\dot{m}}\phi$. %Thus, we can neglect the terms in $(\nabla\phi)^3$ and higher. %
At order $u_0$, Eq.~(\ref{eq:phi:quadrat}) reduces to 
\be \label{eq:phi:linear}
\dot{m}\partial_m\phi -V'(m) \partial_{\dot{m}}\phi =0.
\ee
The general solution of Eq.~(\ref{eq:phi:linear}) reads \cite{guislain_nonequil2023}
\be \label{eq:phi:fH}
\phi(m,\dot{m}) = f\big(H(m,\dot{m})\big) + f_0
\ee
where the function $H(m,\dot{m})$ takes a form similar to a Hamiltonian,
\be \label{eq:def:H}
H(m, \dot{m}) = \frac{\dot{m}^2}{2} + V(m)\,.
\ee
The minimum value of $V(m)$ is set to $V=0$, and $f$ is at this stage an arbitrary function, satisfying for convenience $f(0)=0$.
The constant $f_0$ in Eq.~(\ref{eq:phi:fH}) ensures that the minimal value of $\phi(m,\dot{m})$ is zero.
Contributions of order $u_0^2$ to Eq.~(\ref{eq:phi:quadrat}) yield a condition determining the derivative $f'(H)$ (see \cite{guislain_nonequil2023} for a detailed derivation)
\begin{widetext}
\be \label{eq:f'H}
f'(H) = -\frac{\int_{m_1}^{m_2}dm\, \sqrt{2[H-V(m)]}\, g\left(m, \sqrt{2[H-V(m)]}\right)}
  {\int_{m_1}^{m_2}dm \,\bigg[ D_{11}\frac{V'(m)^2}{\sqrt{2[H-V(m)]}}+2D_{12} V'(m)+D_{22} \sqrt{2[H-V(m)]}\bigg]}\,,
\ee
\end{widetext}
where $m_1$ and $m_2$ are such that
$V(m_1)=V(m_2)=H$ and $V(m)\le H$ for $m_1\le m\le m_2$.

The form Eq.~(\ref{eq:phi:fH}) of the large deviation function $\phi(m,\dot{m})$ can be interpreted as giving a statistical weight to deterministic trajectories determined by the Hamiltonian dynamics 
\be
\frac{dm}{dt}=\frac{\partial H}{\partial \dot{m}}, \qquad \frac{d\dot{m}}{dt}=-\frac{\partial H}{\partial m},
\ee
valid at order $\ve$, where the Hamiltonian $H(m, \dot{m})$ is defined in Eq.~(\ref{eq:def:H}).
Denoting $m_0$ a minimum of $V(m)$ (we assume here for simplicity that $V(m)$ has a single minimum or two symmetric minima)
the case $H=V(m_0)$ corresponds to a fixed point ($m_0, 0$) of the deterministic dynamics, whereas values
$H > V(m_0)$ correspond to closed orbits, and thus to oscillations. 
The most probable value of $H$, and thus the macroscopically observed behavior, is determined by the global minimum $H^*$ of $f(H)$.
%The local minima of $f(H)$ coincide with the fixed points or stable orbits obtained in the deterministic limit. 
Note that the method used here follows similar lines as the determination of nonequilibrium potentials in dissipative dynamical systems \cite{Graham_nonequilibrium1987, graham.tel.1984a, graham_1989}.

When the three phases meet at the critical point $(T_c, \mu_c)$, the ferromagnetic points, noted $\pm m_0$, have a small amplitude ($m_0\leq \ve$), so that $g(m_0, 0)\sim \ve$ with $\ve=(T_c-T)/T_c$. Therefore, close to the critical point, i.e., for small $\ve$, the framework described above can be used to obtain the large deviation function $\phi(m, \dot{m})$ and thus the probability density $P_N(m, \dot{m})$ for large $N$.

\section{Continuous transition from a paramagnetic phase to an oscillating phase}% at $T=T_c$ (for $\mu>\mu_c$ ou $\mu>\mu_l$)}
\label{sec:cas0}
In this section, we briefly recall and extend results presented in \cite{guislain_nonequil2023} for the continuous transition observed when $J_1$ and $J_2$ are positive, at $T=T_c$, for $\mu >\mu_c$, from a high-$T$ paramagnetic phase to a low-$T$ oscillating phase [vertical green line in Fig.~\ref{fig:phase:diagrams}(a), (b)], corresponding to a Hopf bifurcation at the deterministic limit.
Above $T_c$ ($\ve<0$), the system is in a paramagnetic phase, whereas below $T_c$ ($\ve>0$) it is in an oscillating phase.

\subsection{Transition from a paramagnetic phase to an elliptic limit cycle}

\subsubsection{Large deviation function}
\label{sec:largedev:para}

\begin{figure}[t]
    \centering
    \includegraphics{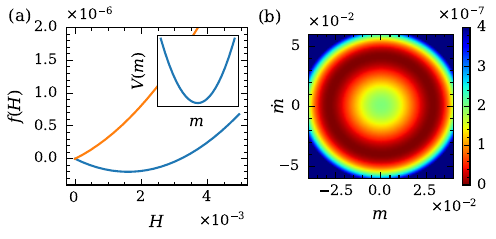}
    \caption{Large deviation function around the paramagnetic-oscillating transition. (a) Examples of $f(H)$ for $\ve=10^{-3}$ (blue curve) and for $\ve=-10^{-3}$ (orange curve). The inset represents the shape of $V(m)$. (b) Colormap of $\phi(m, \dot{m})$ in the space $(m,\dot{m})$ for $\ve=10^{-3}$. Scalings with $\ve$: $m\sim \ve^{1/2}$, $\dot{m}\sim \ve^{1/2}$, $H\sim \ve$ and $f=\phi\sim \ve^2$. Parameters: $J_1=0.6$, $J_2=0.4$, $(\mu-\mu_c)/\mu_c=1$.}
    \label{fig:cas0:fH:V:Phi}
\end{figure}
As in the paramagnetic phase $m$ and $\dot{m}$ are small, we use a power-series expansion of $V(m)$ and $g(m, \dot{m})$ in $m$ and $\dot{m}$. 
At the lowest order required to describe the transition, one has
\begin{align}
\label{eq:cas0:expansion:v}
&V(m) = \frac{\mu-\mu_l(T)}{2T^2}m^2,\\
\label{eq:cas0:expansion:g}
&g(m, \dot{m}) = a_0 \ve-a_1m^2-a_2m\dot{m}-a_3\dot{m}^2,
\end{align}
where $\ve=(T_c-T)/T_c$; $\mu_l(T)$ and $a_0$ were introduced previously in Eq.~(\ref{eq:g(m,0):small:m}) and their expressions are recalled in Appendix~\ref{appendix:values:coeff} along with the expressions of $a_1$ and $a_2$, which are all positive quantities.
Compared to Eq.~(\ref{eq:vpm:3}), we only keep the quadratic term in $V(m)$, which is positive around $T_c$ for $\mu>\mu_c=\mu_l(T_c)$.
Higher order terms are necessary only when the quadratic term is negative, in order to describe ferromagnetic order, or when the quadratic term is small, which is discussed in the next section. 
%Compared to the expression of $g(m, 0)$ given in Eq.~(\ref{eq:g(m,0):small:m}), which gives the stability of fixed point, we also need to consider terms in $\dot{m}$ as they are necessary to obtain the large deviation function [see Eq.~(\ref{eq:f'H})].
An illustration of the quadratic potential $V(m)$ is plotted in the inset of Fig.~\ref{fig:cas0:fH:V:Phi}(a). 

In the deterministic limit, the oscillating phase appears for $\ve>0$ when the paramagnetic point $(0,0)$ looses stability. The small perturbative parameter $u_0$ introduced in
Sec.~\ref{sec:method} [see Eq.~(\ref{eq:def:u0})] is proportional to $\ve$, since here $m_0=0$ and $u_0=g(0,0)=a_0 \ve$ from Eq.~(\ref{eq:cas0:expansion:g}). Hence the formalism introduced in the previous section to obtain the large deviation function $\phi(m, \dot{m})=f(H)$ can be used to describe the phase transition for small $\ve$, when the perturbative approach is valid. 

We find from Eq.~(\ref{eq:f'H}), after integration,
\be \label{eq:fH:elliptic} f(H)=-\ve aH+bH^2,\ee
with 
\be \label{eq:cas0:elliptic:a}a=\frac{T^2a_0}{T^2D_{22}+D_{11}[\mu-\mu_l(T)]}\ee
and 
\be\label{eq:cas0:elliptic:b} b=\frac{a_1 T^4+3T^2(\mu-\mu_l)a_3}{4(\mu-\mu_l)[T^2D_{22}+D_{11}(\mu-\mu_l)]}.\ee
When $\ve<0$, $f(H)$ is minimal for $H=0$, which corresponds to the paramagnetic phase. When $\ve>0$, $f(H)$ has a minimum in $H^*=\ve a/2b$, see Fig.~\ref{fig:cas0:fH:V:Phi}(a) for examples of $f(H)$ around the paramagnetic-oscillating transition.  The equation $H(m, \dot{m})=H^*$ describes an ellipse in the phase space $(m, \dot{m})$ as depicted in Fig.~\ref{fig:cas0:fH:V:Phi}(b).
The period $\tau$ of a limit cycle described by $V(m)+\dot{m}^2/2=H^*$ is obtained as
\be \label{eq:period}\tau =2\int_{-m^*}^{m^*}\frac{dm}{\dot{m}}=\sqrt{2}\int_{-m^*}^{m^*}\frac{dm}{\sqrt{H^*-V(m)}}\,,\ee
where $m^*$ is such that $H^*=V(m^*)$. Using expression (\ref{eq:cas0:expansion:v}) of $V(m)$, we find
\be \tau =\frac{2\sqrt{2}\pi T}{\sqrt{\mu-\mu_l(T)}}.\ee

\subsubsection{Order parameters}
The paramagnetic to oscillating phase transition is characterized by two order parameters, $\langle m^2\rangle$ and $\langle \dot{m}^2\rangle$, where 
$\langle x\rangle=\int dm d\dot{m} P_N(m, \dot{m})\,x(m, \dot{m})$, and where the observable $x$ stands for $m^2$ or $\dot{m}^2$ [or, more generally, any even function $x(m, \dot{m})$].
Using Eqs.~(\ref{eq:large:deviation:form}) and (\ref{eq:phi:fH}), $P_N(m,\dot{m})$ can be approximated by its properly normalized large deviation form,
\be
P_N(m,\dot{m}) \approx \frac{\exp\big[-Nf\big(H(m,\dot{m})\big)\big]}{\int dm' d\dot{m}'\, \exp\big[-Nf\big(H(m',\dot{m}')\big)\big]}.
\ee
Then from Eq.~(\ref{eq:def:H}), $\int d\dot{m}$ can be replaced by $\int dH/\sqrt{2[H-V(m)]}$, so that $\langle x\rangle$ becomes, making the integration intervals explicit:
\be \label{eq:def:av:x}
\langle x \rangle=\frac{\int_{-1}^{1} dm \int_{V(m)}^{\infty} \frac{dH}{\sqrt{H-V(m)}} \,x\, e^{-Nf(H)} }{\int_{-1}^{1} dm \int_{V(m)}^{\infty} \frac{dH}{\sqrt{H-V(m)}}e^{-Nf(H)}}\, .
\ee
From Eqs.~(\ref{eq:cas0:expansion:v}) and (\ref{eq:fH:elliptic}) one can then compute the values of $\langle m^2\rangle$ and $\langle \dot{m}^2\rangle$ using Eq.~(\ref{eq:def:av:x}).
In the limit of large system sizes, in the paramagnetic phase ($\ve<0$), $\langle \dot{m}^2\rangle=1/(|\ve|aN)$ which vanishes in the limit $N\to\infty$.
By contrast, in the oscillating phase ($\ve<0)$, $\langle \dot{m}^2\rangle=H^*\sim \ve$ is constant in the limit $N\to\infty$. At the transition, $\ve=0$, $\langle\dot{m}^2\rangle=1/\sqrt{\pi bN}$. 
For smaller system sizes $N\ll |\ve|^{-2}$, $f(H)$ can be approximated as $f(H)=bH^2$ so that one finds
\be \label{eq:scaling:mdot:finiteN}
\langle \dot{m}^2\rangle \sim N^{-1/2},
\ee
which is independent of $\ve$. %for all $\ve$.  

\subsection{Non-elliptic limit cycle near the tricritical point}
\label{sec:nonelliptic}

\subsubsection{Large deviation function}
\begin{figure}[t]
    \centering
    \includegraphics{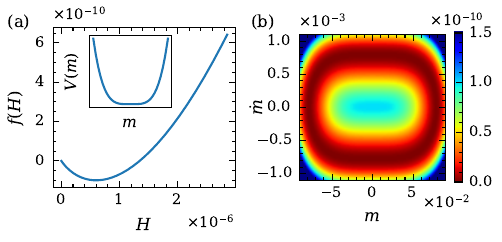}
    \caption{Large deviation function close to the tricritical point. (a) Example of $f(H)$ from Eq.~(\ref{eq:cas1:fH:mu0}); the inset represents the shape of $V(m)$. (b) Colormap of $\phi(m, \dot{m})$ in the space $(m,\dot{m})$. Note the different scalings $m\sim \ve^{1/2}$ and $\dot{m}\sim \ve$. Parameters: $J_1=0.6$, $J_2=0.4$, $\mu=\mu_c$ and $\ve=10^{-3}$.}
    \label{fig:cas0bis:fH:V:Phi}
\end{figure}

The three phases (paramagnetic, oscillating and ferromagnetic) meet at the tricritical point $T=T_c$ and $\mu=\mu_c$. We now look at how the limit cycle changes when approaching the tricritical point at $\mu=\mu_c$ for $T<T_c$. 
We have 
\be\label{eq:mul:muc:eps2} \mu_c- \mu_l(T)=T_c^2\ve^2\,,\ee
such that at $\mu=\mu_c$, the quadratic term $\propto (\mu_c-\mu_l)m^2$ in the expression of $V(m)$ scales as $\ve^2m^2$. Keeping only the quadratic term in $V(m)$, we would find $m\sim \ve^{1/2}$ by following the same reasoning as in Sec.~\ref{sec:largedev:para}.
Hence, the quadratic term becomes of order $\ve^3$ whereas the next-order term $\propto m^4$ scales as $\ve^2$, so that the assumption of neglecting the $m^4$-term in the expansion of $V(m)$ is inconsistent.
To obtain the right behavior at $\mu=\mu_c$, it is thus necessary to include the contribution of $m^4$ in $V(m)$. At $\mu=\mu_c$, we have
\be \label{eq:expansion:Vm:non-elliptic}
V(m)=\frac{\ve^2T_c^2}{2T^2}m^2+\frac{v_1(T, \mu)}{4}m^4,
\ee
with $v_1(T, \mu)$ given in Appendix~\ref{appendix:values:coeff}.
In the following, we assume that the scaling $m\sim \ve^{1/2}$ remains valid close to $\mu_c$, and we check below that the assumption is consistent.
[Note that the reason why the scaling $m\sim \ve^{1/2}$ will eventually prove valid is different from the one mentioned in the previous paragraph, which relies only on the quadratic term in $V(m)$].
Under this assumption, the term in $\ve^2 m^2 \sim \ve^3$ in Eq.~(\ref{eq:expansion:Vm:non-elliptic}) is negligible compared to the term of order $m^4$. Hence, to leading order, we can use the simple form
\be \label{eq:Vm:quartic}
V(m)=\frac{v_1(T, \mu)}{4}m^4.
\ee
We also consider that $v_1(T_c, \mu_c)>0$ as discussed in Sec.~\ref{sec:subsec:deter} (see Fig.~\ref{fig:sign:v1:J1:J2} for possible values of $J_1$, $J_2$) and that $v_1(T, \mu_c)$ remains positive for small $\ve$. An example of $V(m)$ is plotted in the inset of Fig.~\ref{fig:cas0bis:fH:V:Phi}(a). Using the same expression of $g(m, \dot{m})$ as before, Eq.~(\ref{eq:cas0:expansion:g}), we obtain from Eq.~(\ref{eq:f'H}) after %expansion in $H$ and 
integration,
\be \label{eq:cas1:fH:mu0}
f(H)=-\frac{\ve a_0}{D_{22}} H +c H^{3/2}
\ee
with
\be\label{eq:cas0:bis:c}
c=\frac{8\, \Gamma\left(\frac{3}{4}\right)^4 a_1}{5\pi^2 D_{22}\sqrt{v_1}}\,
\ee
where $\Gamma(x)=\int_0^{\infty}t^{x-1}e^{-t}dt$.
One finds that $f(H)$ has a minimum for
\be \label{eq:def:Hstar}
H^*=\left( \frac{2\ve a_0}{3c D_{22}}\right)^2
\ee
[see Fig.~\ref{fig:cas0bis:fH:V:Phi}(a)] corresponding to a limit cycle in the phase space $(m, \dot{m})$. However, $H$ is no longer quadratic in $m$, because of the quartic form (\ref{eq:Vm:quartic}) of $V(m)$.
Hence the limit cycle is no longer elliptic, see Fig.~\ref{fig:cas0bis:fH:V:Phi}(b) for a colormap of $\phi(m, \dot{m})$ in the phase space $(m, \dot{m})$. We recall that for $\mu\gg \mu_c$, one had $m\sim \dot{m}\sim \ve^{1/2}$, whereas now one finds distinct scalings $m \sim \ve^{1/2}$ and $\dot{m}\sim \ve$.
These scalings are obtained by using $H^*\sim\ve^2$ from Eq.~(\ref{eq:def:Hstar}), and the expression (\ref{eq:def:H}) of $H(m,\dot{m})$ together with the quartic form (\ref{eq:Vm:quartic}) of $V(m)$.
Finally, evaluating the oscillation period using Eq.~(\ref{eq:period}), we find
\be
\tau=\frac{4\sqrt{\pi}\Gamma\!\left(\frac{5}{4}\right)}{\Gamma\left(\frac{3}{4}\right) (v_1 H^*)^{1/4}}\,.
\ee
As $H^* \sim \ve^2$, the period diverges as $\ve^{-1/2}$ when $\ve \to 0$.

\begin{figure}[t]
    \centering
    \includegraphics{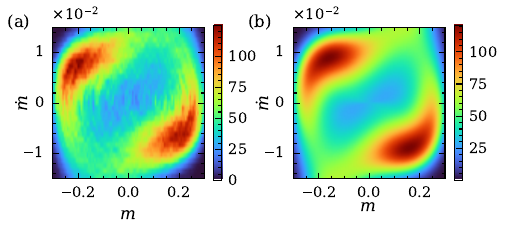}
    \caption{(a) $P_N(m, \dot{m})$ obtained from stochastic numerical simulations. (b) Theoretical $P_N(m, \dot{m})$ evaluated by including leading-order corrections, as given in Eq.~(\ref{eq:Pmdotm:higher:corr}).
    %analytically obtained up to order 2 in $\phi(m, \dot{m})$ and order 1 of $C(m, \dot{m})$: $P_N(m, \dot{m})=\exp[-N(\phi_1+\phi_2)+C_0+C_1]$.
    Parameters: $\ve=10^{-2}$, $\mu=\mu_l(T)$, $N=10^7$.}
    \label{fig:cas1_P_simu_mu0_analytic}
\end{figure}

\subsection{Comparison with stochastic simulations and need for higher order corrections}
\label{sec:cas0bis:corrections}

The method used to obtain analytically the large deviation function, developed in Sec.~\ref{sec:method}, relies on two main assumptions: $N$ is large and $\ve$ is small.  
We now compare the analytical results with numerical simulations of the stochastic spin model. 
We use the Gillespie algorithm \cite{gillespie_stochastic_2007} to simulate the stochastic dynamics with the rates given by Eq.~(\ref{eq:transition:rate}) for a time-interval $\tau$. The initial condition for the simulations are $m=0$ and $h=0$. 
In this algorithm, time-steps are of $O(N^{-1})$ such that the number of steps required to have $t=1$ is of $O(N)$. 
To observe an non-elliptic limit cycle close to $T_c$, one needs to have $N|f(H^*)|\gg1$ which corresponds to $N\ve^3\gg1$ [see Eqs.~(\ref{eq:cas1:fH:mu0}) and Eq.~(\ref{eq:def:Hstar})]. For example, a value $\ve=10^{-3}$ would require simulations with at least $N\sim10^{11}$. % to observe the phase transition. 
To obtain data with converged statistics depicting the transition, we make simulations for larger $\ve$ where the approximations made in Sec~\ref{sec:method} are no longer expected to be quantitatively valid. We now discuss the notable differences observed in numerical simulations due to larger $\ve$ values and smaller system sizes $N$.

In Fig.~\ref{fig:cas1_P_simu_mu0_analytic}, we plot $P_N(m, \dot{m})$ obtained from stochastic simulations % on the left and $P_N(m, \dot{m})$ using the Hamiltonian approximation at the lowest order (Eq.~\ref{eq:cas1:fH:mu0}) 
for $(T_c-T)/T_c=\ve=10^{-2}$ and for $\mu=\mu_l(T)$. We take $\mu=\mu_l(T)$ (instead of $\mu=\mu_c$) so that the term in front of $m^2$ in $V(m)$ is exactly zero.
Significant discrepancies are observed between the simulation results and the theoretical predictions of the perturbative approach described in Sec.~\ref{sec:method}.
From leading order calculations in $\ve$, we obtain that $\phi(m, \dot{m})=f(H)$ with $H=V(m)+\dot{m}^2/2$ which has, in particular, two consequences. First, $m$ and $\dot{m}$ are decoupled, and the symmetries $m\,\mapsto\,-m$ and $\dot{m}\,\mapsto\,-\dot{m}$ hold independently. Second, the probability $P_N(m,\dot{m})$ is uniform along the limit cycle, corresponding in a dynamical view (and in the deterministic limit) to a constant `speed' along the limit cycle. Both of these characteristics are not observed in the stochastic simulations, see Fig.~\ref{fig:cas1_P_simu_mu0_analytic}(a).

These differences come from higher order corrections, in $\ve$ %(the small parameter in control of the Hamiltonian approximation, here $\ve=(T_c-T)/T_c$) 
and in $N$ in the probability density $P_N(m, \dot{m})$. Similar corrections were studied in \cite{Graham_nonequilibrium1987} in the context of noisy dynamical systems. 
We now give an example of the first corrections for $\mu=\mu_l(T)$. The detailed steps of the derivation are given in \cite{SM}.
To perform a systematic $\ve$-expansion, we introduce rescaled variables $\tilde{H}=H/H^*$ and $x=m/m_0$ with $H^*$ given in Eq.~(\ref{eq:def:Hstar}) and $m_0=(4H^*/v_1)^{1/4}$ consistently with Eq.~(\ref{eq:def:m0}). 
At lowest order in $\ve$, from Eq.~(\ref{eq:cas1:fH:mu0}), one finds $\phi(m, \dot{m})\sim \ve^3 \tilde{\phi}_1(x, \tilde{H})$, where $\tilde{\phi}_1$ is a rescaled function independent of $\ve$. 
We introduce $C(m, \dot{m})$ the correction of $O(N^0)$ to $\ln P_N(m, \dot{m})$,
\be
\label{eq:def:C:m:dotm}
P_N(m, \dot{m}) \propto \exp[-N\phi(m, \dot{m})+C(m, \dot{m})],
\ee
and we expand $\phi(m, \dot{m})$ and $C(m, \dot{m})$ in power series of $\ve^{1/2}$ \cite{SM}, 
\be \phi(m, \dot{m})=\ve^{3}\sum_{i=1}^{\infty} \ve^{(i-1)/2}\tilde{\phi}_i(x, \tilde{H})\ee
and 
\be C(m, \dot{m})=\sum_{i=0}^{\infty} \ve^{i/2}\tilde{C}_i(x, \tilde{H}).\ee
Injecting these expressions into the master equation on $P_N(m, \dot{m})$, one finds equations on $\tilde{\phi}_i$ and $\tilde{C}_i$ at each order $i$ \cite{SM}. 
For the lowest order, we find $\phi_1(m, \dot{m})=\ve^{3 }\tilde{\phi}_1(x, \tilde{H})=f(H)$ [Eq.~(\ref{eq:cas1:fH:mu0})] and $C_0(m, \dot{m})=\ve^{1/2}\tilde{C}_0(x, \tilde{H})=\ve^{1/2}c_0$ a constant given by the normalization of $P_N(m, \dot{m})$. For $\tilde{\phi}_2(x, \tilde{H})$ and $\tilde{C}_1(x, \tilde{H})$ one finds:
\begin{widetext}
\be \begin{aligned}
\tilde{\phi}_2(x, \tilde{H})&=A\left(1-\sqrt{\tilde{H}}\right)\tilde{H} x \left[a_0\, {}_{2}F_1\left(-\frac{1}{2}, \frac{1}{4},\frac{5}{4}, x^4\right)-\frac{2}{3}\frac{D_{22}}{\alpha}x^2 \, {}_2F_1\left(-\frac{1}{2}, \frac{3}{4},\frac{7}{4}, x^4\right)\right]\,,\\
\tilde{C}_1(x, \tilde{H})&=B\tilde{H}^{1/4}\left[-6a_0x\sqrt{1-x^4}+30a_0\sqrt{v_1}x\, {}_2F_1\left(-\frac{1}{2}, \frac{1}{4}, \frac{5}{4}, x^4\right)-8\frac{D_{22}}{\alpha}x^3{}_2F_1\left(\frac{1}{2}, \frac{5}{4}, \frac{7}{4}, x^4\right)\right]\,,
\end{aligned} \ee
\end{widetext}
where $\alpha$, $A$ and $B$ depend on $v_1$, $a_0$ and $a_1$ and are given in Appendix~\ref{appendix:values:coeff}, and ${}_2F_1(a, b, c, x)$ denotes the hypergeometric function. 
The correction $\tilde{\phi}_2$ changes the orientation and the shape of the limit cycle, whereas the correction $\tilde{C}_1$ breaks the uniformity of the probability $P_N(m, \dot{m})$ along the limit cycle. 
We plot in Fig.~\ref{fig:cas1_P_simu_mu0_analytic} the following expression of $P_N(m, \dot{m})$ that includes leading corrections,
\be \label{eq:Pmdotm:higher:corr}
P_N(m, \dot{m})=\exp[-N(\phi_1+\phi_2)+C_0+C_1]
\ee
with the definitions $\phi_i(m, \dot{m})=\ve^{3+(i-1)/2}\tilde{\phi_i}(x, \tilde{H})$ and $C_i(m, \dot{m})=\ve^{i/2}\tilde{C}_i(x, \tilde{H})$.
The main features of the probability density obtained from the simulations are captured by these leading corrections.

%%%%%%%%%%%%%%%%%%%%%%%%%%%%%%%%%%%%%%%%%%%%%%%%%%
%%%%%%%%%%%%%%%%%%%%%%%%%%%%%%%%%%%%%%%%%%%%%%%%%%

\section{Type-I discontinuous transition between ferromagnetic and oscillating phases}
\label{sec:cas1}
In this section, we investigate the properties, near the tricritical point $(T_c, \mu_c)$, of the ferromagnetic to oscillating phase transition where a limit cycle appears around the ferromagnetic points, called coexistence of Type I. This case corresponds to $v_1(T_c, \mu_c)>0$, see Fig.~\ref{fig:sign:v1:J1:J2} as well as the phase diagram of Fig.~\ref{fig:phase:diagrams}(a) and the trajectories displayed in Figs.~\ref{fig:phase:diagrams}(c) and \ref{fig:phase:diagrams}(e). %The figures in this section are obtained for $J_1=0.6$ and $J_2=0.4$.%, and if not stated other wise, the data are obtained from numerical integration of Eq.~\ref{eq:f'H} and $V(m)$, $g(m, \dot{m})$ from Eq.~?....

\subsection{Large deviation function and phase diagram}

%\subsubsection{On the validity of the method of Sec.~\ref{sec:method}}
\subsubsection{Validity of the perturbative approach}

We start from the generic expansion of the potential $V(m)$ given in Eq.~(\ref{eq:vpm:3}), recalled here for clarity,
\be \label{eq:cas1:expansion:v}V(m)=\frac{\mu-\mu_l(T)}{2T^2}m^2+\frac{v_1(T, \mu)}{4}m^4+V_0,\ee
where $V_0$ is chosen such that $V(m)\geq 0$ and its minimal value is zero. An example of $V(m)$ for $\mu<\mu_l(T)$ is given in Fig.~\ref{fig:cas1:fH:V:Phi}.
We recall that $g(m, \dot{m})$ is given in Eq.~(\ref{eq:cas0:expansion:g}).
As discussed in Sec.~\ref{sec:tricrit:deter}, ferromagnetic points $m_0^2=(\mu_l(T)-\mu)/T^2v_1$ exist for $\mu<\mu_l(T)$, and are locally stable for $\mu\leq\mu_F$ with $\mu_F(T)=\mu_l(T)-\ve a_0T^2v_1/a_1$. 
In this section, we focus on the region where the ferromagnetic points loose stability ($\mu \approx \mu_F$), thus:
\be\label{eq:cas1:mu:mul:eps} \mu_l(T)-\mu\sim \ve\, \ee
and one has $m_0^2 \sim \ve$ and thus $u_0=g(m_0, 0)\sim \ve$. For small $\ve=(T_c-T)/T_c$, the main assumption made in Sec.~\ref{sec:method}, i.e., that $u_0$ is small, is verified, and we can use the method developed in this section to obtain the large deviation function and study the phase transition from a ferromagnetic phase to an oscillating phase.

\subsubsection{Typical $f(H)$ and phase diagram}%Numerical computation of $f(H)$}

\begin{figure}[t]
    \centering
    \includegraphics{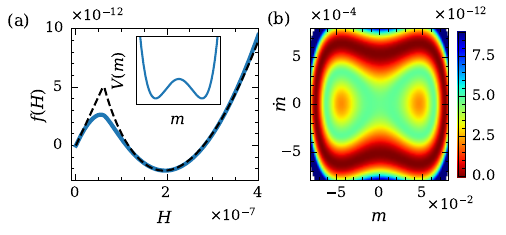}
    \caption{Large deviation function for Type-I discontinuous transition between ferromagnetic and oscillating phases. (a) $f(H)$ determined numerically from Eq.~(\ref{eq:f'H}) (full line); the dashed line corresponds to local approximations described in Sec.~\ref{sec:cas1:subsec:local:expressions}. Inset: shape of $V(m)$ with two minima. (b) Colormap of $\phi(m, \dot{m})$ in the plane $(m,\dot{m})$. Note the different scalings $m\sim \ve^{1/2}$ and $\dot{m}\sim \ve$.
    Parameters: $J_1=0.6$, $J_2=0.4$, $(\mu-\mu_c)/\mu_c=-3.18\times10^{-5}$ and $\ve=10^{-3}$.}
    \label{fig:cas1:fH:V:Phi}
\end{figure}

In general, except for particular cases as the one described in the previous section, one cannot obtain explicit analytical expressions of $f'(H)$ from Eq.~(\ref{eq:f'H}), and one needs to perform a numerical integration of the integrals in Eq.~(\ref{eq:f'H}) to determine $f(H)$. 
An example of $f(H)$, for $\mu\leq \mu_l(T)$, numerically obtained from Eq.~(\ref{eq:f'H}), is plotted in Fig.~\ref{fig:cas1:fH:V:Phi}(a). We observe that $f(H)$ has two local minima: one in $H=0$ corresponding to the ferromagnetic points $m=m_0$ and $\dot{m}=0$ [since $V(m_0)=0$], and one for $H=H^*>0$ corresponding to a limit cycle in the phase space $(m, \dot{m})$.
We numerically obtain (not shown) that 
\be \label{eq:cas1:H*:eps2} H^*\sim \ve^2.\ee
An example of colormap of $\phi(m,\dot{m})=f\big(H(m,\dot{m})\big)$ in the phase space $(m,\dot{m})$ is displayed in Fig.~\ref{fig:cas1:fH:V:Phi}(b). Here, the most stable phase is the oscillating phase as $f(0)>f(H^*)$. Contrary to Sec.~\ref{sec:cas0}, no analytical expression of $H^*$ is available in the present case.

\begin{figure}[t]
    \centering
    \includegraphics{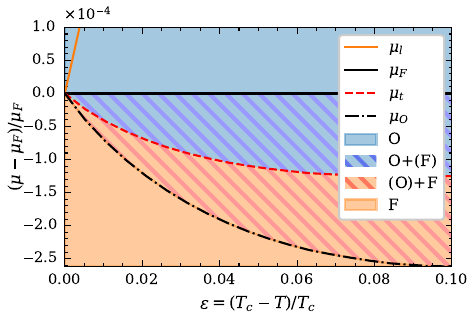}
    \caption{Close up of the phase diagram of Fig.~\ref{fig:phase:diagrams}(a) near the tricritical point, for $J_1=0.6$ and $J_2=0.4$, in the reduced parameters $\ve=(T_c-T)/T_c$ and $(\mu-\mu_F(T))/\mu_F(T)$, where $\mu_F$ is defined in Eq.~(\ref{eq:def:muF}) and $(\mu_c-\mu_F(T))/\mu_c\sim \ve$. O corresponds to the oscillating phase, F to the ferromagnetic phase. In the hatched area, both phases are locally stable. The most stable phase, given by the global minima of $f(H)$, is either the ferromagnetic phase [(O)+F] or the oscillating phase [O+(F)], separated by the transition line $\mu_t$.
    $\mu_l(T)$ corresponds to the limit of existence of the ferromagnetic points in the deterministic limit, $\mu_F$ to their linear stability limit, and $\mu_O$ to the limit of existence of the oscillating state at deterministic level.}
    \label{fig:phase_diagram:1}
\end{figure}

The transition from the ferromagnetic phase to the oscillating phase takes place when $f(0)=f(H^*)$; we note $\mu_t(T)$ the value of $\mu$ at the transition. The value of $H$ jumps from $H=0$ to the nonzero value $H^*$ at the transition, meaning that the latter is discontinuous. We obtain numerically that $(\mu_c-\mu_t)/\mu_c\sim \ve$ where $\ve=(T_c-T)/T_c$ [see Fig.~\ref{fig:phase_diagram:1} with $(\mu_c-\mu_F)/\mu_c\sim \ve$]. 

From the numerical determination of $f(H)$, one obtains a phase diagram in the space $(\ve, \mu)$ with the determination of the different phases: ferromagnetic phase (F), oscillating phase (O) or the phase where both coexist, with one being more stable than the other. A close up on the phase diagram near the tricritical point for $J_1=0.6$ and $J_2=0.4$ is plotted in Fig.~\ref{fig:phase_diagram:1}, where we represent $(\mu-\mu_F)/\mu_F$, with $\mu_F$ given in Eq.~(\ref{eq:def:muF}), in order to visualize the different phases. The limits of existence of the ferromagnetic and oscillating states can also be obtained in the deterministic limit, but for the determination of $\mu_t$ (which characterizes the most stable phase) it is necessary to consider finite system sizes using the large deviation approach.

\subsubsection{Local analytical expressions of $f(H)$}
\label{sec:cas1:subsec:local:expressions}

In most cases, keeping only the first orders of the series expansions of $V(m)$ and $g(m, \dot{m})$ is not enough to obtain an analytical expression of $f'(H)$. Still, local approximations can be obtained.
Near a minimum $m_0$ of $V(m)$, a quadratic expansion of $V$ gives
\be \label{eq:cas1:fH:ferro}
f(H) = f_F(H) \equiv -\frac{g(m_0, 0)}{D_{22}+2v_1m_0^2D_{11}}H,
\ee
where $f_F(H)$ stands for the local approximate expression of $f(H)$ in the ferromagnetic state.
We recover that the point $(m, \dot{m})=(m_0,0)$ is stable when $g(m_0, 0)<0$.
This expression of $f(H)$ is valid for small $H$ only. 
For $H \gg H^*$, we recover that $f(H)\approx c H^{3/2}$, with $c$ defined in Eq.~(\ref{eq:cas0:bis:c}).
For intermediate values of $H$ ($H\sim H^*$), we do not have an analytical expression of $f(H)$. However the regime $H\gg H^*$ is similar to the one obtained for $\mu=\mu_c$, which suggests that the form of $f(H)$ obtained for $\mu\approx \mu_c$ in Eq.~(\ref{eq:cas1:fH:mu0}) remains approximately valid up to a redefinition of coefficient values. One can perform a local fit of the form $f(H)=f_O(H)+f(H^*)$ with
\be\label{eq:cas1:fH:fit}
f_O(H)=\tilde{c} \left(H^{3/2} -\frac{3}{2}\sqrt{H^*}H + \frac{1}{2}H^{*3/2}\right),   %+f(H^*)
\ee
where the parameters $\tilde{c}$ and $H^*$ are fitted on the numerically evaluated $f(H)$ to get a local approximation of $f(H)$ near $H^*$ (see Fig.~\ref{fig:cas1:fH:V:Phi} for an example of a fit of $f(H)$ close to its minimum). The functional form (\ref{eq:cas1:fH:fit}) provides a reasonable description of the large $H$ behavior of $f(H)$, and is more accurate than a simple parabolic fit around the minimum $H=H^*$.

%\subsection{Large-$N$ and low-$N$ scalings at the transition close to the tricritical point}
\subsection{Scalings of order parameters with system size at the transition}

\subsubsection{Large-$N$ scaling at $\mu=\mu_t$}
Using the two local approximations of $f(H)$ given in Eqs.~(\ref{eq:cas1:fH:ferro}) and (\ref{eq:cas1:fH:fit}), we study the behaviors of $\langle m^2\rangle$ and $\langle \dot{m}^2\rangle$ in the large-$N$ limit when approaching the critical point $\ve=0$ where the three phases meet.
In the ferromagnetic phase ($\mu<\mu_t$), using Eq.~(\ref{eq:cas1:fH:ferro}) and $g(m_0, 0)=a_0\ve -a_1m_0^2$, one finds in the large-$N$ limit, 
\begin{align} \label{eq:cas1:tricritical:m2:md2:F}\langle m^2\rangle&=m_0^2=\frac{\mu_l(T)-\mu}{T^2v_1}, \\
 \langle\dot{m}^2\rangle&=\frac{D_{22}+2v_1m_0^2D_{11}}{(a_1 m_0^2-a_0\ve )N}. \end{align}
We recover the results of the deterministic limit for $N\to\infty$, $\langle m^2\rangle=m_0^2\sim \ve$ and $\langle \dot{m}^2\rangle =0$. For large but finite $N$, we obtain that  $\langle \dot{m}^2\rangle\sim \ve^{-1}N^{-1}$. 
In the oscillating phase ($\mu>\mu_t$), $f(H)$ is minimal in $H^*$, so that for large enough $N$ we can replace $e^{-Nf(H)}$ by $\delta(H-H^*)$
%\be \underset{N\to\infty}{\lim} e^{-Nf(H)}= \delta(H-H^*),\ee
in Eq.~(\ref{eq:def:av:x}), yielding
\begin{align} \label{eq:cas1:tricritical:m2:md2:O}
\langle m^2\rangle &= \frac{\int_{-m^*}^{m^*} dm \frac{m^2}{\sqrt{H^*-V(m)}} }{\int_{-m^*}^{m^*} dm \frac{1}{\sqrt{H^*-V(m)}}},\\
\langle \dot{m}^2\rangle &= \frac{\int_{-m^*}^{m^*} dm \sqrt{2(H^*-V(m))}}{\int_{-m^*}^{m^*} dm \frac{1}{\sqrt{2(H^*-V(m))}}},
\end{align} 
where $m^*$ is such that $H^*=V(m^*)$.
Both $\langle m^2\rangle$ and $\langle \dot{m}^2\rangle$ reach constant values at large $N$. 
We obtain numerically that $\langle m^2\rangle\sim \ve$ and $\langle \dot{m}^2\rangle\sim \ve^2$, which are the same scalings as the one observed for the nonelliptic limit cycle for $\mu=\mu_c$
(see Sec.~\ref{sec:nonelliptic}).

\subsubsection{Moderate-$N$ scaling at $\mu=\mu_t$}
\label{sec:cas1:lowN}

\begin{figure}[t]
    \centering\includegraphics{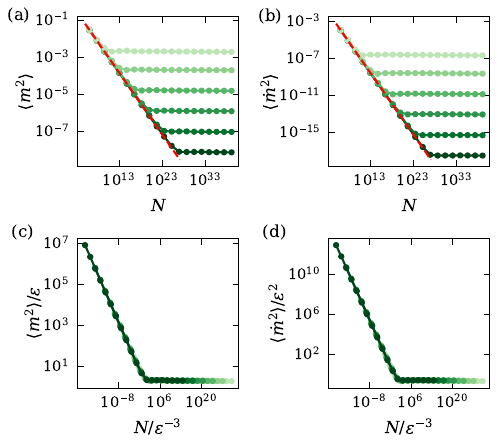}
    \caption{Order parameters $\langle m^2\rangle$ and $\langle \dot{m}^2\rangle$ as a function of system size $N$, for $\ve\in[10^{-3}, 10^{-4}, 7\times10^{-6}, 6\times10^{-7}, 4.6\times10^{-8}, 3.6\times10^{-9}$] at $\mu=\mu_t(\ve)$. Order parameters are evaluated numerically from Eqs.~(\ref{eq:f'H}) and ~(\ref{eq:def:av:x}). % the approximation of $f(H)$ given in Eqs.~(\ref{eq:cas1:fH:ferro}) and (\ref{eq:cas1:fH:fit}). %$1-\mu_t(\ve)/\mu_c\in[3.25\times10^{-5}, 3.22\times10^{-6}, 2.5\times10^{-7}, 1.9\times10^{-8}, 1.48\times10^{-9}, 1.15\times10^{10}]$. . 
    (a) $\langle m^2\rangle$ vs $N$. (b) $\langle \dot{m}^2\rangle$ vs $N$.
    The red dashed line corresponds to the scaling predictions for moderate values of $N$ from Eqs.~(\ref{eq:cas1:m2:lowN}) and (\ref{eq:cas1:md2:lowN}), with $\langle m^2\rangle \sim N^{-1/3}$ and $\langle \dot{m}^2\rangle \sim N^{-2/3}$. (c) $\langle m^2\rangle/\ve$ and (d) $\langle \dot{m}^2\rangle\ve^{2}$ vs the rescaled system size $N/\ve^{-3}$, showing data collapse. Parameters: $J_1=0.6$, $J_2=0.3$.}
    \label{fig:cas1:m2:md2:eps:rescaled}
\end{figure}

Unlike for large values of $N$, for intermediate values of $N$, $\langle m^2\rangle$ and $\langle \dot{m}^2\rangle$ are found not to depend much on $\ve$ and on $\mu$.

In Figs.~\ref{fig:cas1:m2:md2:eps:rescaled}(a) and \ref{fig:cas1:m2:md2:eps:rescaled}(b), we plot $\langle m^2\rangle$ and $\langle \dot{m}^2\rangle$ for different $\ve$ at the transition, $\mu=\mu_t(\ve)$. We observe that $\langle m^2\rangle \sim N^{-1/3}$ and $\langle \dot{m}^2\rangle\sim N^{-2/3}$ for moderate $N$ values. These scaling behaviors can be understood as follows.
The expression of the average $\langle x\rangle$ contains an integral over $H$ with the factor $\exp[-Nf(H)]$, see Eq.~(\ref{eq:def:av:x}).
For moderate $N$, the integral is dominated by the $H^{3/2}$-term in $f(H)$. This can be justified by performing the change of variable
$z=N H^{3/2}$ in the integral. One then finds that higher order powers of $H$ in $f(H)$ are negligible when $N \gg 1$, while linear contributions in $H$ are also negligible as long as $\sqrt{H^*} N^{1/3} \ll 1$, i.e., $N \ll (H^*)^{-3/2}$, where $H^*$ is small for small $\ve$
[see Eq.~(\ref{eq:cas1:H*:eps2})]. The integral is then dominated by the contribution of the region $z\sim 1$, i.e., which corresponds to $H\gg H^*$.
In addition, Eq.~(\ref{eq:def:av:x}) also contains an integral over $m$.
Due to the factor $[H-V(m)]^{-1/2}$ in the integrals, values of $m$ which contribute the most are where $V(m)\approx v_1 m^4/4$. 
In a similar way as above, this can be justified by performing the change of variable $z'=m/H^{1/4}$ in the integral with the expression of $V(m)$ given in Eq.~(\ref{eq:cas1:expansion:v}). One then finds that higher order powers of $m$ are negligible when $H\ll 1$ and the quadratic order is negligible when $v_0(\mu-\mu_l)\ll H^{1/4}$ which is verified for $H\gg H^*$ as $\mu-\mu_l\sim \ve$ [Eq.~(\ref{eq:cas1:mu:mul:eps})] and $H^*\sim \ve^2$ [Eq.~(\ref{eq:cas1:H*:eps2})].
Using these two approximations on $f(H)$ and $V(m)$, we find:
\be \label{eq:cas1:m2:lowN} \langle m^2\rangle=\frac{4\Gamma\left(\frac{5}{6}\right)\Gamma\left(\frac{3}{4}\right)^4}{\pi^{5/2}\sqrt{v_1}c^{1/3}}N^{-1/3},\ee
and 
\be \label{eq:cas1:md2:lowN}\langle \dot{m}^2\rangle= \frac{4\Gamma\left(\frac{7}{6}\right)}{3\sqrt{\pi}c^{2/3}}N^{-2/3},\ee
where $c$ is given in Eq.~(\ref{eq:cas0:bis:c}).
We plot these quantities in red in Figs.~\ref{fig:cas1:m2:md2:eps:rescaled}(a) and \ref{fig:cas1:m2:md2:eps:rescaled}(b) alongside the numerical values obtained from Eq.~(\ref{eq:def:av:x}) using the numerical evaluation of $f(H)$ from Eq.~(\ref{eq:f'H}).

We note $N^*$ the crossover value of $N$ between the moderate- and large-$N$ regimes. 
The crossover takes place when the value of $\langle m^2\rangle \sim N^{-1/3}$ in the moderate-$N$ approximation is comparable to the one in the large-$N$ approximation, $\langle m^2\rangle \sim \ve$. One thus finds that $N^*$ behaves as
\be N^* \sim \ve^{-3}. \ee
[Note that, according to the integration argument above, $N^* \sim (H^*)^{-3/2}$, implying $H^*\sim \ve^2$, consistently with Eq.~(\ref{eq:cas1:H*:eps2})].
A similar argument for $\dot{m}$ yields the same scaling for $N^*$: in the moderate-$N$ regime ($N\ll N^*$), $\langle \dot{m}^2\rangle \sim N^{-2/3}$, while in the large-$N$ approximation ($N\gg N^*$), one has $\langle \dot{m}^2\rangle \sim \ve^2$ in the oscillating phase and $\langle \dot{m}^2\rangle \sim \ve^{-1}N^{-1}$ in the ferromagnetic phase, which both give a crossover at $N^*\sim\ve ^{-3}$.

Focusing on the oscillating phase, these scaling behaviors of $\langle m^2\rangle$ and $\langle \dot{m}^2\rangle$ can be encompassed into two scaling functions
\be
\langle m^2\rangle = \ve \, \mathcal{F}_{m}(\ve^3 N), \quad \langle \dot{m}^2\rangle = \ve^2 \, \mathcal{F}_{\dot{m}}(\ve^3 N),
\ee
with asymptotic behaviors $\mathcal{F}_{m}(x) \sim x^{-1/3}$ and $\mathcal{F}_{\dot{m}}(x) \sim x^{-2/3}$ for $x\to 0$, while both functions go to constant values for  $x\to \infty$.
Figs.~\ref{fig:cas1:m2:md2:eps:rescaled}(c) and \ref{fig:cas1:m2:md2:eps:rescaled}(d) show the data collapse obtained by plotting the rescaled variables $\langle m^2\rangle/\ve$ and $\langle \dot{m}^2\rangle/\ve^2$ versus the rescaled system size $N/\ve^{-3}$, for different values of $\ve$ at $\mu=\mu_t(\ve)$.

%\subsection{Phase coexistence for a finite system size}
\subsection{Detailed study of the crossover regime}
\label{sec:crossover}
%influence of the metastable phase 

We reported above two distinct scaling regimes $N \ll N^*$ and $N \gg N^*$ of the observables $\langle m^2\rangle$ and $\langle \dot{m}^2\rangle$ as a function of system size $N$ for $\mu=\mu_t(\ve)$, and we identified the scaling with $\ve$ of the crossover size $N^*$. We now investigate in more details the behavior of these observables in the crossover regime $N\sim N^*$, now focusing on the effect of the variations of $\mu$ close to the transition value $\mu_t$, for a fixed $\ve$.
We find in particular that in the crossover regime, $\langle m^2\rangle$ has a non-monotonic behavior as a function of $N$, whose details significantly depend on $\mu$. The behavior of $\langle \dot{m}^2\rangle$, while monotonic as a function of $N$, is found to strongly depend on $\mu$.

\subsubsection{Influence of $\mu$ on the crossover regime}

\begin{figure}[t]
    \centering
    \includegraphics{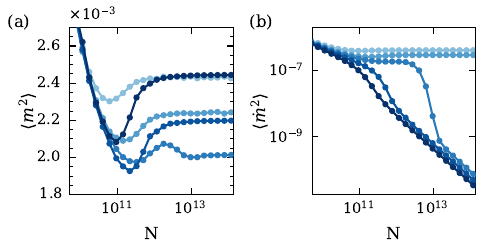}
    \caption{(a) $\langle m^2\rangle$ and (b) $\langle \dot{m}^2\rangle$ as a function of system size $N$, for different values of $\mu$ close to $\mu_c$, corresponding to $(\mu_c-\mu)/\mu_c\in[4.0, 3.6, 3.3, 3.1, 2.5]\times10^{-5}$
    %$1-\mu/\mu_l\in[3.98, 3.58, 3.28, 3.08, 2.48]\times 10^{-5}$ 
    from darker to lighter colors. The transition between ferromagnetic and oscillating states takes place at $\mu_t$ given by $(\mu_c-\mu_t)/\mu_c\approx 3.2\times 10^{-5}$. Parameters: $J_1=0.6$, $J_2=0.4$, and $\ve=10^{-3}$. Order parameters are evaluated numerically in the same way as in Fig.~\ref{fig:cas1:m2:md2:eps:rescaled}.} 
    \label{fig:cas1:m2:md2:mu}
\end{figure}

As mentioned above, the observables $\langle m^2\rangle$ and $\langle \dot{m}^2\rangle$ are seen to have a very weak dependence on $\ve$ and $\mu$ in the moderate-$N$ regime.
For large $N$, the value of $\langle \dot{m}^2\rangle$ is found to be significantly different in the oscillating phase $(\mu>\mu_t)$ where $\langle \dot{m}^2\rangle\sim \ve^2$ and in the ferromagnetic phase $(\mu<\mu_t)$ where $\langle \dot{m}^2\rangle \sim \ve^{-1}N^{-1}$. In contrast, the value of $\langle m^2\rangle$ is similar in both phases, with $\langle m^2\rangle \sim \ve$. 
%We now investigate the behavior of these two observables in the crossover regime $N\sim N^*$.
In Fig.~\ref{fig:cas1:m2:md2:mu}, $\langle m^2\rangle$ and $\langle\dot{m}^2\rangle$ are plotted as a function of system size $N$ in the crossover regime $N\sim N^*$, for different values of $\mu$ across the transition, keeping $\ve$ fixed. %The limit of large $N$ expected is verified: in the ferromagnetic phase $\langle m^2\rangle$ is nonzero and $\langle \dot{m}^2\rangle \sim N^{-1}$ and in the oscillating phase both are nonzero. 
%For moderate values of $N$, both $\langle m^2\rangle$ and $\langle \dot{m}^2\rangle$ are almost the same for different values of $\mu$ and decrease with $N$, as explained in Sec.~\ref{sec:cas1:lowN}.
After an initial decay for $N \ll N^*$, we observe that $\langle m^2\rangle$ slightly increases before reaching a constant value, as expected in the limit $N\to\infty$.  For some values of $\mu$, like for $(\mu-\mu_c)/\mu_c=-3.3\times10^{-5}$, a second decay is observed before reaching the asymptotic constant value.
The behavior of $\langle \dot{m}^2\rangle$ is significantly different from that of $\langle m^2\rangle$
as the large-$N$ limit yields two different behaviors in the ferromagnetic or in the oscillating phase. We observe that for $\mu<\mu_t$, $\langle \dot{m}^2\rangle$ first reaches a plateau for a significant range of $N$, before steeply decreasing to eventually reach the large-$N$ scaling $\langle \dot{m}^2\rangle \sim N^{-1}$.

\subsubsection{Interpretation as a finite-size phase coexistence}

The observed non-trivial behaviors can be given a simple interpretation in terms of finite-size phase coexistence and metastability. For a finite-size system, a metastable state has a finite probability to be visited, and this probability decreases exponentially with system size.
Based on this idea, we introduce a simple decomposition of average values into contributions of each phase, and show that such a decomposition is sufficient to account for most of the observed behaviors.

\begin{figure}[t]
    \centering
    \includegraphics{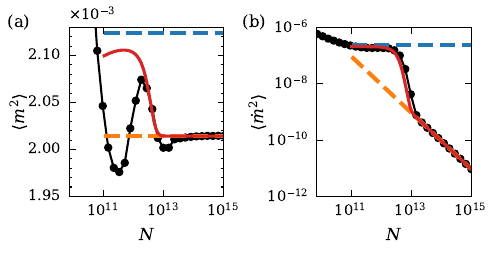}%\includegraphics{cas1_m2_approx.pdf}
    \caption{(a) $\langle m^2\rangle$ and (b) $\langle \dot{m}^2\rangle$ vs system size $N$. In black, data from Fig.~\ref{fig:cas1:m2:md2:mu} for $(\mu_c-\mu)/\mu_c=3.3\times 10^{-5}$. The dashed lines correspond to $\langle m^2\rangle$ and $\langle \dot{m}^2\rangle$ computed for $f=f_H$ [Eq.~(\ref{eq:cas1:fH:ferro})] (orange) and for $f=f_O$ [Eq.~(\ref{eq:cas1:fH:fit})] (blue). The plain red lines correspond to $\langle x(m, H)\rangle_{\mathrm{approx}}$ defined in Eq.~(\ref{eq:cas1:approx}). Here, $f(H^*)=1.1\times 10^{-12}$. }
    \label{fig:cas1:m2:md2:approx}
\end{figure}

Starting from the expression of an average observable given in Eq.~(\ref{eq:def:av:x}),
we split the semi-axis $H \ge 0$ into two regions, separated by the value $H_0$ corresponding to the local maximum of $f(H)$. 
For small $H$, i.e., near the ferromagnetic points, $f(H)$ is linear [see Eq.~(\ref{eq:cas1:fH:ferro})].
For $H$ around $H^*$, we write $f(H)=f_O(H)+f(H^*)$ with $f_O$ from Eq.~(\ref{eq:cas1:fH:fit})
where $\tilde{c}$, $H^*$ and $f(H^*)$ are parameters fitted on the numerically evaluated $f(H)$. 
We consider $N$ large enough so that the integration interval can be extended to the entire real axis due to the rapidly decaying factor $\exp[-Nf(H)]$.
Then, for any quantity $x(m, H)$, we use the approximate expression of the average value $\langle x(m, H)\rangle$,
\be \label{eq:cas1:approx}
\langle x(m, H)\rangle_{\mathrm{approx}} = \frac{\langle x\rangle_F +\langle x\rangle_O C_1\sqrt{N} e^{-Nf(H^*)} }{1 + C_1\sqrt{N} e^{-Nf(H^*)}}
\ee
with
\be \label{eq:cas1:approx:C}
C_1=\frac{1}{\sqrt{N}}\frac{\int_{-1}^{1} dm \int_{V(m)}^{\infty}dH \frac{1}{\sqrt{H-V(m)}}e^{-Nf_O(H)}}{\int_{-1}^{1} dm \int_{V(m)}^{\infty}dH \frac{1}{\sqrt{H-V(m)}}e^{-Nf_F(H)}}.
\ee
In Eq.~(\ref{eq:cas1:approx}), $\langle x\rangle_F$ (resp.~$\langle x\rangle_O$) corresponds to the `pure-state' average computed in the ferromagnetic
state with $f(H)=f_F(H)$ defined in Eq.~(\ref{eq:cas1:fH:ferro}) [resp.~$f(H)=f_O(H)$ in the oscillating state, see Eq.~(\ref{eq:cas1:fH:fit})].
We obtain
\be
C_1=-\frac{4g(m_0, 0)m_0\sqrt{v_1}H^{*1/4}}{\sqrt{3\pi \tilde{c}}(D_{22}+2v_1m_0^2D_{11})} \int_{m*}^{1}\frac{dm}{\sqrt{H^*-V(m)}},
\ee  
with $m^*$ such that $V(m^*)=H^*$. 
The oscillating phase has a contribution weighted with the factor $C_1\sqrt{N}e^{-Nf(H^*)}/(1+C_1\sqrt{N}e^{-Nf(H^*)})$. When $f(H^*)>0$ (i.e., the ferromagnetic phase is the most stable one), the contribution of the oscillating phase disappears at large $N$ but is non-negligible for $N\sim f(H^*)$.
In Fig.~\ref{fig:cas1:m2:md2:approx}, the `pure-state' averages $\langle m^2\rangle$ and $\langle \dot{m}^2\rangle$ evaluated using either $f=f_H$ for the ferromagnetic state, or $f=f_O$ for the oscillating state, as well as the `mixed-state' approximation Eq.~(\ref{eq:cas1:approx}) are compared to the values of $\langle m^2\rangle$ and $\langle \dot{m}^2\rangle$ obtained from the numerically evaluated $f(H)$ (same data as on Figs.~\ref{fig:cas1:m2:md2:eps:rescaled} and \ref{fig:cas1:m2:md2:mu}). %for one value of $\mu$.
$\langle m^2\rangle$ increases due to the influence of the oscillating phase as $\langle m^2\rangle$ is higher for $f=f_O$ than for $f=f_H$;
then for larger $N$, it decreases to its expected value in the ferromagnetic state.
When $f(H^*)<0$ (i.e., the oscillating phase is the most stable one), we observe a monotonous increase between the moderate-$N$ decay the asymptotic large-$N$ value (see Fig.~\ref{fig:cas1:m2:md2:mu}). 
The influence of the oscillating phase is even more pronounced for $\langle \dot{m}^2\rangle$, as we observe that for moderate $N$, $\langle \dot{m}^2\rangle$ is almost constant and equal to the value expected in the oscillating state (obtained using $f=f_O$), before eventually steeply decreasing when $Nf(H^*)\sim 1$.

\subsubsection{Comparison with stochastic simulations}
\begin{figure}[t]
    \centering
    \includegraphics{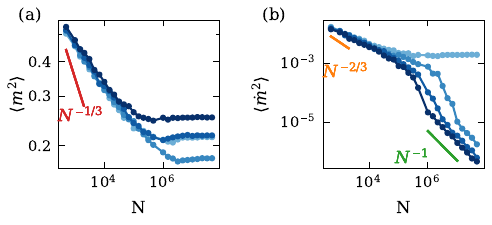}
    \caption{Comparison with stochastic simulations of the spin model. (a) $\langle m^2\rangle$ and (b) $\langle \dot{m}^2\rangle$ obtained from stochastic numerical simulations, as a function of $N$, for $(\mu_c-\mu)/\mu_c\in[4.44, 4.04, 3.64, 3.24]\times10^{-3}$ %1-\mu/\mu_l\in[2.4, 2.0, 1.6, 1.2]\times 10^{-3}$ %$\mu\in[1.0055, 1.0059, 1.0063, 1.0067]$
    (from darker to lighter colors). Parameters: $J_1=0.6$, $J_2=0.4$, and $\ve=10^{-1}$.}
    \label{fig:cas1:m2:md2:simu}
\end{figure}
In Fig.~\ref{fig:cas1:m2:md2:simu}, $\langle m^2\rangle$ and $\langle\dot{m}^2\rangle$ obtained from stochastic simulations of the spin model are plotted for different system sizes $N$, for $\ve =(T_c-T)/T_c=10^{-1}$.
Qualitatively, the moderate- and large-$N$ regimes are visible on the data. However, the decay of $\langle m^2\rangle$ in the moderate-$N$ regime is significantly slower than the theoretically predicted behavior $\langle m^2\rangle \sim N^{-1/3}$. This is most likely due to the fact that $\ve$ is not small enough to enter the asymptotic low-$\ve$ regime, as discussed below.
The moderate-$N$ decay of $\langle \dot{m}^2\rangle$ seems better described by the theoretical prediction $\langle \dot{m}^2\rangle \sim N^{-2/3}$,
although significant deviations are also visible. In the large-$N$ regime, $\langle \dot{m}^2\rangle$ reaches a constant value $\sim \ve^2$ in the oscillating phase ($\mu>\mu_t$), or decreases as $\ve^{-1}N^{-1}$ in the ferromagnetic phase ($\mu<\mu_t$). The transition between the moderate- and large-$N$ regimes takes place around $N^*\approx 10^3 \approx \ve^{-3}$, as expected.
%The qualitative behavior is correctly predicted by the analytical approximation. However, we observe slightly smaller exponents for the power laws for moderate $N$ values than the expected $\langle m^2\rangle\sim N^{-1/3}$ and $\langle \dot{m}^2\rangle \sim N^{-2/3}$. 

To understand the discrepancies found between stochastic simulations data and theoretical predictions, we note that the main approximation made to obtain the power laws $\langle m^2\rangle\sim N^{-1/3}$ and $\langle \dot{m}^2\rangle \sim N^{-2/3}$ is the assumption $\sqrt{H^*}N^{1/3}\ll 1$ (see Sec.~\ref{sec:cas1:lowN}), which is not valid here for moderate $N$ values. Moreover, we showed in Sec.~\ref{sec:cas0bis:corrections} that there are discrepancies between the theory and the simulations for low values of $\ve$ when $N$ is not large enough. We discuss this issue in more details in the next subsection. %Sec.~\ref{sec:cas1:discussion:limit:ve}.

\subsubsection{Discussion on the low-$\ve$ and large-$N$ approximations}

\label{sec:cas1:discussion:limit:ve}
\begin{figure}[t]
    \centering
    \includegraphics{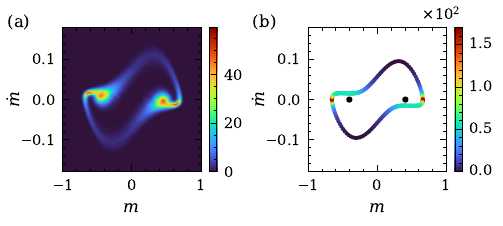}
    \caption{(a) $P_N(m, \dot{m})$ obtained from stochastic simulations of the spin model for $N=5\times10^5$. (b) Trajectories ($m(t), \dot{m}(t))$ in the phase space $(m, \dot{m})$ obtained in the deterministic limit. The color corresponds to $v(m, \dot{m})^{-1}$, where $v(m, \dot{m})$ is the local speed along the limit cycle, and can be interpreted as the local density $p(m, \dot{m})\propto v(m, \dot{m})^{-1}$. The two black dots are added for visual purpose and correspond to the ferromagnetic points. Parameters: $J_1=0.6$, $J_2=0.4$, $\ve=10^{-1}$ and $(\mu_c-\mu)/\mu_c=3.5\times 10^{-3}$. At the deterministic limit, both the limit cycle and the ferromagnetic points are linearly stable.}
    \label{fig:cas1:simu:hist}
\end{figure}

In Fig.~\ref{fig:cas1:m2:md2:simu}, we compared $\langle m^2\rangle$ and $\langle \dot{m}^2\rangle$ obtained from the theoretical results of Sec.~\ref{sec:method} with stochastic simulations of the spin model. 
We observed that the behavior is qualitatively the same, but we did not obtain quantitative results. % in the high-$\ve$ and moderate-$N$ regime.  
We now discuss the low-$\ve$ approximation and its consequences. 
We plot in Fig.~\ref{fig:cas1:simu:hist}(a) an example of $P_N(m,\dot{m})$ obtained from numerical simulations for $N=5\times 10^5$, $\ve=10^{-1}$ and $(\mu_c-\mu)/\mu_c=3.5\times 10^{-3}$.
In the deterministic limit, both the limit cycle and the ferromagnetic points are linearly stable for these parameter values. We observe a significant difference with the results obtained in this section, similar to what was observed in Sec.~\ref{sec:cas0bis:corrections}: there is no individual symmetry $m\,\mapsto\,-m$ or $\dot{m}\, \mapsto\,-\dot{m}$, and the probability density is not constant along the limit cycle. 
In Sec.~\ref{sec:cas0}, we showed that higher order corrections [both in $\phi(m, \dot{m})$ and $C(m, \dot{m})$] to the large deviation form of $P_N(m, \dot{m})$ may account for discrepancies between analytical predictions and numerical results of stochastic simulations. 
We recall that corrections in $\ve$ to the large deviation function $\phi$ lead to changes in the shape of the limit cycle, and are responsible for the breaking of the individual reversal symmetry in $m$ and $\dot{m}$. In contrast, corrections in $N^{-1}$ and $\ve$ to the large deviation function, given by the function $C(m, \dot{m})$ [Eq.~(\ref{eq:def:C:m:dotm})], break the uniformity of the probability density along the limit cycle. 
In Sec.~\ref{sec:cas0}, we computed the first corrections analytically for $\mu=\mu_l(T)$. However, these corrections are more complicated to compute for any $\mu$, and we thus propose here a different way to determine corrections to the large deviation function. 
In the deterministic limit $N \to \infty$, Eq.~(\ref{eq:def:C:m:dotm}) can be rewritten in the form
\be
P_N(m, \dot{m})\underset{N\to\infty}{\to} \exp[C(m, \dot{m})]\, \delta(\phi(m, \dot{m})),
\ee
and the probability density on the limit cycle can be obtained from the local speed
\be
v(m, \dot{m})=[\left(dm/dt\right)^2+\left(d\dot{m}/dt\right)^2]^{1/2},
\ee
since $P(m, \dot{m})\propto v(m, \dot{m})^{-1}$. 
The deterministic limit provides information on the location of the minima of the function $\phi(m, \dot{m})$, which corresponds to the limit cycle, as well as the value of $C(m, \dot{m})$ on the limit cycle. 
To obtain these corrections, we compute the trajectory $(m(t), \dot{m}(t))$ in the deterministic limit [using Eq.~(\ref{eq:deter:equations})] and we plot, in Fig.~\ref{fig:cas1:simu:hist}(b) the trajectories in the phase space $(m,\dot{m})$ where the color can be interpreted as the local density $P(m, \dot{m})\propto v(m, \dot{m})^{-1}$ along the limit cycle. We observe the same shape of limit cycle as in the stochastic simulations, and we recover a higher probability density near the axis $\dot{m}=0$ (close to the ferromagnetic points), in qualitative agreement with numerical results.

\subsection{Entropy production}
\begin{figure}[t]
    \centering
    \includegraphics{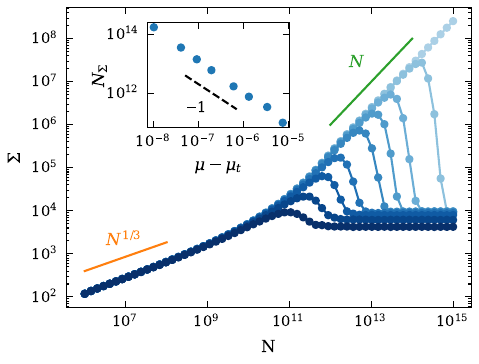}
    \caption{Entropy production $\Sigma$ vs.~$N$, for different values of $\mu$ such that $(\mu_c-\mu)/\mu_c\in[4.0, 3.6, 3.4, 3.33, 3.29, 3.28, 3.275, 3.272, 3.271]\times10^{-5}$ %$1-\mu/\mu_l\in [3.975, 3.575, 3.375,3.305, 3.265, 3.255, 3.250, 3.246]\times 10^{-5}$ 
    from darker to lighter colors. The inset represents $N_{\Sigma}$, the value of $N$ for which $\Sigma$ is maximal, as a function of $\mu-\mu_t$. The dashed line indicates an exponent $-1$.
    The entropy production is evaluated numerically from Eq.~(\ref{eq:entropy:prod}) where the averages are computed as in Fig.~\ref{fig:cas1:m2:md2:eps:rescaled}.
    Parameters: $J_1=0.6$, $J_2=0.4$, and $\ve=10^{-3}$; $(\mu_c-\mu_t)/\mu_c=3.271\times 10^{-5}$.}
    \label{fig:cas1:entropy:prod}
\end{figure}
Beyond the order parameter $\langle \dot{m}^2 \rangle$, the transition to an oscillating phase may also be characterized thermodynamically as a transition from microscopic to macroscopic irreversibility \cite{xiao_entropy_2008,seara_irreversibility_2021}, by introducing the entropy production density $\sigma=\Sigma/N$ in the limit $N\to\infty$, where the steady-state entropy production $\Sigma$ identifies with the entropy flux \cite{schnakenberg_1976,gaspard_time-reversed_2004},
\be
\Sigma = \frac{1}{2} \sum_{\mC,\mC'} \big[ W(\mC'|\mC)P(\mC)-W(\mC|\mC')P(\mC')\big]\, \ln \frac{W(\mC'|\mC)}{W(\mC|\mC')} \,.
\ee
We briefly investigate the influence of the bistability of the system on the entropy production.
In the large-$N$ and small-$\ve$ limits, one has (see Appendix~\ref{appendix:entropy:production} and \cite{guislain_nonequil2023})
\be\label{eq:entropy:prod}
\frac{\Sigma}{N}= \left[1+(T-J_1)^2\right]\left\langle\dot{m}^2\right\rangle +T^2 \left\langle V'(m)^2\right\rangle.
\ee
In the paramagnetic phase or ferromagnetic phase, one finds $\Sigma=O(N^0)$, and in an oscillating phase $\Sigma=O(N)$. 
Using Eq.~(\ref{eq:entropy:prod}), we compute the entropy production numerically for different system sizes; the results are plotted in Fig.~\ref{fig:cas1:entropy:prod}.
For large $N$, we recover that $\Sigma$ is independent of $N$ in the ferromagnetic phase, whereas $\Sigma \sim N$ in the oscillating phase.
However, for moderate values of $N$, one has $\Sigma \sim N^{1/3}$, due to the scaling $\langle\dot{m}^2\rangle \sim N^{-2/3}$ obtained in Sec.~\ref{sec:crossover}.
Note that this scaling is different from the scaling of the usual transition to an oscillating phase with an elliptic limit cycle, where one finds for moderate $N$, $\Sigma\sim N^{1/2}$, as a consequence of the scaling $\langle \dot{m}^2 \rangle \sim N^{-1/2}$ [see Eq.~(\ref{eq:scaling:mdot:finiteN})].
In the ferromagnetic phase, due to the influence of the oscillating phase, the entropy production increases before having a steep decrease to its constant value. Interestingly, this `overshoot' effect is still present for $\mu$ slightly below $\mu_O$ (see Fig.~\ref{fig:phase_diagram:1}), that is when the limit cycle no longer exists at the deterministic level. In this situation, the fluctuations described by the large deviation function keep track of the nearby existence of the limit cycle in parameter space, and are still able to generate a non-monotonous behavior.
We introduce $N_{\Sigma}$ the value of $N$ where the entropy production is maximal. The evolution of $N_{\Sigma}$ with $\mu-\mu_t$, where $\mu_t$ is the value of $\mu$ at the transition between the oscillating and the ferromagnetic phases, is plotted in the inset of Fig.~\ref{fig:cas1:entropy:prod} over a range of small values of $\mu-\mu_t$. Numerical data can be approximately described by a power-law decay $N_{\Sigma}\sim (\mu-\mu_t)^{-1}$, although no theoretical prediction is available to support this scaling relation. Accordingly, for $N$ larger than $N_{\Sigma}$, the entropy production drops by an amount $\Delta \Sigma\sim N_{\Sigma}\sim (\mu-\mu_t)^{-1}$, before reaching its asymptotic constant value.
%[commentaire]

%%%%%%%%%%%%%%%%%%%%%%%%%%%%%%%%%%%%%%%%%%%%%%%%%%%%%%%%%%%%%%%%%%%%%%%%%%%%%
%%%%%%%%%%%%%%%%%%%%%%%%%%%%%%%%%%%%%%%%%%%%%%%%%%%%%%%%%%%%%%%%%%%%%%%%%%%%%

\section{Type-II discontinuous transition between ferromagnetic and oscillating phases}
\label{sec:cas2}
In this section, we investigate the properties of the transition of Type II between the ferromagnetic and oscillating phases, near the tricritical point $(T_c, \mu_c)$. In this case, obtained for $J_1=J_2$, a small, almost elliptic limit cycle around the center is observed. The Type II scenario is illustrated in the phase diagram of Fig.~\ref{fig:phase:diagrams}(b) and on the trajectories of Figs.~\ref{fig:phase:diagrams}(d) and \ref{fig:phase:diagrams}(f).
All figures in this section are obtained with $J_1=J_2=0.5$.

\subsection{Large deviation function and phase diagram}
\label{sec:cas2:largedev}

%\subsubsection{On the validity of the method of Sec.~\ref{sec:method}}% on the deterministic limit and regions of the parameter space where the method of Sec.~\ref{sec:method} is valid}
\subsubsection{Validity of the perturbative approach}

For $J_1=J_2$, one has $\mu_c=1$ and $T_c=J_1$. 
The main difference with the previous case is that $v_1(T, \mu)$, the factor in front of $m^4$ in $V(m)$ [Eq.~(\ref{eq:cas1:expansion:v})] vanishes at $(T_c, \mu_c)$.
We have to leading order in an expansion in $\ve$ and $\mu-\mu_c$,
\be \label{eq:v1:leading}
v_1(T, \mu)=-\frac{J^2}{12}\ve -\frac{1}{4 J^2}(\mu-\mu_c) %+ o(\mu-\mu_c, \ve)
\ee
with $\ve=(T_c-T)/T_c$. Corrections to Eq.~(\ref{eq:v1:leading}) include terms proportional to $(\mu-\mu_c)^2$, $\ve(\mu-\mu_c)$ and $\ve^2$.
Numerically, we observe that the transition between the oscillating and ferromagnetic phases takes place for $(\mu-\mu_c)\ll \ve$ so that the term in $\mu-\mu_c$ can be neglected and we write $v_1=-\alpha\ve$ with $\alpha=J^2/12$.
As $v_1<0$, higher order terms in the expansion of $V(m)$ are necessary to compensate for the term $-\ve m^4$ and thus to describe the ferromagnetic points. Numerically, we observe ferromagnetic points whose amplitude goes to zero with $\ve$.
The coefficients of the terms proportional to $m^6$ and $m^8$ in the expansion of $V(m)$ obtained from Eq.~(\ref{eq:full:expr:Vm:appB}) scale as $\ve$ for small $\ve$, which would give ferromagnetic points independent of $\ve$ in this limit, if only these terms were retained.
Expanding $V(m)$ further, we find that the coefficient of the term proportional to $m^{10}$ is independent of $\ve$.
Assuming that the ferromagnetic point $m_0$ results from the balance of the terms in $m^4$ and in $m^{10}$, i.e., $\ve m_0^4\sim m_0^{10}$, yields $m_0 \sim \ve^{1/6}$.
%thus, it is necessary to go up to order $10$ in $V(m)$.
We now check a posteriori that the assumption to neglect the terms in $m^6$ and $m^8$ was valid.
For $m\sim \ve^{1/6}$, one has $\ve m^6\sim \ve^2$ and $\ve m^8\sim \ve^{7/3}$ which are both much smaller than the term $m^{10}\sim \ve^{5/3}$ for $\ve \to 0$, so that neglecting the terms in $m^6$ and $m^8$ was justified
for $m \sim m_0$. We thus write the following minimal form for $V(m)$,
\be \label{eq:cas2:vm}
V(m)=\frac{\mu-\mu_l(T)}{2T^2}m^2-\frac{\alpha\ve}{4}m^4+\frac{v_4}{10}m^{10}+V_0
\ee
where $v_4=J^2/81$. $V_0$ is such that $V(m)\geq0$ and the minimal value of $V(m)$ is zero. An example of the shape of $V(m)$ is given in Fig.~\ref{fig:cas2:V:fH:Phi}(a). 
Until now, we have considered only $V(m)$ with one or two minima, whereas now it can have three of them. As we now show, this has important consequences which makes this case of interest, and quite different from the previous ones. 
We found that the ferromagnetic points are $m_0\sim \ve^{1/6}$ when $\mu-\mu_c\ll \ve$, so that $u_0=g(m_0, 0)\sim \ve^{1/3}$ is small. Thus, for small $\ve$ and close to $\mu_c$, one can use the perturbative method described in Sec.~\ref{sec:method} to obtain the large deviation function.

%\subsubsection{Numerical computation of $f(H)$}
\subsubsection{Typical $f(H)$ and phase diagram}
\begin{figure}[t]
    \centering
    \includegraphics{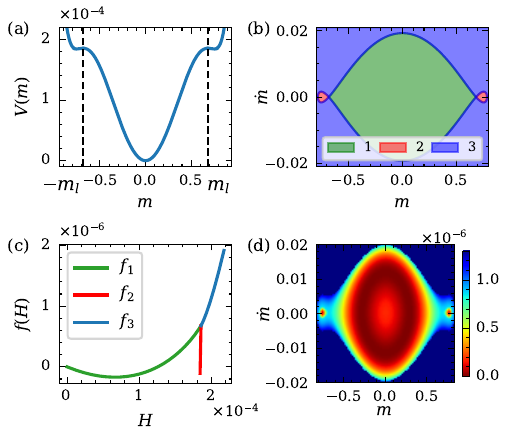}
    \caption{Type-II transition between ferromagnetic and oscillating states. (a) Potential $V(m)$; the dashed lines correspond to $m=\pm m_l$. (b) Representation of the different areas in the plane $(m, \dot{m})$. (c) $f(H)$ in the different areas defined in Eq.~(\ref{eq:cas2:f:zones}). (d) Colormap of $\phi(m, \dot{m})$ in the plane $(m,\dot{m})$. Note the different scalings $m\sim \ve^{1/2}$ and $\dot{m}\sim \ve$. Parameters: $J_1=J_2=0.5$, $(\mu-\mu_l)/\mu_l=4.55\times10^{-4}$ and $\ve=10^{-2}$.}
    \label{fig:cas2:V:fH:Phi}
\end{figure}

The main difference with the previous case is as follows. As illustrated in Fig.~\ref{fig:cas2:V:fH:Phi}(a), the condition $V(m)=H$ may correspond to six values of $m$ instead of only two or four previously, when considering values of $H$ close to the local minima of $V(m)$ with $m\ne 0$ (ferromagnetic points).
In the definition of $f'(H)$ given in Eq.~(\ref{eq:f'H}), we integrate over $m_1(H)$ and $m_2(H)$ such that $V(m_1)=V(m_2)=H$ and $V(m)\leq H$ for $m\in[m_1, m_2]$. Thus, for a given value of $H$, $f$ can have different values depending on the range of values of $m$ over which the integral is computed. We note $m_l$ the positive value of $m$ where $V(m)$ has a local maximum. 
In the phase space $(m, \dot{m})$ there are four different areas, which are represented in Fig.~\ref{fig:cas2:V:fH:Phi}(b). A first area around the center corresponds to small $m$ and $\dot{m}$, where $|m|<|m_l|$ and $V(m)<V(m_l)$, which is denoted area 1 and is represented in green. Two symmetric domains situated around the ferromagnetic points, where $|m|>m_l$ and $V(m)<V(m_l)$, correspond to area 2 and are represented in red. 
A last area for higher values of $H$, is denoted area 3 and is represented in blue.
We define three different functions $f$, one for each area:
\be\label{eq:cas2:f:zones}
\phi(m, \dot{m})\!=\!
\left\{\!
    \begin{array}{ll}
        f_1(H(m,\dot{m})) \! & \! \text{if } |m|\!<\!m_l \text{ and } V(m)\!<\!V(m_l)\\
        f_2(H(m, \dot{m})) \!&\! \text{if } |m|\!>\!m_l \text{ and } V(m)\!<\!V(m_l)\\
        f_3(H(m, \dot{m})) \! & \!\text{otherwise.}
  \end{array}
\right.
\ee
As $f$ is defined up to a constant in every area, we impose that $f_1(0)=0$ and we assume $f(H)$ to be continuous at the border between two different areas. %in $\pm m_1$. 
In Fig.~\ref{fig:cas2:V:fH:Phi}(c), (d), examples of $f(H)$ and $\phi(m, \dot{m})$ are plotted. A limit cycle around the center, and two ferromagnetic points are locally stable. Once again, the most stable phase is given by the global minima of $f$, here the oscillating phase. 

From the numerical determination of $f(H)$ and its minima, one obtains the phase diagram in the parameter space ($T, \mu$). We plot in Fig.~\ref{fig:phase_diagram:2} the phase diagram for $J_1=J_2$ close to the tricritical point $(T_c, \mu_c)$. 
We introduce the line $\mu_F$ indicating the existence of the ferromagnetic points and the line $\mu_O$ indicating the existence of the oscillating state. 
We also introduce $\mu_t(T)$ the value of $\mu$ at the transition, such that $f_2(H(m_0, 0))=f_1(H^*)$ (where $m_0$ corresponds to the ferromagnetic point and $H^*$ is where $f_1(H)$ is minimal).
Numerically, we obtain (see Fig.~\ref{fig:phase_diagram:2}) that 
\be \label{eq:cas2:mut:muc:eps}(\mu_t-\mu_c)/\mu_c\sim \ve^{4/3}.\ee
Indeed, the transition almost takes place when the ferromagnetic points disappear, meaning that $\mu_t\approx \mu_F$. The ferromagnetic points disappear when the term in $m^2$ balances the $m^4$ term in $V(m)$ at $m_0\sim\ve^{1/6}$, so that $(\mu_F-\mu_l)m_0^2\sim \ve m_0^4$. As $\mu_l\sim \mu_c$ [Eq.~(\ref{eq:mul:muc:eps2})], this gives $(\mu_F-\mu_c)/\mu_c\sim \ve^{4/3}$. 
We observe that when the ferromagnetic phase and the oscillating phase coexist, the ferromagnetic phase is almost always the most stable one. 

\subsubsection{Approximate local analytical expressions of $f(H)$}

To go beyond the numerical evaluation of $f(H)$, we now try to obtain an approximate analytical expression of $f(H)$, which will be helpful in particular to determine the scaling regimes of the order parameters $\langle m^2\rangle$ and $\langle \dot{m}^2\rangle$.
As in Sec.~\ref{sec:cas1}, one cannot obtain a full analytic expression of $f(H)$, and we thus focus on local approximations. For an expression of $f(H)$ around the ferromagnetic points, a quadratic expansion of $V$ around one of its local minima $m_0$ gives $f_2(H)=f_F(H)+\bar{f}_2$ where
\be\label{eq:cas2:fH:ferro}
f_F(H)=-\frac{g(m_0, 0)}{D_{22}+6v_4m_0^8D_{11}}\left[H-V(m_0)\right],
\ee
and $\bar{f}_2$ is a constant such that $f$ is continuous in $H=V(m_l)$.
We note here that $H(m_0, 0)\neq0$ for the ferromagnetic points unlike in the previous section as now $V(m_0)$ can be nonzero if the global minimum of $V$ is for $m=0$. 

For the area around the center, the leading term of $V(m)$ is the $m^2$ term, leading to the same $f(H)$ as for the Hopf bifurcation [see Eq.~(\ref{eq:fH:elliptic})], $f_{1}(H)=f_O(H)+\bar{f}_1$ where 
\be \label{eq:cas2:fH:lc}
f_O(H) = -\ve a H +b H^2 + \frac{(\ve a)^2}{b}
\ee
with $a$ and $b$ given in Eqs.~(\ref{eq:cas0:elliptic:a}) and (\ref{eq:cas0:elliptic:b}), and $\bar{f}_1$ a constant chosen such that $f(H)$ is continuous in $H=V(m_l)$.
%\be b=\frac{a_1+3a_3(\mu-\mu_l)/2T^2}{2(\mu-\mu_l)(D_{22}+D_{11}v_0)}\ee
%\be \label{eq:cas2:fH:lc}f(H)=bH^2-\frac{g(0,0)}{D_{22}}H\ee  with 
%\be b=\frac{a_1T^2}{2D_{22}(\mu-\mu_l)}\ee%enlève les termes en mu-mu0 d'ordre supérieurs
%which is similar to the one obtained for the  transition from a paramagnetic phase to an oscillating phase for high $\mu$, corresponding to a Hopf bifurcation \textbf{on a enlever $D_11v_0$ pourquoi?}. 
The minimum of $f_O(H)$ corresponds to an elliptic limit cycle around the center, as depicted in Fig.~\ref{fig:cas2:V:fH:Phi}(d), with
\be \label{eq:cas2:hstar}
H^*=\frac{a\ve}{2b}\sim \ve[\mu-\mu_l(T)]\,,
\ee
as $b\sim[\mu-\mu_l(T)]^{-1}$ from Eq.~(\ref{eq:cas0:elliptic:b}). %, for small enough $\mu-\mu_l(T)$.
At the transition ($\mu=\mu_t$), we find [Eqs.~(\ref{eq:mul:muc:eps2}) and (\ref{eq:cas2:mut:muc:eps})],
\be \label{eq:cas2:mut:mul:ve} \mu_t-\mu_l(T)\sim \ve^{4/3}\ee
and thus $H^*\sim \ve^{7/3}$. 

\begin{figure}[t]
    \centering
    \includegraphics{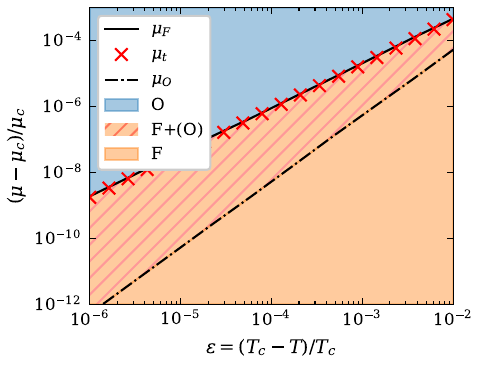}
    \caption{Close up of the phase diagram of Fig.~\ref{fig:phase:diagrams}(b) for $J_1=J_2=0.5$ for $T<T_c$. $\mu_F$ corresponds to the limit of existence of the ferromagnetic points (obtained from $V(m)$), and $\mu_O$ of the limit cycle (obtained in the deterministic limit). The transition between the F and O phases takes place for $\mu=\mu_t$. In the hatched area both the ferromagnetic points and the limit cycle are local minima of $f(H)$. As $\mu_F$ and $\mu_t$ are very close to each other, in most of the coexistence region, the ferromagnetic phase is the most stable one. Scalings of transition lines: $(\mu_t-\mu_c)/\mu_c\sim \ve^{4/3}$, $(\mu_F-\mu_c)/\mu_c\sim \ve^{4/3}$ and $(\mu_O-\mu_c)/\mu_c\sim \ve^2$, with $\ve=(T_c-T)/T_c$.  }
    \label{fig:phase_diagram:2}
\end{figure}

In the phase diagram of Fig.~\ref{fig:phase_diagram:2}, we observe that the transition line $\mu_t$ is very close to the line $\mu_F$ indicating the limit of existence of the ferromagnetic points. Hence in most of the coexistence region, the ferromagnetic phase is the most stable phase. 
This can be explained with the following argument.  Around the ferromagnetic points, $f'(H)\sim \ve^{1/3}$ whereas near the limit cycle, $f'(H)\sim\ve$. The slope of $f$ near the ferromagnetic points is much steeper that around the limit cycle (see Fig.~\ref{fig:cas2:V:fH:Phi} for an example of $f$). When the area around the ferromagnetic points exists, it rapidly becomes the global minimum of $f$ when varying $\mu$ at fixed $\ve$.

\subsection{Multiple scalings of order parameters with $N$ at the transition}

In Figs.~\ref{fig:cas2:m2:md2:eps:rescaled}(a) and \ref{fig:cas2:m2:md2:eps:rescaled}(b) we plot $\langle m^2\rangle$ and $\langle \dot{m}^2\rangle$ as a function of $N$ at the transition, for $\mu=\mu_t(\ve)$. We observe three different regimes depending on the value of $N$ and $\ve$, that are described below.

\subsubsection{Large-$N$ scaling at $\mu=\mu_t$}

Using the local approximations, we obtain the behaviors of $\langle m^2\rangle$ and $\langle \dot{m}^2\rangle$ in the large-$N$ limit when approaching the critical point $\ve=0$ where the three phases meet. 
In the ferromagnetic phase, using Eq.~(\ref{eq:cas1:fH:ferro}), we find that $\langle m^2\rangle=m_0^2\sim \ve^{1/3}$ and
\be \langle\dot{m}^2\rangle=\frac{D_{22}+6v_4m_0^8D_{11}}{-g(m_0, 0)N}, \ee
%$\langle \dot{m}^2\rangle\sim g(m_0, 0)^{-1}N^{-1}\sim \ve^{-1/3}N^{-1}$. 
so that $\langle \dot{m}^2\rangle\sim \ve^{-1/3}N^{-1}$ as $|g(m_0, 0)| \sim m_0^2\sim \ve^{-1/3}$.
Due to the $m^{10}$ term in $V(m)$, we obtain power laws in $\ve$ with critical exponents quite different from the corresponding values previously obtained. 
In the oscillating phase, one finds $\langle m^2\rangle \sim \ve$ similarly to the elliptic limit cycle obtained in Sec.~\ref{sec:cas0}, and $\langle \dot{m}^2\rangle \sim \ve [\mu_t-\mu_l(T)]$. One has $\mu_t-\mu_l\sim\ve^{4/3}$ [Eq.~(\ref{eq:cas2:mut:mul:ve})] such that one finds $\langle \dot{m}^2\rangle\sim \ve^{7/3}$ in the oscillating phase.

\subsubsection{Moderate-$N$ scaling at $\mu=\mu_t$}
\begin{figure}[t]
    \centering\includegraphics{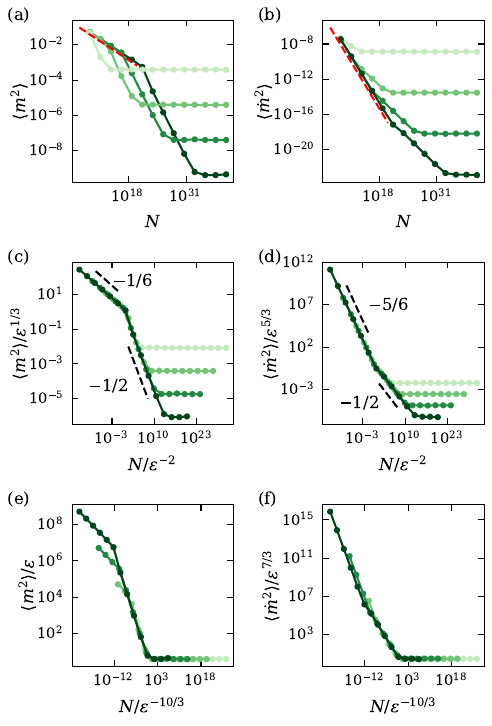}
    \caption{Order parameters $\langle m^2\rangle$ and $\langle \dot{m}^2\rangle$ as a function of the system size $N$, for $\ve\in[10^{-4}, 10^{-6},10^{-8}, 10^{-10}]$ from lighter to darker colors at $\mu=\mu_t(\ve)$. Order parameters are evaluated numerically from Eqs.~(\ref{eq:f'H}) and (\ref{eq:def:av:x}).  
    (a) $\langle m^2\rangle$ vs $N$. (b) $\langle \dot{m}^2\rangle$ vs $N$. The red dashed line corresponds to the theoretical prediction for low $N$ [Eqs.~(\ref{eq:cas2:lowN:m2}) and (\ref{eq:cas2:lowN:md2})], with a scaling of $N^{-1/6}$ for $\langle m^2\rangle$ and $N^{-5/6}$ for $\langle \dot{m}^2\rangle$. (c) $\langle m^2\rangle$ rescaled by $\ve^{1/3}$ and (d) $\langle \dot{m}^2\rangle$ rescaled by $\ve^{5/3}$ versus rescaled system size $N/\ve^{-2}$, highlighting the crossover between the moderate and intermediate-$N$ regimes. (e) $\langle m^2\rangle$ rescaled by $\ve$ and (f) $\langle \dot{m}\rangle$ rescaled by $\ve^{7/3}$ versus rescaled system size $N/\ve^{-10/3}$, highlighting the crossover to the large-$N$ regime. Parameters: $J_1=J_2=0.5$.}
    \label{fig:cas2:m2:md2:eps:rescaled}
\end{figure}

For moderate values of $N$, we observe that $\langle m^2\rangle$ and $\langle \dot{m}^2\rangle$ decrease with $N$ and are independent of $\ve$. 
Similarly to Sec.~\ref{sec:cas1:lowN}, for low values of $N$ we can keep only the first $\ve$-independent term in $V(m)$, namely here $V(m)=v_4m^{10}/10$.
Using Eq.~(\ref{eq:f'H}), this approximation gives $f(H)=AH^{5/6}$, leading to
\begin{align} \label{eq:cas2:lowN:m2}\langle m^2\rangle \sim N^{-1/6},\\
\label{eq:cas2:lowN:md2}\langle \dot{m}^2\rangle \sim N^{-5/6},\end{align}
where the exact asymptotic relations including prefactors are given in Appendix~\ref{appendix:cas2:v10}, and are plotted in dashed red lines in Fig.~\ref{fig:cas2:m2:md2:eps:rescaled}(a,b).

\subsubsection{Intermediate-$N$ scaling at $\mu=\mu_t$}

For intermediate values of $N$, we observe that for all $\ve$, $\langle m^2\rangle \sim N^{-1/2}$ and $\langle \dot{m}^2\rangle\sim N^{-1/2}$. 
Indeed, for values of $N$ such that $f(H(m_l))\sim N^{-1}$ where $m_l$ is the positive junction point between the different areas (see Fig.~\ref{fig:cas2:V:fH:Phi}), because the ferromagnetic areas are small, the main contribution to the integrals corresponds to $m$ in area 1 (in the center). In this area, one has $f(H)\sim f_O(H)$ where $f_O(H)$ is given in Eq.~(\ref{eq:cas2:fH:lc}). 
The leading correction in $N$ of $\langle m^2\rangle$ and $\langle \dot{m}^2\rangle$ is given by 
\be \langle\dot{m}^2\rangle=\frac{\mu-\mu_l(T)}{T^2}\langle m^2\rangle=H^*+\frac{e^{-NbH^{*2}}}{2\sqrt{\pi b N}}, \ee%+O(\frac{e^{-NbH^{*2}}}{N^{3/2}}).\ee
with $H^*=a\ve/b$. We recall that $b\sim (\mu-\mu_l)^{-1}\sim \ve^{-4/3}$ [Eq.~(\ref{eq:cas2:mut:mul:ve})] at the transition, thus we find $\langle m^2\rangle \sim \ve^{-2/3}N^{-1/2}$ and  $\langle \dot{m}^2\rangle \sim \ve^{2/3}N^{-1/2}$ for $N\ve^2\ll 1$. 
%For high values of $H$, one can keep only the quadratic contribution $f(H)\sim bH^2$ where we recall that $b\sim (\mu-\mu_l)^{-1}\sim \ve^{-4/3}$ at the transition, thus $f(H)\sim \ve^{-4/3}H^2$. %Thus one finds that the main contribution in the integrals is for $H \sim \ve^{2/3}N^{-1/2}$. As $\dot{m}\sim H$ and $m\sim \ve^{-4/3}H$,
%which gives $\langle m^2\rangle \sim \ve^{-2/3}N^{-1/2}$ and  $\langle \dot{m}^2\rangle \sim \ve^{2/3}N^{-1/2}$. 
We note $N_1$ the crossover value of $N$ between those the moderate- and intermediate-$N$ regimes, and $N_2$ the crossover value between the intermediate- and large-$N$ regimes.

For $N\ll N_1$, $\langle m^2\rangle \sim N^{-1/6}$ while for $N_1 \ll N \ll N_2$, $\langle m^2\rangle \sim \ve^{-2/3}N^{-1/2}$, so that $N_1\sim \ve^{-2}$. The same argument holds for $\langle \dot{m}^2\rangle$: for $N\ll N_1$, $\langle \dot{m}^2\rangle\sim N^{-5/6}$ and for $N_1 \ll N \ll N_2$, $\langle \dot{m}^2\rangle \sim \ve^{2/3}N^{-1/2}$, also implying $N_1 \sim \ve^{-2}$.
These different scaling behaviors for $\langle m^2\rangle$ around the first crossover regime $N \sim N_1$ can be encompassed into a single scaling function
\be
\langle m^2\rangle = \ve^{1/3}\, \mathcal{F}_{m,1}(\ve^2 N),
\ee
where $\mathcal{F}_{m,1}(x)$ asymptotically behaves as $\mathcal{F}_{m,1}(x) \sim x^{-1/6}$ for $x\to 0$ and $\mathcal{F}_{m,1}(x)\sim x^{-1/2}$ for $x\to \infty$.
In a similar way, $\langle \dot{m}^2\rangle$ can be expressed in terms of a scaling function,
\be
\langle \dot{m}^2\rangle = \ve^{5/3} \, \mathcal{F}_{\dot{m},1}(\ve^2 N),
\ee
with asymptotic behaviors $\mathcal{F}_{\dot{m},1}(x) \sim x^{-5/6}$ for $x\to 0$ and $\mathcal{F}_{\dot{m},1}(x)\sim x^{-1/2}$ for $x\to \infty$.
We plot in Fig.~\ref{fig:cas2:m2:md2:eps:rescaled}(c,d) $\langle m^2\rangle/\ve^{1/3}$ and $\langle \dot{m}^2\rangle/\ve^{5/3}$ as a function of $N/\ve^{-2}$, which is proportional to the rescaled system size $N/N_1$. As expected, the different curves corresponding to different values of $\ve$ collapse for moderate up to intermediate values of $N$. 

We now turn to the second crossover $N\sim N_2$ between the intermediate- and large-$N$ regimes. 
For $N_1 \ll N \ll N_2$, one has $\langle m^2\rangle \sim \ve^{-2/3}N^{-1/2}$ while for $N \gg N_2$, $\langle m^2\rangle \sim \ve$. Balancing the two contributions thus gives $N_2\sim \ve^{-10/3}$ (note that $N_2 \gg N_1$). The same argument holds for $\langle \dot{m}^2\rangle$: for $N_1 \ll N \ll N_2$, one finds $\langle \dot{m}^2\rangle \sim \ve^{2/3}N^{-1/2}$ and for $N \gg N_2$, $\langle \dot{m}^2\rangle \sim \ve^{7/3}$, which also gives $N_2 \sim \ve^{-10/3}$.
These scaling behaviors of $\langle m^2\rangle$ and $\langle \dot{m}^2\rangle$ can be encompassed into two scaling functions
\be
\langle m^2\rangle = \ve \, \mathcal{F}_{m,2}(\ve^{10/3} N), \quad \langle \dot{m}^2\rangle = \ve^{7/3} \, \mathcal{F}_{\dot{m},2}(\ve^{10/3} N),
\ee
with asymptotic behaviors $\mathcal{F}_{m,2}(x) \sim \mathcal{F}_{\dot{m},2}(x) \sim x^{-1/2}$ for $x\to 0$, while both functions go to constant values for  $x\to \infty$.
In Fig.~\ref{fig:cas2:m2:md2:eps:rescaled}(e,f), we plot $\langle m^2\rangle/\ve$ and $\langle \dot{m}^2\rangle/\ve^{7/3}$ as a function of the rescaled system size $N/\ve^{-10/3} \sim N/N_2$. As expected, for different values of $\ve$, the different curves collapse for intermediate up to large values of $N$.

\subsection{Crossover between intermediate and large $N$ regimes}
We reported above three distinct scaling regimes, separated by $N_1$ and $N_2$, for the observables $\langle m^2\rangle$ and $\langle \dot{m}^2\rangle$ as a function of the system size $N$ for $\mu=\mu_t(\ve)$, and we identified the scalings with $\ve$ of the crossover sizes $N_1$ and $N_2$.  
We now investigate in more details the behavior of the observables in the crossover regimes, focusing on the effect of the variable $\mu$ close to the transition value $\mu_t$ for a fixed $\ve$. 
We numerically find that in the moderate- and intermediate-N regimes, the two observables only weakly depend on $\mu$, unlike for large-$N$ values. Thus, we now focus on the dependence of the observables $\langle m^2\rangle$ and $\langle \dot{m}^2\rangle$ on $\mu$ for a fixed $\ve$, in the crossover regime between intermediate- and large-$N$ values ($N\sim N_2$).  

\subsubsection{Influence of $\mu$ on the second crossover regime}

\begin{figure}[t]
    \centering
    \includegraphics{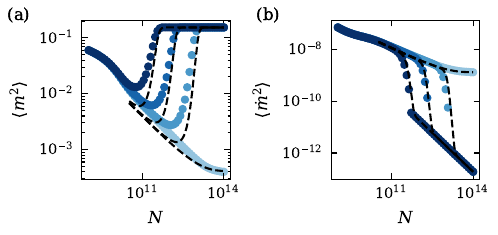}
    \caption{(a) $\langle m^2\rangle$ and (b) $\langle \dot{m}^2\rangle$ as a function of the system size $N$, for $(\mu-\mu_c)/\mu_c\in[8.310, 8.340, 8.348, 8.350]\times10^{-7}$ from darker to lighter colors. Parameters: $J_1=J_2=0.5$ and $\ve=10^{-4}$. The transition takes place at $(\mu_t-\mu_c)/\mu_c\approx 8.349\times 10^{-7}$.
    Order parameters are evaluated numerically in the same way as in Fig.~\ref{fig:cas2:m2:md2:eps:rescaled}.
    The black dashed line correspond to $\langle x(m, H)\rangle_{\mathrm{approx}}$ defined in Eq.~(\ref{eq:cas2:approx}).}
    \label{fig:cas2:m2:md2:mu}
\end{figure}

In the large-$N$ limit, the observables $\langle m^2\rangle$ and $\langle \dot{m}^2\rangle$ are discontinuous at the transition, whereas for moderate and intermediate-$N$ values, they do not depend much on the value of $\mu$. In Fig.~\ref{fig:cas2:m2:md2:mu}, $\langle m^2\rangle$ and $\langle\dot{m}^2\rangle$ are plotted as a function of system size $N$ for different values of $\mu$ along the transition, at fixed $\ve$. Like in Sec.~\ref{sec:cas1}, we observe jumps in $\langle m^2\rangle$ and $\langle \dot{m}^2\rangle$ in the ferromagnetic phase ($\mu<\mu_t$) at a finite system size, which takes place for higher values of $N$ when approaching the transition.
The two values of $\langle m^2\rangle$ in the different phases are nonzero whereas $\langle \dot{m}^2\rangle$ goes from a nonzero value (in the oscillating phase) to a value decreasing as $N^{-1}$. 
The main difference with Sec.~\ref{sec:cas1} is that the jump in $\langle m^2\rangle$ is now much more pronounced because of its different dependence on $\ve$. Indeed, we had in Sec.~\ref{sec:cas1} that $\langle m^2\rangle\sim \ve$ in both phases, whereas now $\langle m^2\rangle\sim \ve$ in the oscillating phase and $\langle m^2\rangle\sim \ve^{1/3}$ in the ferromagnetic phase, leading for small $\ve$ to a strong mismatch of $\langle m^2\rangle$ between the two phases.

\subsubsection{Approximation in terms of phase coexistence}
Here again, the observed behaviors can be given a simple interpretation in terms of finite-size phase coexistence and metastability.
Hence, as in Sec.~\ref{sec:cas1}, we introduce a simple decomposition of average values into contributions of each phases.
For any quantity $x(m, H)$, we introduce the approximate average value obtained by taking into account the statistical weight of each phase,
\be \label{eq:cas2:approx}\langle x(m, H)\rangle_{\rm approx} = \frac{\langle x\rangle_F +\langle x\rangle_O C_2\sqrt{N} e^{-N(\bar{f}_2-\bar{f}_1)} }{1 + C_2\sqrt{N} e^{-N(\bar{f}_2-\bar{f}_1)}}\ee
where the `pure-state' averages $\langle x\rangle_F$ and $\langle x\rangle_O$ are respectively obtained from the ferromagnetic state large deviation function
$f_F(H)$ given in Eq.~(\ref{eq:cas2:fH:ferro}), and from the oscillating state large deviation function $f_O(H)$ given in Eq.~(\ref{eq:cas2:fH:lc});
$\bar{f}_i$ is the minimum of $f_i(H)$ ($i=1$, $2$).
The constant $C_2$, whose expression is similar to that of the constant $C_1$ given in Eq.~(\ref{eq:cas1:approx:C}), but with different expressions for $f_F$ and $f_O$ from the ones found in Sec.~\ref{sec:cas1}, now becomes
\be C_2=\frac{-g(m_0, 0)}{(D_{22}+6v_4 m_0^8D_{11})} \sqrt{\frac{6\pi v_4}{bv_0}}m_0^4\,.\ee

In Sec.~\ref{sec:cas1}, we could not have a local expression of $f_O(H)$ in the oscillating phase, so we used fitting parameters for the coefficients $\tilde{c}$ and $H^*$. Here, both $f_F$ and $f_O$ are known analytically. The only quantity which is not known analytically and is a fitted parameter, obtained from the numerical evaluation of $f(H)$, is $\bar{f}_2-\bar{f}_1$. 

This decomposition is plotted for different values of $\mu$ in Fig.~\ref{fig:cas2:m2:md2:mu} in black dashed lines. The jump in $\langle m^2\rangle$ and $\langle \dot{m}^2\rangle$ is well described by this simple decomposition. When increasing $N$, the influence of the oscillating phase dominates until $f(H^*)\sim N^{-1}$.

\subsubsection{Comparison with stochastic simulations}
\begin{figure}[t]
    \centering
    \includegraphics{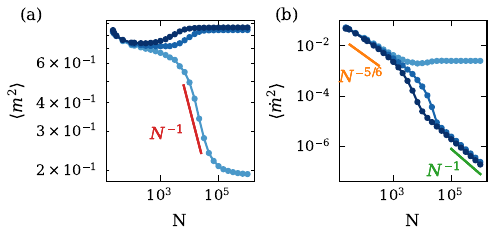}
    \caption{Comparison with stochastic simulations of the spin model. (a) $\langle m^2\rangle$ and (b) $\langle \dot{m}^2\rangle$ obtained from numerical simulations as a function of the system size $N$, for $(\mu-\mu_c)/\mu_c\in[2, 3, 4]\times10^{-3}$ (from darker to lighter colors). Parameters: $J_1=J_2=0.5$ and $\ve=5\times 10^{-2}$.}
    \label{fig:cas2:m2:md2:simu}
\end{figure}
We plot in Fig.~\ref{fig:cas2:m2:md2:simu} $\langle m^2\rangle$ and $\langle \dot{m}^2\rangle$ computed from stochastic simulations for different system sizes and different $\mu$, with $\ve=(T_c-T)/T_c=5\times 10^{-2}$. We observe the different expected behaviors described above. At large $N$, a jump of $\langle m^2\rangle$ is observed between the ferromagnetic phase (high values of $\langle m^2\rangle\sim \ve^{1/3}$ for low values of $\mu-\mu_c$) and the oscillating phase (low values of $\langle m^2\rangle\sim \ve$ for higher values of $\mu-\mu_c$). A jump of $\langle \dot{m}^2\rangle$ is also observed: 
$\langle \dot{m}^2\rangle$ is constant and of order $\ve^{7/3}\sim 10^{-3}$ for sufficiently high values of $\mu-\mu_c$, while it decreases as $N^{-1}$ for lower values of $\mu-\mu_c$. 

Similarly to Sec.~\ref{sec:cas1}, we are able to describe the qualitative behavior of the observables $\langle m^2\rangle$ and $\langle \dot{m}^2\rangle$. However, for the restricted range of values of $\ve$ and $N$ accessible in stochastic simulations, we are not able to reach a quantitative agreement with analytical predictions obtained in the small-$\ve$, large-$N$ limit. We now briefly discuss the effect of considering finite values of $\ve$ and $N$.

\label{sec:cas2:discussion:limit:ve}
\begin{figure}[t]
    \centering
    \includegraphics{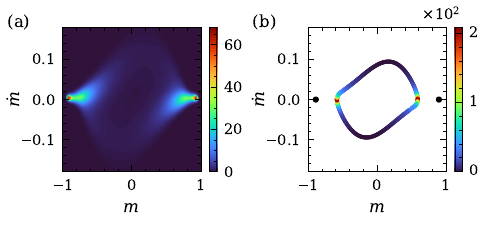}
    \caption{(a) $P_N(m, \dot{m})$ obtained from stochastic simulations for $N=2\times10^4$. (b) Trajectories ($m(t), \dot{m}(t))$ in the phase space $(m, \dot{m})$ obtained in the deterministic limit. The color corresponds to $v(m, \dot{m})^{-1}$ where $v(m, \dot{m})$ is the local velocity, as the local density $p(m, \dot{m})\sim v(m, \dot{m})^{-1}$. The black red dots are added for visual purpose and correspond to the ferromagnetic points. Parameters: $J_1=J_2=0.5$, $\ve=5\times 10^{-2}$ and $(\mu-\mu_c)/\mu_c=3.981\times10^{-3}$.}
    \label{fig:cas2:simu:hist}
\end{figure}

%We compared in Fig.~\ref{fig:cas2:m2:md2:simu} $\langle m^2\rangle$ and $\langle \dot{m}^2\rangle$ obtained from the perturbative approach presented in Sec.~\ref{sec:method} with stochastic simulations. We observe that the behavior is qualitatively the same, but we do not obtain a quantitative agreement with the theoretical result obtained in the high-$\ve$, low-$N$ limit. 

In Fig.~\ref{fig:cas2:simu:hist}(a), we plot an example of $P_N(m, \dot{m})$ obtained from stochastic simulations for $\ve=10^{-1}$ and $N=5\times 10^5$, for a value of $\mu$ where both the limit cycle and the ferromagnetic points are linearly stable in the deterministic limit. We observe important discrepancies with the theory described in Sec.~\ref{sec:cas2:largedev}, which assumed a small elliptic limit cycle in the center with uniform probability along the cycle, and ferromagnetic points outside the cycle.
From a dynamical viewpoint, a non-uniform probability along the cycle in an ensemble approach means that an individual system goes along the cycle at a non-uniform speed.
Here, as shown in Fig.~\ref{fig:cas2:simu:hist}(a), the limit cycle is hardly visible, meaning the probability is strongly non-uniform along the cycle. To understand this result, we plot in Fig.~\ref{fig:cas2:simu:hist}(b) trajectories, obtained in the deterministic limit, in the phase space $(m,\dot{m})$ where the color codes for $v(m, \dot{m})^{-1}$, with $v(m, \dot{m})=[\left(dm/dt\right)^2+\left(d\dot{m}/dt\right)^2]^{1/2}$
the local speed on the cycle. The quantity $v(m, \dot{m})^{-1}$ is proportional the local probability density $p(m, \dot{m})$ along the limit cycle.
We observe that the limit cycle is not elliptic, and that the speed $v(m, \dot{m})$ is far from uniform along the cycle.
The probability density $p(m, \dot{m})$ is much higher close to the ferromagnetic points, in qualitative agreement with stochastic simulations.
This discrepancy with theoretical predictions derived in Sec.~\ref{sec:cas2:largedev} comes from both the small-$\ve$ and large-$N$ approximations made to obtain analytical results. The small-$\ve$ approximation gives an elliptic limit cycle, and the large-$N$ one gives a constant speed along the limit cycle [see Sec.~\ref{sec:cas0bis:corrections} and Sec.~\ref{sec:cas1}].

\subsection{Entropy production}
\begin{figure}[t]
    \centering
    \includegraphics{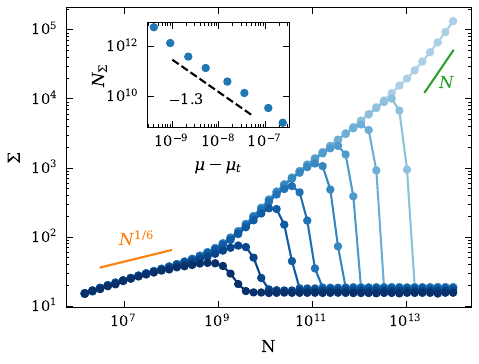}
    \caption{Entropy production $\Sigma$ vs.~$N$, for $(\mu-\mu_c)/\mu_c\in [6, 7.2, 8.0, 8.2, 8.3, 8.330, 8.343, 8.348, 8.352]\times 10^{-7}$ from darker to lighter colors. The entropy production is evaluated numerically from Eq.~(\ref{eq:entropy:prod}). Inset: $N_{\Sigma}$, the value of $N$ where $\Sigma$ is maximal, as a function of $\mu-\mu_t$, with $(\mu_t-\mu_c)/\mu_c=8.352\times 10^{-7}$. A slope $-1.3$ is indicated by a dashed line, as a guide to the eye. Parameters: $J_1=J_2=0.5$ and $\ve=10^{-4}$.}
    \label{fig:cas2:entropy:prod}
\end{figure}

Similarly to Sec.~\ref{sec:cas1}, we discuss the behavior of the entropy production Eq.~(\ref{eq:entropy:prod}) with system size at the transition, which is plotted in Fig.~\ref{fig:cas2:entropy:prod} for different values of $\mu$ across the transition ($\mu \approx \mu_t$). 
For low values of $N$, the entropy production increases as $N^{1/6}$ as $\langle \dot{m}^2\rangle\sim N^{-5/6}$. For larger values of $N$ and for $\mu>\mu_t$, $\Sigma\sim N$ as expected in an oscillating phase.
For $\mu<\mu_t$, as in Sec.~\ref{sec:cas1}, we observe that $\Sigma$ increases like in the oscillating phase before steeply decreasing to a constant value. This behavior is similar to the one observed in Sec.~\ref{sec:cas1} and is a consequence of the proximity of the oscillating phase.
We again denote as $N_{\Sigma}$ the value of $N$ when $\Sigma$ is maximal. In the inset of Fig.~\ref{fig:cas2:entropy:prod}, we plot $N_{\Sigma}$ as a function of the distance to the transition $\mu-\mu_t$, showing an approximate power-law divergence of $N_{\Sigma}$ when $\mu-\mu_t \to 0$.
As a result, getting closer to the transition, the drop of $\Sigma$ takes place for larger $N$ and thus becomes bigger, since it eventually decays to approximately the same asymptotic large-$N$ value for all $\mu-\mu_t$.

\subsection{Comment on the continuous transition for $J_2=\pm2+J_1$}
\label{sec:cas2:discussion:heteroclinic}
In this section, we investigated the properties of the discontinuous transition of Type II between the ferromagnetic phase and the oscillating phase taking place for $J_1=J_2$ where $v_1(T_c, \mu_c)=0$. 
As seen in Fig.~\ref{fig:sign:v1:J1:J2}, the condition $v_1(T_c, \mu_c)=0$ is also satisfied for $J_2=\pm 2+J_1$, and it would thus be natural to also study the transition in this case.
However, in the deterministic limit we find for $J_2=\pm 2+J_1$ that the transition between the ferromagnetic phase and the oscillating phase is of a different type: the transition is continuous, and the ferromagnetic points turn into a limit cycle with a infinite period at the transition. We now comment further on the difference between the two cases $J_1=J_2$ and $J_2=\pm 2+J_1$ and why the method presented in this paper does not allow for a characterization of continuous transitions between ferromagnetic points and a limit cycle with infinite period. 

In Sec.~\ref{sec:cas2:largedev}, we computed the large deviation locally using Eq.~(\ref{eq:f'H}) and assumed that the large deviation function is continuous in $m=m_l$ in order to obtain the large deviation function in all three areas, and thus in the whole plane ($m,\dot{m}$). 
However, an important issue is that at the point where the different areas meet, $m=m_l$, we have $V'(m_l)=0$ and $g(m_l, 0)\neq 0$. Thus, the assumption $V'(m)\gg \dot{m}g(m, \dot{m})$ made to split the different orders of Eq.~(\ref{eq:phi:quadrat}) is not valid close to $m=m_l$. 
Therefore, we expect to get corrections close to $m_l$ which are not taken into account here. When the ferromagnetic points and the limit cycle are well separated and far from $m_l$, which is true for $J_1=J_2$ as we have $m\sim \ve^{1/6}$ for the ferromagnetic phase and $m\sim \ve^{1/2}$ for the oscillating phase, the corrections close to $m_l$ do not affect the qualitative behavior at the transition, and thus the method presented in this section describes well the phase transition. However, if one or both of them are close to $m_l$, corrections, that are not taken into account in this paper and which would require a different approach, are necessary.

For $J_2=\pm 2+J_1$, we find that the $m^6$-term in $V(m)$ is independent of $\ve$ (whereas for $J_1=J_2$ it scales as $\ve$) and thus is enough to describe the ferromagnetic points, which turn out to scale as $m\sim\ve^{1/2}$ (as seen by balancing the terms in $\ve m^4$ and $m^6$).
If we blindly apply the method described in this section, we find a limit cycle with $m\sim\ve^{1/2}$ in between the ferromagnetic points, which have the same scaling. The limit cycle is very close to the point $m=m_l$ where $V'(m_l)=0$ and where the theory breaks downs. 
Furthermore, from the definition of the period of the limit cycle Eq.~(\ref{eq:period}), we find that when $H^*\approx V(m_l)$, the period is very large, and diverges when $H^*=V(m_l)$.
Thus, by applying the method without enough care, we would still recover some qualitative properties of the transition: the ferromagnetic points and the limit cycle are very close to each other and the period of the limit cycle is very large.
Yet, this would not be a correct description of the transition as we would find a discontinuous transition instead of a continuous one as observed numerically.

%%%%%%%%%%%%%%%%%%%%%%%%%%%%%%%%%%%%%%%%%%%%%%%%%%%%%%%%%%%%%%%%%%%%%%%
%%%%%%%%%%%%%%%%%%%%%%%%%%%%%%%%%%%%%%%%%%%%%%%%%%%%%%%%%%%%%%%%%%%%%%%

\section{Conclusion}
\label{sec:conclusion}

We have considered a mean-field spin model with a dynamics breaking detailed balance due to the non-reciprocal couplings between spins and auxiliary dynamic fields.
The presence of ferromagnetic interactions between spins on one side, and between dynamic fields on the other side, allows for the presence of both ferromagnetic and spontaneously oscillating phases. We have characterized in details the transition between these two phases, showing that it is discontinuous with the coexistence of both ferromagnetic and oscillating states, one state being stable and the other one metastable. The relative stability of both states is determined by a large deviation function, generalizing the Landau free energy, that we evaluated explicitly in different cases thanks to a perturbative framework. Two main scenarios are discussed, whether the ferromagnetic points turn out to be inside or outside the limit cycle.
In addition, we found that the entropy production is peaked as a function of system size, leading to a maximally dissipative system for an optimal finite system size.

A natural generalization of this work may be to try to extend these results beyond mean-field, by considering finite-dimensional systems, with the goal to formulate a Ginzburg-Landau theory extending the present Landau framework based on a large deviation principle. Such a theory might then be amenable to a renormalization group treatment,
extending the results of \cite{risler_universal_2004,risler_universal_2005} which considered the synchronization of coupled oscillators. Here, we have started from more basic ingredients, in the sense that the microscopic degrees of freedom of the model (i.e., the spins and dynamic fields) do not oscillate in the absence of interaction. Connecting these types of models to previous results obtained on the synchronization transition is thus an interesting challenge for future work.

\acknowledgments
L. G. acknowledges funding from the French Ministry of Higher Education and Research.

\appendix
\section{Derivation of the deterministic evolution equations}
\label{appendix:deterministic:equations}
In this appendix, we derive the deterministic evolution equations Eqs.~(\ref{eq:deter:equations:m}) and (\ref{eq:deter:equations:h}) from the microscopic spin and field dynamics. 
The dynamics of the system is determined from the master equation [see Eq.~(\ref{eq:master:equation})].
As the average is defined as $\langle x\rangle=\sum_{\mC}x(\mC)P(\mC)$, we find after rearranging terms,
\begin{align} \label{eq:def:mdot:av:SM}
d_t\langle m \rangle=\sum_\mC  P(\mC)\sum_{\mC'\neq \mC} \left[m(\mC')-m(\mC)\right]W(\mC'| \mC), \\
 \label{eq:def:hdot:av:SM}
d_t\langle h\rangle=\sum_\mC  P(\mC)\sum_{\mC'\neq \mC} \left[h(\mC')-h(\mC)\right]W(\mC'| \mC),
\end{align}
with the shorthand notation $d_t \equiv d/dt$.
From a configuration $\mC=\{s_1, ..., s_N, h_1, ..., h_N\}$ with magnetization $m=N^{-1}\sum_{i=1}^N s_i$ and
average field $h=N^{-1}\sum_{i=1}^N h_i$, there are $(1\pm m)N/2$ possibilities to flip a spin $s_i=\pm 1$ and $(1\pm h)N/2$ possibilities to flip a field $h_i=\pm 1$. For a flip of a spin $s_i=\pm 1$, $m(\mC')-m(\mC)=\mp 2/N$ and for a flip of a field $h_i=\pm 1$, $h(\mC')-h(\mC)=\mp 2/N$. Thus, using the definition of the transition rates given in the main text [Eq.~(\ref{eq:transition:rates})], we find:
%\begin{align} d_t\langle m\rangle&=\left\langle\frac{1-m}{1+e^{-\beta(J_1m+h)}}- \frac{1+m}{1+e^{\beta(J_1m+h)}}\right\rangle,\\ d_t\langle h\rangle&=\left\langle \frac{1-h}{1+e^{-\beta(J_2 h+(1-\mu)m)}}-\frac{1+h}{1+e^{\beta(J_2 h+(1-\mu)m)}}\right\rangle.\end{align}
%Thus, 
\begin{align} d_t\langle m\rangle&=\big\langle -m+\tanh[\beta (J_1m+ h)]\,\big\rangle,\\ 
d_t\langle h\rangle&=\big\langle-h+\tanh[\beta(J_2h+(1-\mu)m)]\, \big\rangle .\end{align}
Assuming that the law of large numbers applies in the limit $N\to\infty$, $m$ and $h$ obey the following deterministic equations:
\begin{align}
  \label{eq:dyn:m}
  d_tm &= -m+\tanh[\beta(J_1 m+h)],\\
  \label{eq:dyn:h}
  d_th  &= -h+\tanh[\beta(J_2 h+(1-\mu)m)].
\end{align}
These deterministic equations can be used to determine the macroscopic phase when a single solution exists for given values of the control parameters $\beta=T^{-1}$ and $\mu$. When two solutions exist, the most stable one has to be determined from the large deviation function approach, as explained in the main text.

\section{Values of the different functions and coefficients of the model}
\label{appendix:values:coeff}
In this appendix, we give the values of the different functions and coefficients introduced in the main text. %for the specific spin model.
The function $Y(m, \dot{m})$ introduced in Eq.~(\ref{eq:deter:equations}) as $d_t\dot{m}=Y(m, \dot{m})$ has been split into a $\dot{m}$-independent part, $V'(m)=Y(m, 0)$ and a $\dot{m}$-dependent part $\dot{m}g(m, \dot{m})=Y(m, \dot{m})-Y(m, 0)$. From Eqs.~(\ref{eq:deter:equations:m}) and (\ref{eq:deter:equations:h}), we have:
\begin{widetext}
\begin{equation}
\begin{aligned} \label{eq:Ymdotm:appA}
Y(m, \dot{m})=&\beta J_1m+(-1+\beta J_1)\dot{m}-\tanh^{-1}(m+\dot{m})+\beta \tanh[J_2\tanh^{-1}(m+\dot{m})+\beta(1-\mu-J_1J_2)m]\\&+(m+\dot{m})^2\left[\tanh^{-1}(m+\dot{m})-\beta\tanh[J_2\tanh^{-1}(m+\dot{m})+\beta(1-\mu-J_1J_2)m]-\beta J_1(m+\dot{m})\right]\,,\end{aligned}\end{equation}
\begin{equation} \label{eq:full:expr:Vm:appB}
V'(m)=-\beta J_1m +\beta J_1m^3+(1-m^2)\tanh^{-1}(m)-\beta(1-m^2)\tanh[J_2\tanh^{-1}(m)+\beta (1-\mu-J_1J_2)m]\,.
\end{equation}
\end{widetext}
%where we use the notation $\beta=T^{-1}$. 
%
In the main text, we introduced the following expansions of $V(m)$ and $g(m, \dot{m})$ [see Eqs.~(\ref{eq:vpm:3}) and (\ref{eq:g(m,0):small:m})],
\be V(m)=\frac{v_0}{2}m^2+\frac{v_1}{4}m^4,\ee
and
\be g(m, \dot{m})=a_0\ve -a_1m^2-a_2m\dot{m}-a_3\dot{m}^2.\ee
The coefficients appearing in these expansions are given by
\begin{align} &v_0=(\mu-1)/T^2+(1-J_1/T)(1- J_2/T),\\
& \begin{aligned}v_1=&-2/3+(2J_2+3J_1)/3T-(\mu-1+J_1J_2)/T^2\\
&-(\mu-1-J_2T+J_1J_2)^3/3T^4,\end{aligned}\\
&\ve=(T_c-T)/T_c,\\
&a_{0}=2 T_c/T,\\
&\begin{aligned}a_{1}&=-2 + (2 J_2 + J_2^3 + 3 J_1)/T \\
&+ J_2 (-1 + J_1 J_2 +\mu)^2/T^3\\
&-  2(1 + J_2^2) (-1 + J_1 J_2 +\mu)/T^2,\end{aligned}\\
&\begin{aligned}a_2&=-2+2\beta J_2+\beta J_2^3+3\beta J_1\\&+\beta^2(1+J_2^2)(-1+J_1J_2+\mu),\end{aligned}\\
&a_{3}=-2/3 +(2J_2 + J_2^3 + 3J_1)/3T.\end{align}

In addition, we introduced in Eq.~(\ref{eq:def:D}) the coefficients $D_{11}(m, \dot{m})$, $D_{12}(m, \dot{m})=D_{21}(m, \dot{m})$ and $D_{22}(m, \dot{m})$, which read:
\begin{align}
&D_{11}=1-m(m+\dot{m}),\\
&D_{12}=[1-m(m+\dot{m})][-1+\beta J_1(1-(m+\dot{m})^2)]\\
&\begin{aligned}D_{22}&=[1-m(m+\dot{m})][-1+\beta J_1(1-(m+\dot{m})^2)]^2\\&+[1-h(h+\dot{h})]\beta^2[1-(m+\dot{m})^2]^2\end{aligned}
\end{align}
with 
\be h=T\tanh^{-1}(m+\dot{m})-J_1m\ee
and 
\be \label{eq:def:hdot} \dot{h}(m, \dot{m}) = T\frac{Y(m, \dot{m})+\dot{m}}{1-(m+\dot{m})^2}-J_1\dot{m}.\ee
In the main text, we use the coefficients evaluated at $m=0$ and $\dot{m}=0$, which simplify to:
\begin{align}& D_{11}(0,0)=1,\\
& D_{12}(0, 0)=-1+J_1/T,\\
& D_{22}(0,0)=1/T^2+(J_1/T-1)^2.\end{align}

\section{Entropy production}
\label{appendix:entropy:production}
In steady-state, the entropy production can be identified with the entropy flux, which is defined from the microscopic configurations $\mC=\{s_1, ..., s_N, h_1, ...,h_N\}$ as 
\be \label{eq:def:Sigma:app}
\Sigma = \sum_{\mC,\mC'}  W(\mC'|\mC)P(\mC)\, \ln \frac{W(\mC'|\mC)}{W(\mC|\mC')}\,.
\ee
Note that Eq.~(\ref{eq:def:Sigma:app}) is equivalent to the definition (\ref{eq:entropy:prod}) given in the main text, up to a symmetrization of expression (\ref{eq:def:Sigma:app}).
To compute the entropy production, we aim at changing the sum over the configurations into integrals over $m$ and $\dot{m}$.
We thus replace $\sum_{\mC}$ by $\int dm d\dot{m}\sum_{\mC\in S(m, \dot{m})}$, where $S(m, \dot{m})$ denotes the ensemble of configurations $\mC$ with $m(\mC)=m$ and $\dot{m}(\mC)=\dot{m}$. We transform the integral over $\mC'$ into a sum over all possible transitions.
We recall that spin reversals are labelled with $k=1$ and field reversals with $k=2$, keeping track of the sign $\sigma=\pm$ of the variable prior to reversal. We denote as $W_{k}^{\sigma}$ the coarse-grained transition rates given in Eq.~(\ref{eq:transition:rate}) and $n_k^{\sigma}$ the fraction of possible transitions, 
\be
n_1^{\pm} = \frac{1}{2}(1\pm m), \quad n_2^{\pm} = \frac{1}{2}(1\pm h).
\ee
We find
\be
\Sigma=N\sum_{k, \sigma}\left\langle  W_k^{\sigma}(\mathbf{x})\ln\left[\!\frac{W_k^{\sigma}(\mathbf{x})n_k^{-\sigma}(\mathbf{x}+\frac{\sigma \mathbf{d}_k}{N})}{W_k^{-\sigma}(\mathbf{x}+\frac{\sigma \mathbf{d}_k}{N})n_k^{\sigma}(\mathbf{x})} \right]\!\right\rangle
\ee
with the shorthand notation $\mathbf{x}=(m, \dot{m})$. To further simplify notations, we make the dependence on $\mathbf{x}$ implicit in what follows.
At leading order in $N$, we find
\be \Sigma=N\sum_{k}\left\langle \left(W_k^{+}-W_k^{-}\right)\ln \left[\frac{W_k^{+}n_k^{-}}{W_k^{-}n_k^{+}}\right]\right\rangle.\ee
We define 
\be
\Sigma_k=\left\langle \left(W_k^{+}-W_k^{-}\right)\ln \left[\frac{W_k^{+}n_k^{-}}{W_k^{-}n_k^{+}}\right]\right\rangle.
\ee
For $k=1$, we have $W_1^+-W_1^-=\dot{m}/2$ and
\be
\ln \left[\frac{W_k^{+}n_k^{-}}{W_k^{-}n_k^{+}}\right] = 2\tanh^{-1}(m+\dot{m}),
\ee
thus we find that
\be
\Sigma_1 = N \left\langle \dot{m}\tanh^{-1}(m+\dot{m})\right\rangle.
\ee
In the main text, we showed that close to a transition, one generically has a scaling behavior $\dot{m}\sim \ve ^{\alpha}$ (with $\alpha>0$ an exponent depending on the specific transition considered) and $\langle m \dot{m}\rangle=0$. Thus, keeping the lowest order in $\ve$, we find 
\be \Sigma_1/N=\langle \dot{m}^2\rangle.\ee

For $k=2$, we recall that we have $h(m, \dot{m})=-J_1m+T\tanh^{-1}[m+\dot{m}]$ and we introduce $\dot{h}=-h+\tanh[\beta(J_2h+(1-\mu)m)]$, the equivalent of $\dot{m}$ for the fields variables $h_i$, given in Eq.~(\ref{eq:def:hdot}) as a function of $m$ and $\dot{m}$.
We find $\Sigma_2=N\langle \dot{h}\tanh^{-1}(h+\dot{h})\rangle$. As spins and fields play symmetric roles, one also has $\langle h\dot{h}\rangle=0$ close to a transition. Using that $Y(m, \dot{m})=-V'(m)$ and $\langle V'(m)\dot{m}\rangle=0$ at the lowest order in $\ve$, we find
\be
\frac{\Sigma_2}{N} =\langle  T^2 V'(m, \dot{m})^2+(T-J_1)^2\dot{m}^2\rangle.
\ee
Gathering contributions, one finds
\be
\frac{\Sigma}{N} = \left[1+(T-J_1)^2\right]\left\langle\dot{m}^2\right\rangle +T^2 \left\langle V'(m)^2\right\rangle
\ee
at leading order in $\ve$ and $N$.

\section{Moderate-$N$ approximation of Sec.~\ref{sec:cas2}}
\label{appendix:cas2:v10}
In this appendix, we give the exact expressions of $\langle m^2\rangle$ and $\langle\dot{m}^2\rangle$ for moderate values of $N$ for the Type-II discontinuous transition of Sec.~\ref{sec:cas2}.
For intermediate values of $N$, only high values of $V(m)$ contribute, thus we consider that $V(m)=v_4m^{10}/10$. Using Eq.~(\ref{eq:f'H}), we find
\be f(H)=AH^{6/5}\,,\ee
with
\be A=\frac{9\Gamma\left(\frac{3}{10}\right)^2a_1}{(2^65^4v_4)^{1/5}\sqrt{\pi}\Gamma\left(\frac{1}{10}\right)}\,.\ee
Thus, using the definition of $\langle m^2\rangle$ and $\langle \dot{m}^2\rangle$ from Eq.~(\ref{eq:def:av:x}) we find:
%[écrire en fonction de A, simplifier les $\Gamma$]
\begin{align} &\langle m^2\rangle=\frac{c_m}{(a_1v_4)^{1/6}}N^{-1/6},\\ &\langle \dot{m}^2\rangle=c_{\dot{m}}\frac{v_4^{1/6}}{a_1^{5/6}} N^{-5/6},\end{align}
with 
\begin{align}
&c_m=\frac{5^{1/3}\Gamma\left(\frac{3}{10}\right)\Gamma\left(\frac{2}{3}\right)}{3\times 2^{2/5}\pi}\left(\frac{\Gamma\left(\frac{13}{10}\right)}{\Gamma\left(\frac{11}{10}\right)}\right)^{5/6} \! \left(\frac{\Gamma\left(\frac{9}{5}\right)}{\Gamma\left(\frac{8}{5}\right)}\right)^{1/6} \!\approx 0.53, \\
&c_{\dot{m}}=\frac{5^{2/3}\Gamma\left(\frac{3}{5}\right)\Gamma\left(\frac{4}{3}\right)\left(\Gamma\left(\frac{11}{10}\right)\Gamma\left(\frac{9}{5}\right)\right)^{5/6}}{2\sqrt{\pi}\Gamma\left(\frac{8}{5}\right)^{11/6}\Gamma\left(\frac{13}{10}\right)^{5/6}}\approx 1.33.\end{align}

\newpage

\

\newpage
%\maketitle

\begin{widetext} 
  \begin{center}\textbf{\large Supplementary Information: Discontinuous phase transition from ferromagnetic to oscillating states in a nonequilibrium mean-field spin model}\end{center}
\end{widetext}

\setcounter{equation}{0}
\setcounter{figure}{0}

\onecolumngrid

\section*{Corrections to the large deviation function for $\mu=\mu_l$}
Stochastic simulations are made for values of $\ve$ and $N$ for which the assumptions of small $\ve$ and large $N$ are not fully satisfied. We introduce here the first corrections in $\ve$ and in $N^{-1}$ to the results obtained in the main text, which describe well the observations made in the stochastic simulations: the loss of the individual symmetries $m\mapsto-m$ and $\dot{m}\mapsto-\dot{m}$, and the non-uniformity of the probability density along the limit cycle. 
We compute the corrections for $\mu=\mu_l(T)$ for which we can obtain analytical results from series expansion of $V(m)$ and $g(m, \dot{m})$.

\subsection{Introducing higher-order corrections}
The dynamics of the system is controlled by a master equation, which can be expressed in terms of $m$ and $\dot{m}$ as
\be \label{eq:eq:p}
\partial_t P_N(m, \dot{m})=N\sum_{k,\sigma} \bigg[ 
-W_k^{\sigma}(m, \dot{m})P_N(m, \dot{m})+W_k^{\sigma}\left((m, \dot{m})-\frac{\sigma\textbf{d}_{k}}{N}\right)P_N\left((m, \dot{m})-\frac{\sigma\textbf{d}_{k}}{N}\right)\bigg],
 \ee
where $k=1, 2$, $\sigma=\pm$, $\mathbf{d}_k$ are the jumps in $m$ and $\dot{m}$ when flipping a spin or a field, and $W_k^{\sigma}$ are the coarse-grained transition rates introduced in the main text, see Eq.~(26).

A first assumption is that the probability density $P_N(m, \dot{m})$ follows a large deviation principle $P_N(m, \dot{m})\sim \exp[-N\phi(m, \dot{m})]$, with $\phi(m, \dot{m})$ the large deviation function. 
Under this assumption, a limit cycle in the phase space $(m, \dot{m})$ corresponds to a minimum of $\phi(m, \dot{m})$ and has a uniform probability at leading order in $N$. 
To obtain a non-uniform probability along the limit cycle, it is necessary to introduce corrections in $N$. We introduce the function $C(m, \dot{m})$ as   
\be \label{eq:PN:corr:SM}
P_N(m, \dot{m}) \propto \exp[-N\phi(m, \dot{m})+C(m, \dot{m})]\,.
\ee
Injecting this expression into Eq.~(\ref{eq:eq:p}) and expanding to lowest order in $N$, we get the following equation:
\be\label{eq:phi:total}  0=\dot{m}\partial_m\phi+[-V'(m)+\dot{m}g(m, \dot{m})]\partial_{\dot{m}}\phi+D_{ij}\partial_i\phi\partial_j\phi,\ee
where the functions $V'(m)$, $g(m, \dot{m})$ and the coefficients $D_{ij}$ were introduced in the main text, and using implicit summation over repeated indices: $D_{ij}\partial_i\phi\partial_j\phi=D_{11}(\partial_m\phi)^2+D_{22}(\partial_{\dot{m}}\phi)^2+2D_{12}\partial_m\phi\partial_{\dot{m}}\phi$.
Eq.~(\ref{eq:PN:corr:SM}) corresponds to the same equation on $\phi(m, \dot{m})$ as the one used in the main text, Eq.~(29).
At the next order in $N$, we obtain an equation on $C(m, \dot{m})$ which depends on the derivatives of $\phi(m, \dot{m})$:
\be \label{eq:C:total} 0=\dot{m}\partial_mC +[-V'(m)+\dot{m}g(m, \dot{m})]\partial_{\dot{m}}C+2D_{ij}\partial_i\phi\partial_j C+D_{ij}\partial_i\partial_j\phi+\partial_{\dot{m}}\dot{m}g(m,\dot{m}).\ee

\subsection{Change of variables}
As suggested by the expression of $\phi(m,\dot{m})$ obtained in the main text at the lowest order in $\ve$, we introduce the change of variable $(m, \dot{m})\to (m, H(m, \dot{m}))$ with 
\be H(m, \dot{m})=V(m)+\frac{\dot{m}^2}{2}.\ee
We write $\dot{m}=\sqrt{2[H-V(m)]}>0$ and we introduce 
\be \phi^{\pm}(m, H)=\phi(m, \pm \dot{m}).\ee
The equation on $\phi(m, \dot{m})$, Eq.~(\ref{eq:phi:total}), becomes an equation on $\phi^{\pm}(m, H)$:
\be \label{eq:phi:total:m:H} 0=\partial_m \phi^{\pm}\pm \dot{m} g(m, \dot{m})\partial_H \phi^{\pm}\pm D_{22}\dot{m}(\partial_H\phi^{\pm})^2+2D_{12}(\partial_m\phi^{\pm}+V'(m)\partial_H\phi^{\pm})\partial_{H}\phi^{\pm}+D_{11}(\partial_m \phi^{\pm}+V'(m)\partial_H\phi^{\pm})^2\ee
where we make the dependence on $(m, H)$ of $\phi^{\pm}(m, H)$ and $\dot{m}(m, H)$ implicit. 
The equation on $C(m, \dot{m})$, Eq.~(\ref{eq:C:total}), becomes an equation on $C^{\pm}(m, H)$:
\be \label{eq:C:total:m:h}\begin{aligned} 0=&\pm \dot{m}\partial_m C^{\pm}\pm \dot{m}g( m, \dot{m})\partial_H C^{\pm}+\dot{m}\partial_H[\dot{m}g(m, \dot{m})]+2D_{22}\dot{m}^2\partial_H\phi^{\pm}\,\partial_H C^{\pm}+D_{22}\dot{m}\partial_H(\dot{m}\partial_H\phi^{\pm}) \\
&+2D_{11}[\partial_m+V'(m)\partial_H]\phi^{\pm}\, [\partial_m+V'(m)\partial_H]C^{\pm}+D_{11}[\partial_m+V'(m)\partial_H]^2\phi^{\pm}\\
&\pm 2D_{12}\dot{m}\partial_H C^{\pm} [\partial_m+V'(m)\partial_H]\phi^{\pm}
\pm 2D_{12}\dot{m}\partial_H\phi^{\pm}\, [\partial_m+V'(m)\partial_H]C^{\pm}\\
&\pm D_{12}(\partial_m+V'(m)\partial_H)(\dot{m}\partial_H \phi^{\pm})
\pm D_{12}\dot{m}\partial_H[(\partial_m+V'(m)\partial_H)\phi^{\pm}].\end{aligned}\ee

The second assumption introduced in the main text is that $\phi(m, \dot{m})\sim \ve$ without introducing the dependence in $\ve$ of $m$ and $\dot{m}$. Thus, we expand $\phi^{\pm}(m, H)$ and $C^{\pm}(m, H)$ in power series of $\ve$. To make the dependencies in $\ve$ explicit, we introduce rescaled variables for $m$ and $H$. For $\mu=\mu_l(T)$, we obtained a limit cycle with $m\sim \ve^{1/2}$, $\dot{m}\sim \ve$ and $H=H^*\sim \ve^2$. Thus, we consider the rescaled variables $\tilde{H}=H/H^*$ and $x=m/m_0$ with $H^*=(2\ve a_0/3cD_{22})^2\sim \ve^2$ and  $m_0=(4H^*/v_1)^{1/4}\sim \ve^{1/2}$. 
At the lowest order in $\ve$ we have $\phi(m, \dot{m})=-\ve a_0 H/D_{22}+cH^{3/2}$ (see Eq.~(52) of the main text), thus $\phi(m, \dot{m})\sim \ve^{3}$ close to its minimum. 
From the stochastic simulations, we expect corrections in $x=m/m_0\sim\ve^{1/2}$ to the large deviation function.
Thus, we expand $\phi^{\pm}(m, H)$ and $C^{\pm}(m,H)$ in power series of $\ve^{1/2}$ and we write:
\begin{align} & \label{eq:phi:dvp}\phi^{\pm}(m,H)=\ve^{3}\sum_{i=1}^{\infty} \ve^{(i-1)/2}\tilde{\phi}^{\pm}_i(x, \tilde{H})\\
& \label{eq:C:dvp}C^{\pm}(m, H)=\sum_{i=0}^{\infty} \ve^{i/2}\tilde{C}^{\pm}_i(x, \tilde{H}).\end{align}
For later convenience, we also introduce functions of the non-rescaled variables:
\begin{align}&\phi^{\pm}_i(m,H)=\ve^{3+(i-1)/2}\tilde{\phi}^{\pm}_i(x, \tilde{H})\\&C_i^{\pm}(m, H)=\ve^{i/2}\tilde{C}^{\pm}_i(x, \tilde{H}).\end{align}

\subsection{Correction in $\ve$ of $\phi(m,H)$}
We aim in this section at obtaining the first correction in $\ve$ of the large deviation function $\phi^{\pm}(m, H)$. Thus, we look at the different orders in $\ve$ of Eq.~(\ref{eq:phi:total:m:H}). 
The lowest order of Eq.~(\ref{eq:phi:total:m:H}) is of order $\ve^{7/2}$ and reads:
\be \partial_m \phi_1^{\pm}(m, H)=0.\ee
Thus, one finds $\phi_1^{\pm}(m, H)=f(H)$ as obtained in the main text, with $f(H)$ an arbitrary function for now. 
The next order of Eq.~(\ref{eq:phi:total:m:H}) is of order $\ve^{4}$ and reads:
\be  \label{eq:phi2}\partial_m \phi^{\pm}_2\pm \dot{m}g(m, \dot{m})\partial_H\phi^{\pm}_1\pm D_{22}\dot{m}(\partial_H\phi^{\pm}_1)^2=0,\ee
where we consider that $D_{22}$ is constant and non-zero at lowest order in $\ve$. 
We introduce $m_1(H)$ and $m_2(H)$ such that for $m\in[m_1, m_2]$ one has $V(m)\leq H$ and $V(m_1)=V(m_2)=H$. Using the continuity of $\phi(m, \dot{m})$ in $m_1$ and $m_2$, we have
\be \label{eq:integration}\int_{m_1}^{m_2}dm\, \partial_m\phi^++\int_{m_2}^{m_1}dm\, \partial_m\phi^-=0.\ee
Integrating the remaining term of Eq.~(\ref{eq:phi2}), we find the expression of $f'(H)$ given in the main text, 
\be \label{eq:fpH}\partial_H\phi^{\pm}_1(m, H)=f'(H)=-\frac{\int_{m_1}^{m_2}dm\, \dot{m}g(m, \dot{m})}{D_{22}\int_{m_1}^{m_2}dm \,\dot{m}},\ee
which then gives $\phi_1^{\pm}(m, H)=\int dH f'(H)+f_1^{\pm}$. As $\phi_1$ is continuous in $\dot{m}=0$, one finds $f_1^{+}=f_1^{-}$.
For $\mu=\mu_l(T)$, we have at the lowest order in $\ve$,
\begin{align}&\label{eq:v}V(m)=\frac{v_1}{4}m^4\,,\\
&\label{eq:g}g(m, \dot{m})=\ve a_0-a_1m^2.\end{align}
Thus, we find
\be f(H)=-\ve a H+cH^{3/2},\ee 
with $a=a_0/D_{22}$, $c=2a_1\alpha/ 3D_{22}\sqrt{v_1}$ and 
\be \alpha=\frac{\Gamma(7/4)\Gamma(3/4)}{\Gamma(9/4)\Gamma(1/4)}.\ee

We now use Eq.~(\ref{eq:phi2}) and higher orders of Eq.~(\ref{eq:phi:total:m:H}) to obtain an expression for $\phi_2^{\pm}(m, H)$.
Using Eq.~(\ref{eq:fpH}) for $f'(H)$, the equation on $\phi^{\pm}_2$ [Eq.~(\ref{eq:phi2})] becomes:
\be \partial_m\phi^{\pm}_2=\pm \frac{\left(\int_{m_1}^{m_2}dm'\dot{m}g(m', \dot{m}) \right)^2}{D_{22}\int_{m_1}^{m_2}dm'\dot{m}}\left(\frac{\dot{m}g(m, \dot{m})}{\int_{m_1}^{m_2}dm'\dot{m}g(m', \dot{m})}-\frac{\dot{m}}{\int_{m_1}^{m_2}dm'\dot{m}}\right).\ee
Integrating this equation from $m_1$ to an arbitrary value $m$, we obtain:
\be \label{eq:phi2:result:temp} \phi^{\pm}_2\left(m,H\right)=\pm \frac{\left(\int_{m_1}^{m_2}dm'\dot{m}g(m', \dot{m}) \right)^2}{D_{22}\int_{m_1}^{m_2}dm'\dot{m}}\left(\frac{\int_{m_1}^mdm'\dot{m}g(m', \dot{m})}{\int_{m_1}^{m_2}dm'\dot{m}g(m', \dot{m})}-\frac{\int_{m_1}^m dm'\dot{m}}{\int_{m_1}^{m_2}dm'\dot{m}}\right)+f^{\pm}_2(H)\ee
where $f^{\pm}_2(H)$ is independent of $m$ and is to be determined.
By continuity in $\dot{m}=0$, i.e., in $m=m_1$ or $m=m_2$, we obtain that $f_2^+(H)=f_2^-(H)$. %$=f_2(H)$.

To obtain $f_2^+(H)$, we consider the next order in $\ve$ of Eq.~(\ref{eq:phi:total}) which is of order $\ve^{9/2}$ and gives an equation involving $\phi_3$ and $\phi_2$:
\be \label{eq:phi2:1}0=\partial_m \phi^{\pm}_3\pm \dot{m}g(m, \dot{m})\partial_H\phi_2^{\pm}\pm D_{22}\dot{m}(\partial_H\phi_2^{\pm})^2+2D_{12}V'(m)(\partial_H\phi_1^{\pm})^2.\ee
Performing the same integration as in Eq.~(\ref{eq:integration}), we find that the first term cancels out due to the continuity of $\phi_3$ in $m_1$ and $m_2$, the last term gives a zero contribution as $V'(m)$ is odd in $m$, and the linear term in $\partial_H \phi_2^{\pm}$ in Eq.~(\ref{eq:phi2:1}) does not contribute as it is odd in $\dot{m}$. Thus, one finally finds that ${f^+_2}'(H)$ verifies the same equation as $f'(H)$ in Eq.~(\ref{eq:fpH}).

Using Eqs.~(\ref{eq:v}) and (\ref{eq:g}) in Eq.~(\ref{eq:phi2:result:temp}), we obtain the expression of $\phi(x, H)$ [or $\tilde{\phi}_2(x, \tilde{H})$ in the rescaled variables] given in the main text in Eq.~(59):
\be \tilde{\phi}_2(x, \tilde{H})=A\left(1-\sqrt{\tilde{H}}\right)\tilde{H} x \left[a_0\, {}_{2}F_1\left(-\frac{1}{2}, \frac{1}{4},\frac{5}{4}, x^4\right)-\frac{2}{3}\frac{D_{22}}{\alpha}x^2 {}_2F_1\left(-\frac{1}{2}, \frac{3}{4},\frac{7}{4}, x^4\right)\right]+\tilde{f}(\tilde{H}),\ee 
where \be A=\frac{2v_1^{1/2}a_0^{7/2}}{(a_1\alpha)^{3/2}D_{22}^2},\ee 
and ${}_2F_1$ is a hyper-geometric function, $\phi_2(m, \dot{m})=\ve^{7/2}\tilde{\phi}_2(x, \tilde{H})$ and
with $\tilde{f}(\tilde{H})=-a\tilde{H}+c\tilde{H}^{3/2}$.

\subsection{Corrections in $\ve$ of $C(m, \dot{m})$}
We now aim to find the first order in $\ve$ of the function $C(m,\dot{m})$ corresponding to corrections in $N$ to the large deviation function. 
The leading order in $\ve$ of Eq.~(\ref{eq:C:total:m:h}) is in $\ve^{1/2}$ and gives:
\be 0=\partial_m C_0(m, H).\ee
Thus, we find $C_0(m, H)=C_0(H)$. 
The next order of Eq.~(\ref{eq:C:total:m:h}) is of order $\ve$ and gives an equation on $C_1$, $C_0$ and $f'(H)=\partial_H\phi_1(m, H)$: 
\be\label{eq:C1} \partial_m C_1^{\pm}=- \left[\dot{m}g(m,\dot{m})+2D_{22}\dot{m}f'(H)\right]\partial_HC_0 
\mp\left[\partial_{H}[\dot{m}g(m, \dot{m})]+D_{22}\dot{m}^{-1}f'(H)+D_{22}\dot{m}f''(H)\right].\ee 
From Eq.~(\ref{eq:fpH}) and using $\dot{m}^{-1}=\partial_H \dot{m}^2$, we can write $f''(H)$ as 
\be f''(H)=-\frac{\int_{m_1}^{m_2}dm\,\partial_H[\dot{m}g(m, \dot{m})]\int_{m_1}^{m_2}dm\,\dot{m}-\int_{m_1}^{m_1}dm \,\dot{m}g(m, \dot{m})\int_{m_1}^{m_2}dm \,\dot{m}^{-1}}{D_{22}\left(\int_{m_1}^{m_2}dm \dot{m}\right)^2}.\ee
We perform on Eq.~(\ref{eq:C1}) the same type of integration as in Eq.~(\ref{eq:integration}) and we get:
\be\label{eq:c0:0} \partial_H C_0=0.\ee
Thus, we finally find that $C_0(m, H)=C_0$ is a constant term, which can be obtained from the normalization of $P_N(m,\dot{m})$, i.e.,
$\int_{-1}^{1} dm \, P_N(m,\dot{m})=1$.
Using this result, we can write Eq.~(\ref{eq:C1}) as
\be \begin{split}\partial_m C_1^{\pm}(m,H)&=
\mp \int_{m_1}^{m_1}dm'\partial_H[\dot{m}g(m', \dot{m})]\left(\frac{\partial_H[\dot{m}g(m, \dot{m})]}{\int_{m_1}^{m_2}dm'\partial_H[\dot{m}g(m, \dot{m})]}-\frac{\dot{m}}{\int_{m_1}^{m_2}dm'\dot{m}}\right)\\
&\pm\frac{\int_{m_1}^{m_2}dm'\dot{m}g(m', \dot{m})\int_{m_1}^{m_2}dm '\dot{m}^{-1}}{\int_{m_1}^{m_2}dm' D_{22}\dot{m}}\left(\frac{\dot{m}}{\int_{m_1}^{m_2}dm'\dot{m}}-\frac{\dot{m}^{-1}}{\int_{m_1}^{m_2}dm'\dot{m}^{-1}}\right).\end{split}\ee
Integrating from $m_1$ to an arbitrary value $m$, we find:
\be \begin{split} C_1^{\pm}(m, H)&=\mp\int_{m_1}^{m_1}dm'\partial_H[\dot{m}g(m', \dot{m})]\left(\frac{\int_{m_1}^m dm'\, \partial_H[\dot{m}g(m, \dot{m})]}{\int_{m_1}^{m_2}dm'\,\partial_H[\dot{m}g(m, \dot{m})]}-\frac{\int_{m_1}^m dm'\, \dot{m}}{\int_{m_1}^{m_2}dm'\,\dot{m}}\right)\\
&\pm\frac{\int_{m_1}^{m_2}dm'\dot{m}g(m', \dot{m})\int_{m_1}^{m_2}dm '\dot{m}^{-1}}{\int_{m_1}^{m_2}dm' D_{22}\dot{m}}\left(\frac{\int_{m_1}^m dm'\,\dot{m}}{\int_{m_1}^{m_2}dm'\dot{m}}-\frac{\int_{m_1}^m dm'\,\dot{m}^{-1}}{\int_{m_1}^{m_2}dm'\dot{m}^{-1}}\right)+c_1^{\pm}.\end{split}\ee
The continuity in $\dot{m}=0$ gives $c_1^+ =c_1^-$ and the equation at the next order in $\ve$ (of order $\ve^{3/2}$) gives $c_1^+(H)=0$ in the same way as Eq.~(\ref{eq:c0:0}) was obtained.

For $\mu=\mu_l(T)$, using Eqs.~(\ref{eq:v}) and (\ref{eq:g}), we find the expression given in the main text Eq.~(59):
\be  \tilde{C_1}(x, \tilde{H})=B\tilde{H}^{1/4}\left[-6a_0x\sqrt{1-x^4}+30a_0\sqrt{v_1}x{}_2F_1\left(-\frac{1}{2}, \frac{1}{4}, \frac{5}{4}, x^4\right)-8\frac{D_{22}}{\alpha}x^3{}_2F_1\left(\frac{1}{2}, \frac{5}{4}, \frac{7}{4}, x^4\right)\right]\ee
with $C_1(m, \dot{m})=\ve^{1/2}\tilde{C}_1(x, \tilde{H})$ and 
\be B=\frac{1}{12 D_{22}}\sqrt{\frac{\alpha a_0a_1}{v_1}}.\ee


\begin{thebibliography}{55}%
\makeatletter
\providecommand \@ifxundefined [1]{%
 \@ifx{#1\undefined}
}%
\providecommand \@ifnum [1]{%
 \ifnum #1\expandafter \@firstoftwo
 \else \expandafter \@secondoftwo
 \fi
}%
\providecommand \@ifx [1]{%
 \ifx #1\expandafter \@firstoftwo
 \else \expandafter \@secondoftwo
 \fi
}%
\providecommand \natexlab [1]{#1}%
\providecommand \enquote  [1]{``#1''}%
\providecommand \bibnamefont  [1]{#1}%
\providecommand \bibfnamefont [1]{#1}%
\providecommand \citenamefont [1]{#1}%
\providecommand \href@noop [0]{\@secondoftwo}%
\providecommand \href [0]{\begingroup \@sanitize@url \@href}%
\providecommand \@href[1]{\@@startlink{#1}\@@href}%
\providecommand \@@href[1]{\endgroup#1\@@endlink}%
\providecommand \@sanitize@url [0]{\catcode `\\12\catcode `\$12\catcode
  `\&12\catcode `\#12\catcode `\^12\catcode `\_12\catcode `\%12\relax}%
\providecommand \@@startlink[1]{}%
\providecommand \@@endlink[0]{}%
\providecommand \url  [0]{\begingroup\@sanitize@url \@url }%
\providecommand \@url [1]{\endgroup\@href {#1}{\urlprefix }}%
\providecommand \urlprefix  [0]{URL }%
\providecommand \Eprint [0]{\href }%
\providecommand \doibase [0]{https://doi.org/}%
\providecommand \selectlanguage [0]{\@gobble}%
\providecommand \bibinfo  [0]{\@secondoftwo}%
\providecommand \bibfield  [0]{\@secondoftwo}%
\providecommand \translation [1]{[#1]}%
\providecommand \BibitemOpen [0]{}%
\providecommand \bibitemStop [0]{}%
\providecommand \bibitemNoStop [0]{.\EOS\space}%
\providecommand \EOS [0]{\spacefactor3000\relax}%
\providecommand \BibitemShut  [1]{\csname bibitem#1\endcsname}%
\let\auto@bib@innerbib\@empty
%</preamble>
\bibitem [{\citenamefont {Acebr\'on}\ \emph {et~al.}(2005)\citenamefont
  {Acebr\'on}, \citenamefont {Bonilla}, \citenamefont {P\'erez~Vicente},
  \citenamefont {Ritort},\ and\ \citenamefont
  {Spigler}}]{acebron_kuramoto_2005}%
  \BibitemOpen
  \bibfield  {author} {\bibinfo {author} {\bibfnamefont {J.~A.}\ \bibnamefont
  {Acebr\'on}}, \bibinfo {author} {\bibfnamefont {L.~L.}\ \bibnamefont
  {Bonilla}}, \bibinfo {author} {\bibfnamefont {C.~J.}\ \bibnamefont
  {P\'erez~Vicente}}, \bibinfo {author} {\bibfnamefont {F.}~\bibnamefont
  {Ritort}},\ and\ \bibinfo {author} {\bibfnamefont {R.}~\bibnamefont
  {Spigler}},\ }\href {https://doi.org/10.1103/RevModPhys.77.137} {\bibfield
  {journal} {\bibinfo  {journal} {Rev. Mod. Phys.}\ }\textbf {\bibinfo {volume}
  {77}},\ \bibinfo {pages} {137} (\bibinfo {year} {2005})}\BibitemShut
  {NoStop}%
\bibitem [{\citenamefont {Gupta}\ \emph {et~al.}(2014)\citenamefont {Gupta},
  \citenamefont {Campa},\ and\ \citenamefont {Ruffo}}]{gupta_kuramoto_2014}%
  \BibitemOpen
  \bibfield  {author} {\bibinfo {author} {\bibfnamefont {S.}~\bibnamefont
  {Gupta}}, \bibinfo {author} {\bibfnamefont {A.}~\bibnamefont {Campa}},\ and\
  \bibinfo {author} {\bibfnamefont {S.}~\bibnamefont {Ruffo}},\ }\href@noop {}
  {\bibfield  {journal} {\bibinfo  {journal} {J. Stat. Mech.: Theor. Exp.}\ ,\
  \bibinfo {pages} {R08001}} (\bibinfo {year} {2014})}\BibitemShut {NoStop}%
\bibitem [{\citenamefont {Risler}\ \emph {et~al.}(2004)\citenamefont {Risler},
  \citenamefont {Prost},\ and\ \citenamefont
  {J\"ulicher}}]{risler_universal_2004}%
  \BibitemOpen
  \bibfield  {author} {\bibinfo {author} {\bibfnamefont {T.}~\bibnamefont
  {Risler}}, \bibinfo {author} {\bibfnamefont {J.}~\bibnamefont {Prost}},\ and\
  \bibinfo {author} {\bibfnamefont {F.}~\bibnamefont {J\"ulicher}},\
  }\href@noop {} {\bibfield  {journal} {\bibinfo  {journal} {Phys. Rev. Lett.}\
  }\textbf {\bibinfo {volume} {93}},\ \bibinfo {pages} {175702} (\bibinfo
  {year} {2004})}\BibitemShut {NoStop}%
\bibitem [{\citenamefont {Risler}\ \emph {et~al.}(2005)\citenamefont {Risler},
  \citenamefont {Prost},\ and\ \citenamefont
  {J\"ulicher}}]{risler_universal_2005}%
  \BibitemOpen
  \bibfield  {author} {\bibinfo {author} {\bibfnamefont {T.}~\bibnamefont
  {Risler}}, \bibinfo {author} {\bibfnamefont {J.}~\bibnamefont {Prost}},\ and\
  \bibinfo {author} {\bibfnamefont {F.}~\bibnamefont {J\"ulicher}},\
  }\href@noop {} {\bibfield  {journal} {\bibinfo  {journal} {Phys. Rev. E}\
  }\textbf {\bibinfo {volume} {72}},\ \bibinfo {pages} {016130} (\bibinfo
  {year} {2005})}\BibitemShut {NoStop}%
\bibitem [{\citenamefont {Nicolis}(1986)}]{nicolis_dissipative_1986}%
  \BibitemOpen
  \bibfield  {author} {\bibinfo {author} {\bibfnamefont {G.}~\bibnamefont
  {Nicolis}},\ }\href {https://doi.org/10.1088/0034-4885/49/8/002} {\bibfield
  {journal} {\bibinfo  {journal} {Rep. Prog. Phys.}\ }\textbf {\bibinfo
  {volume} {49}},\ \bibinfo {pages} {873} (\bibinfo {year} {1986})}\BibitemShut
  {NoStop}%
\bibitem [{\citenamefont {Kamino}\ \emph {et~al.}(2017)\citenamefont {Kamino},
  \citenamefont {Kondo}, \citenamefont {Nakajima}, \citenamefont
  {Honda-Kitahara}, \citenamefont {Kaneko},\ and\ \citenamefont
  {Sawai}}]{Kamino_fold2017}%
  \BibitemOpen
  \bibfield  {author} {\bibinfo {author} {\bibfnamefont {K.}~\bibnamefont
  {Kamino}}, \bibinfo {author} {\bibfnamefont {Y.}~\bibnamefont {Kondo}},
  \bibinfo {author} {\bibfnamefont {A.}~\bibnamefont {Nakajima}}, \bibinfo
  {author} {\bibfnamefont {M.}~\bibnamefont {Honda-Kitahara}}, \bibinfo
  {author} {\bibfnamefont {K.}~\bibnamefont {Kaneko}},\ and\ \bibinfo {author}
  {\bibfnamefont {S.}~\bibnamefont {Sawai}},\ }\href@noop {} {\bibfield
  {journal} {\bibinfo  {journal} {Proc. Natl. Acad. Sci. USA}\ }\textbf
  {\bibinfo {volume} {114}},\ \bibinfo {pages} {E4149} (\bibinfo {year}
  {2017})}\BibitemShut {NoStop}%
\bibitem [{\citenamefont {Wang}\ and\ \citenamefont
  {Tang}(2019)}]{Wang_emergence2019}%
  \BibitemOpen
  \bibfield  {author} {\bibinfo {author} {\bibfnamefont {S.-W.}\ \bibnamefont
  {Wang}}\ and\ \bibinfo {author} {\bibfnamefont {L.-H.}\ \bibnamefont
  {Tang}},\ }\href@noop {} {\bibfield  {journal} {\bibinfo  {journal} {Nat.
  Commun.}\ }\textbf {\bibinfo {volume} {10}},\ \bibinfo {pages} {5613}
  (\bibinfo {year} {2019})}\BibitemShut {NoStop}%
\bibitem [{\citenamefont {Saha}\ \emph {et~al.}(2020)\citenamefont {Saha},
  \citenamefont {Agudo-Canalejo},\ and\ \citenamefont
  {Golestanian}}]{saha_scalar_2020}%
  \BibitemOpen
  \bibfield  {author} {\bibinfo {author} {\bibfnamefont {S.}~\bibnamefont
  {Saha}}, \bibinfo {author} {\bibfnamefont {J.}~\bibnamefont
  {Agudo-Canalejo}},\ and\ \bibinfo {author} {\bibfnamefont {R.}~\bibnamefont
  {Golestanian}},\ }\href@noop {} {\bibfield  {journal} {\bibinfo  {journal}
  {Phys. Rev. X}\ }\textbf {\bibinfo {volume} {10}},\ \bibinfo {pages} {041009}
  (\bibinfo {year} {2020})}\BibitemShut {NoStop}%
\bibitem [{\citenamefont {You}\ \emph {et~al.}(2020)\citenamefont {You},
  \citenamefont {Baskaran},\ and\ \citenamefont
  {Marchetti}}]{you_nonreciprocity_2020}%
  \BibitemOpen
  \bibfield  {author} {\bibinfo {author} {\bibfnamefont {Z.}~\bibnamefont
  {You}}, \bibinfo {author} {\bibfnamefont {A.}~\bibnamefont {Baskaran}},\ and\
  \bibinfo {author} {\bibfnamefont {M.~C.}\ \bibnamefont {Marchetti}},\
  }\href@noop {} {\bibfield  {journal} {\bibinfo  {journal} {Proc. Natl. Acad.
  Sci. USA}\ }\textbf {\bibinfo {volume} {117}},\ \bibinfo {pages} {19767}
  (\bibinfo {year} {2020})}\BibitemShut {NoStop}%
\bibitem [{\citenamefont {Cao}\ \emph {et~al.}(2015)\citenamefont {Cao},
  \citenamefont {Wang}, \citenamefont {Ouyang},\ and\ \citenamefont
  {Tu}}]{Cao_free_energy2015}%
  \BibitemOpen
  \bibfield  {author} {\bibinfo {author} {\bibfnamefont {Y.}~\bibnamefont
  {Cao}}, \bibinfo {author} {\bibfnamefont {H.}~\bibnamefont {Wang}}, \bibinfo
  {author} {\bibfnamefont {Q.}~\bibnamefont {Ouyang}},\ and\ \bibinfo {author}
  {\bibfnamefont {Y.}~\bibnamefont {Tu}},\ }\href@noop {} {\bibfield  {journal}
  {\bibinfo  {journal} {Nat. Phys.}\ }\textbf {\bibinfo {volume} {11}},\
  \bibinfo {pages} {772} (\bibinfo {year} {2015})}\BibitemShut {NoStop}%
\bibitem [{\citenamefont {Nguyen}\ \emph {et~al.}(2018)\citenamefont {Nguyen},
  \citenamefont {Seifert},\ and\ \citenamefont {Barato}}]{nguyen_phase_2018}%
  \BibitemOpen
  \bibfield  {author} {\bibinfo {author} {\bibfnamefont {B.}~\bibnamefont
  {Nguyen}}, \bibinfo {author} {\bibfnamefont {U.}~\bibnamefont {Seifert}},\
  and\ \bibinfo {author} {\bibfnamefont {A.~C.}\ \bibnamefont {Barato}},\
  }\href {https://doi.org/10.1063/1.5032104} {\bibfield  {journal} {\bibinfo
  {journal} {J. Chem. Phys.}\ }\textbf {\bibinfo {volume} {149}},\ \bibinfo
  {pages} {045101} (\bibinfo {year} {2018})}\BibitemShut {NoStop}%
\bibitem [{\citenamefont {Aufinger}\ \emph {et~al.}(2022)\citenamefont
  {Aufinger}, \citenamefont {Brenner},\ and\ \citenamefont
  {Simmel}}]{Aufinger_complex2022}%
  \BibitemOpen
  \bibfield  {author} {\bibinfo {author} {\bibfnamefont {L.}~\bibnamefont
  {Aufinger}}, \bibinfo {author} {\bibfnamefont {J.}~\bibnamefont {Brenner}},\
  and\ \bibinfo {author} {\bibfnamefont {F.~C.}\ \bibnamefont {Simmel}},\
  }\href@noop {} {\bibfield  {journal} {\bibinfo  {journal} {Nat. Commun.}\
  }\textbf {\bibinfo {volume} {13}},\ \bibinfo {pages} {2852} (\bibinfo {year}
  {2022})}\BibitemShut {NoStop}%
\bibitem [{\citenamefont {Devailly}\ \emph {et~al.}(2015)\citenamefont
  {Devailly}, \citenamefont {Crauste-Thibierge}, \citenamefont {Petrosyan},\
  and\ \citenamefont {Ciliberto}}]{devailly_phase_2015}%
  \BibitemOpen
  \bibfield  {author} {\bibinfo {author} {\bibfnamefont {C.}~\bibnamefont
  {Devailly}}, \bibinfo {author} {\bibfnamefont {C.}~\bibnamefont
  {Crauste-Thibierge}}, \bibinfo {author} {\bibfnamefont {A.}~\bibnamefont
  {Petrosyan}},\ and\ \bibinfo {author} {\bibfnamefont {S.}~\bibnamefont
  {Ciliberto}},\ }\href@noop {} {\bibfield  {journal} {\bibinfo  {journal}
  {Phys. Rev. E}\ }\textbf {\bibinfo {volume} {92}},\ \bibinfo {pages} {052312}
  (\bibinfo {year} {2015})}\BibitemShut {NoStop}%
\bibitem [{\citenamefont {Andrae}\ \emph {et~al.}(2010)\citenamefont {Andrae},
  \citenamefont {Cremer}, \citenamefont {Reichenbach},\ and\ \citenamefont
  {Frey}}]{andrae_entropy_2010}%
  \BibitemOpen
  \bibfield  {author} {\bibinfo {author} {\bibfnamefont {B.}~\bibnamefont
  {Andrae}}, \bibinfo {author} {\bibfnamefont {J.}~\bibnamefont {Cremer}},
  \bibinfo {author} {\bibfnamefont {T.}~\bibnamefont {Reichenbach}},\ and\
  \bibinfo {author} {\bibfnamefont {E.}~\bibnamefont {Frey}},\ }\href
  {https://doi.org/10.1103/PhysRevLett.104.218102} {\bibfield  {journal}
  {\bibinfo  {journal} {Phys. Rev. Lett.}\ }\textbf {\bibinfo {volume} {104}},\
  \bibinfo {pages} {218102} (\bibinfo {year} {2010})}\BibitemShut {NoStop}%
\bibitem [{\citenamefont {Duan}\ \emph {et~al.}(2019)\citenamefont {Duan},
  \citenamefont {Niu},\ and\ \citenamefont {Wei}}]{Duan_Hopf2019}%
  \BibitemOpen
  \bibfield  {author} {\bibinfo {author} {\bibfnamefont {D.}~\bibnamefont
  {Duan}}, \bibinfo {author} {\bibfnamefont {B.}~\bibnamefont {Niu}},\ and\
  \bibinfo {author} {\bibfnamefont {J.}~\bibnamefont {Wei}},\ }\href@noop {}
  {\bibfield  {journal} {\bibinfo  {journal} {Chaos, Solitons and Fractals}\
  }\textbf {\bibinfo {volume} {123}},\ \bibinfo {pages} {206} (\bibinfo {year}
  {2019})}\BibitemShut {NoStop}%
\bibitem [{\citenamefont {Gualdi}\ \emph {et~al.}(2015)\citenamefont {Gualdi},
  \citenamefont {Bouchaud}, \citenamefont {Cencetti}, \citenamefont {Tarzia},\
  and\ \citenamefont {Zamponi}}]{Gualdi2015}%
  \BibitemOpen
  \bibfield  {author} {\bibinfo {author} {\bibfnamefont {S.}~\bibnamefont
  {Gualdi}}, \bibinfo {author} {\bibfnamefont {J.-P.}\ \bibnamefont
  {Bouchaud}}, \bibinfo {author} {\bibfnamefont {G.}~\bibnamefont {Cencetti}},
  \bibinfo {author} {\bibfnamefont {M.}~\bibnamefont {Tarzia}},\ and\ \bibinfo
  {author} {\bibfnamefont {F.}~\bibnamefont {Zamponi}},\ }\href@noop {}
  {\bibfield  {journal} {\bibinfo  {journal} {Phys. Rev. Lett.}\ }\textbf
  {\bibinfo {volume} {114}},\ \bibinfo {pages} {088701} (\bibinfo {year}
  {2015})}\BibitemShut {NoStop}%
\bibitem [{\citenamefont {Yi}\ \emph {et~al.}(2015)\citenamefont {Yi},
  \citenamefont {Baek}, \citenamefont {Chevereau},\ and\ \citenamefont
  {Bertin}}]{yi_symmetry_2015}%
  \BibitemOpen
  \bibfield  {author} {\bibinfo {author} {\bibfnamefont {S.~D.}\ \bibnamefont
  {Yi}}, \bibinfo {author} {\bibfnamefont {S.~K.}\ \bibnamefont {Baek}},
  \bibinfo {author} {\bibfnamefont {G.}~\bibnamefont {Chevereau}},\ and\
  \bibinfo {author} {\bibfnamefont {E.}~\bibnamefont {Bertin}},\ }\href
  {https://doi.org/10.1088/1742-5468/2015/11/P11001} {\bibfield  {journal}
  {\bibinfo  {journal} {J. Stat. Mech.: Theor. Exp.}\ ,\ \bibinfo {pages}
  {P11001}} (\bibinfo {year} {2015})}\BibitemShut {NoStop}%
\bibitem [{\citenamefont {Collet}\ \emph {et~al.}(2016)\citenamefont {Collet},
  \citenamefont {Formentin},\ and\ \citenamefont
  {Tovazzi}}]{collet_rhythmic_2016}%
  \BibitemOpen
  \bibfield  {author} {\bibinfo {author} {\bibfnamefont {F.}~\bibnamefont
  {Collet}}, \bibinfo {author} {\bibfnamefont {M.}~\bibnamefont {Formentin}},\
  and\ \bibinfo {author} {\bibfnamefont {D.}~\bibnamefont {Tovazzi}},\
  }\href@noop {} {\bibfield  {journal} {\bibinfo  {journal} {Phys. Rev. E.}\
  }\textbf {\bibinfo {volume} {94}},\ \bibinfo {pages} {042139} (\bibinfo
  {year} {2016})}\BibitemShut {NoStop}%
\bibitem [{\citenamefont {De~Martino}\ and\ \citenamefont
  {Barato}(2019)}]{de_martino_oscillations_2019}%
  \BibitemOpen
  \bibfield  {author} {\bibinfo {author} {\bibfnamefont {D.}~\bibnamefont
  {De~Martino}}\ and\ \bibinfo {author} {\bibfnamefont {A.~C.}\ \bibnamefont
  {Barato}},\ }\href {https://doi.org/10.1103/PhysRevE.100.062123} {\bibfield
  {journal} {\bibinfo  {journal} {Phys. Rev. E}\ }\textbf {\bibinfo {volume}
  {100}},\ \bibinfo {pages} {062123} (\bibinfo {year} {2019})}\BibitemShut
  {NoStop}%
\bibitem [{\citenamefont {Dai~Pra}\ \emph {et~al.}(2020)\citenamefont
  {Dai~Pra}, \citenamefont {Formentin},\ and\ \citenamefont
  {Guglielmo}}]{daipra_oscillatory_2020}%
  \BibitemOpen
  \bibfield  {author} {\bibinfo {author} {\bibfnamefont {P.}~\bibnamefont
  {Dai~Pra}}, \bibinfo {author} {\bibfnamefont {M.}~\bibnamefont {Formentin}},\
  and\ \bibinfo {author} {\bibfnamefont {P.}~\bibnamefont {Guglielmo}},\
  }\href@noop {} {\bibfield  {journal} {\bibinfo  {journal} {J. Stat. Phys.}\
  }\textbf {\bibinfo {volume} {179}},\ \bibinfo {pages} {690} (\bibinfo {year}
  {2020})}\BibitemShut {NoStop}%
\bibitem [{\citenamefont {Crawford}(1991)}]{crawford_introduction_1991}%
  \BibitemOpen
  \bibfield  {author} {\bibinfo {author} {\bibfnamefont {J.~D.}\ \bibnamefont
  {Crawford}},\ }\href {https://doi.org/10.1103/RevModPhys.63.991} {\bibfield
  {journal} {\bibinfo  {journal} {Rev. Mod. Phys.}\ }\textbf {\bibinfo {volume}
  {63}},\ \bibinfo {pages} {991} (\bibinfo {year} {1991})}\BibitemShut
  {NoStop}%
\bibitem [{\citenamefont {Fei}\ \emph {et~al.}(2018)\citenamefont {Fei},
  \citenamefont {Cao}, \citenamefont {Ouyang},\ and\ \citenamefont
  {Tu}}]{Fei_design2018}%
  \BibitemOpen
  \bibfield  {author} {\bibinfo {author} {\bibfnamefont {C.}~\bibnamefont
  {Fei}}, \bibinfo {author} {\bibfnamefont {Y.}~\bibnamefont {Cao}}, \bibinfo
  {author} {\bibfnamefont {Q.}~\bibnamefont {Ouyang}},\ and\ \bibinfo {author}
  {\bibfnamefont {Y.}~\bibnamefont {Tu}},\ }\href@noop {} {\bibfield  {journal}
  {\bibinfo  {journal} {Nat. Commun.}\ }\textbf {\bibinfo {volume} {9}},\
  \bibinfo {pages} {1434} (\bibinfo {year} {2018})}\BibitemShut {NoStop}%
\bibitem [{\citenamefont {Gaspard}(2002)}]{gaspard_correlation_2002}%
  \BibitemOpen
  \bibfield  {author} {\bibinfo {author} {\bibfnamefont {P.}~\bibnamefont
  {Gaspard}},\ }\href {https://doi.org/10.1063/1.1513461} {\bibfield  {journal}
  {\bibinfo  {journal} {J. Chem. Phys.}\ }\textbf {\bibinfo {volume} {117}},\
  \bibinfo {pages} {8905} (\bibinfo {year} {2002})}\BibitemShut {NoStop}%
\bibitem [{\citenamefont {Barato}\ and\ \citenamefont
  {Seifert}(2016)}]{barato_cost_2016}%
  \BibitemOpen
  \bibfield  {author} {\bibinfo {author} {\bibfnamefont {A.~C.}\ \bibnamefont
  {Barato}}\ and\ \bibinfo {author} {\bibfnamefont {U.}~\bibnamefont
  {Seifert}},\ }\href@noop {} {\bibfield  {journal} {\bibinfo  {journal} {Phys.
  Rev. X}\ }\textbf {\bibinfo {volume} {6}},\ \bibinfo {pages} {041053}
  (\bibinfo {year} {2016})}\BibitemShut {NoStop}%
\bibitem [{\citenamefont {Barato}\ and\ \citenamefont
  {Seifert}(2017)}]{barato_coherence_2017}%
  \BibitemOpen
  \bibfield  {author} {\bibinfo {author} {\bibfnamefont {A.~C.}\ \bibnamefont
  {Barato}}\ and\ \bibinfo {author} {\bibfnamefont {U.}~\bibnamefont
  {Seifert}},\ }\href@noop {} {\bibfield  {journal} {\bibinfo  {journal} {Phys.
  Rev. E}\ }\textbf {\bibinfo {volume} {95}},\ \bibinfo {pages} {062409}
  (\bibinfo {year} {2017})}\BibitemShut {NoStop}%
\bibitem [{\citenamefont {Oberreiter}\ \emph {et~al.}(2022)\citenamefont
  {Oberreiter}, \citenamefont {Seifert},\ and\ \citenamefont
  {Barato}}]{oberreiter_universal_2022}%
  \BibitemOpen
  \bibfield  {author} {\bibinfo {author} {\bibfnamefont {L.}~\bibnamefont
  {Oberreiter}}, \bibinfo {author} {\bibfnamefont {U.}~\bibnamefont
  {Seifert}},\ and\ \bibinfo {author} {\bibfnamefont {A.~C.}\ \bibnamefont
  {Barato}},\ }\href@noop {} {\bibfield  {journal} {\bibinfo  {journal} {Phys.
  Rev. E}\ }\textbf {\bibinfo {volume} {106}},\ \bibinfo {pages} {014106}
  (\bibinfo {year} {2022})}\BibitemShut {NoStop}%
\bibitem [{\citenamefont {Remlein}\ \emph {et~al.}(2022)\citenamefont
  {Remlein}, \citenamefont {Weissmann},\ and\ \citenamefont
  {Seifert}}]{remlein_coherence_2022}%
  \BibitemOpen
  \bibfield  {author} {\bibinfo {author} {\bibfnamefont {B.}~\bibnamefont
  {Remlein}}, \bibinfo {author} {\bibfnamefont {V.}~\bibnamefont {Weissmann}},\
  and\ \bibinfo {author} {\bibfnamefont {U.}~\bibnamefont {Seifert}},\
  }\href@noop {} {\bibfield  {journal} {\bibinfo  {journal} {Phys. Rev. E}\
  }\textbf {\bibinfo {volume} {105}},\ \bibinfo {pages} {064101} (\bibinfo
  {year} {2022})}\BibitemShut {NoStop}%
\bibitem [{\citenamefont {Sagu\'es}\ \emph {et~al.}(2007)\citenamefont
  {Sagu\'es}, \citenamefont {Sancho},\ and\ \citenamefont
  {Garc\'{\i}a-Ojalvo}}]{Sagues2007}%
  \BibitemOpen
  \bibfield  {author} {\bibinfo {author} {\bibfnamefont {F.}~\bibnamefont
  {Sagu\'es}}, \bibinfo {author} {\bibfnamefont {J.~M.}\ \bibnamefont
  {Sancho}},\ and\ \bibinfo {author} {\bibfnamefont {J.}~\bibnamefont
  {Garc\'{\i}a-Ojalvo}},\ }\href@noop {} {\bibfield  {journal} {\bibinfo
  {journal} {Rev. Mod. Phys.}\ }\textbf {\bibinfo {volume} {79}},\ \bibinfo
  {pages} {829} (\bibinfo {year} {2007})}\BibitemShut {NoStop}%
\bibitem [{\citenamefont {Xu}\ \emph {et~al.}(2020)\citenamefont {Xu},
  \citenamefont {Luo}, \citenamefont {Wu},\ and\ \citenamefont
  {Huang}}]{Xu_Langevin2020}%
  \BibitemOpen
  \bibfield  {author} {\bibinfo {author} {\bibfnamefont {H.-Y.}\ \bibnamefont
  {Xu}}, \bibinfo {author} {\bibfnamefont {Y.-P.}\ \bibnamefont {Luo}},
  \bibinfo {author} {\bibfnamefont {J.-W.}\ \bibnamefont {Wu}},\ and\ \bibinfo
  {author} {\bibfnamefont {M.-C.}\ \bibnamefont {Huang}},\ }\href@noop {}
  {\bibfield  {journal} {\bibinfo  {journal} {Physica D}\ }\textbf {\bibinfo
  {volume} {411}},\ \bibinfo {pages} {132612} (\bibinfo {year}
  {2020})}\BibitemShut {NoStop}%
\bibitem [{\citenamefont {Crochik}\ and\ \citenamefont
  {Tom\'e}(2005)}]{crochik_entropy_2005}%
  \BibitemOpen
  \bibfield  {author} {\bibinfo {author} {\bibfnamefont {L.}~\bibnamefont
  {Crochik}}\ and\ \bibinfo {author} {\bibfnamefont {T.}~\bibnamefont
  {Tom\'e}},\ }\href {https://doi.org/10.1103/PhysRevE.72.057103} {\bibfield
  {journal} {\bibinfo  {journal} {Phys. Rev. E}\ }\textbf {\bibinfo {volume}
  {72}},\ \bibinfo {pages} {057103} (\bibinfo {year} {2005})}\BibitemShut
  {NoStop}%
\bibitem [{\citenamefont {Xiao}\ \emph {et~al.}(2008)\citenamefont {Xiao},
  \citenamefont {Hou},\ and\ \citenamefont {Xin}}]{xiao_entropy_2008}%
  \BibitemOpen
  \bibfield  {author} {\bibinfo {author} {\bibfnamefont {T.~J.}\ \bibnamefont
  {Xiao}}, \bibinfo {author} {\bibfnamefont {Z.}~\bibnamefont {Hou}},\ and\
  \bibinfo {author} {\bibfnamefont {H.}~\bibnamefont {Xin}},\ }\href
  {https://doi.org/10.1063/1.2978179} {\bibfield  {journal} {\bibinfo
  {journal} {J. Chem. Phys.}\ }\textbf {\bibinfo {volume} {129}},\ \bibinfo
  {pages} {114506} (\bibinfo {year} {2008})}\BibitemShut {NoStop}%
\bibitem [{\citenamefont {Xiao}\ \emph {et~al.}(2009)\citenamefont {Xiao},
  \citenamefont {Hou},\ and\ \citenamefont {Xin}}]{xiao_stochastic_2009}%
  \BibitemOpen
  \bibfield  {author} {\bibinfo {author} {\bibfnamefont {T.}~\bibnamefont
  {Xiao}}, \bibinfo {author} {\bibfnamefont {Z.}~\bibnamefont {Hou}},\ and\
  \bibinfo {author} {\bibfnamefont {H.}~\bibnamefont {Xin}},\ }\href
  {https://doi.org/10.1021/jp901610x} {\bibfield  {journal} {\bibinfo
  {journal} {J. Phys. Chem. B}\ }\textbf {\bibinfo {volume} {113}},\ \bibinfo
  {pages} {9316} (\bibinfo {year} {2009})}\BibitemShut {NoStop}%
\bibitem [{\citenamefont {Barato}\ and\ \citenamefont
  {Hinrichsen}(2012)}]{barato_entropy_2012}%
  \BibitemOpen
  \bibfield  {author} {\bibinfo {author} {\bibfnamefont {A.~C.}\ \bibnamefont
  {Barato}}\ and\ \bibinfo {author} {\bibfnamefont {H.}~\bibnamefont
  {Hinrichsen}},\ }\href@noop {} {\bibfield  {journal} {\bibinfo  {journal} {J.
  Phys. A: Math. Theor.}\ }\textbf {\bibinfo {volume} {45}},\ \bibinfo {pages}
  {115005} (\bibinfo {year} {2012})}\BibitemShut {NoStop}%
\bibitem [{\citenamefont {Tom\'e}\ and\ \citenamefont
  {de~Oliveira}(2012)}]{tome_entropy_2021}%
  \BibitemOpen
  \bibfield  {author} {\bibinfo {author} {\bibfnamefont {T.}~\bibnamefont
  {Tom\'e}}\ and\ \bibinfo {author} {\bibfnamefont {M.~J.}\ \bibnamefont
  {de~Oliveira}},\ }\href@noop {} {\bibfield  {journal} {\bibinfo  {journal}
  {Phys. Rev. Lett.}\ }\textbf {\bibinfo {volume} {108}},\ \bibinfo {pages}
  {020601} (\bibinfo {year} {2012})}\BibitemShut {NoStop}%
\bibitem [{\citenamefont {Noa}\ \emph {et~al.}(2019)\citenamefont {Noa},
  \citenamefont {Harunari}, \citenamefont {de~Oliveira},\ and\ \citenamefont
  {Fiore}}]{noa_entropy_2019}%
  \BibitemOpen
  \bibfield  {author} {\bibinfo {author} {\bibfnamefont {C.~E.~F.}\
  \bibnamefont {Noa}}, \bibinfo {author} {\bibfnamefont {P.~E.}\ \bibnamefont
  {Harunari}}, \bibinfo {author} {\bibfnamefont {M.~J.}\ \bibnamefont
  {de~Oliveira}},\ and\ \bibinfo {author} {\bibfnamefont {C.~E.}\ \bibnamefont
  {Fiore}},\ }\href {https://doi.org/10.1103/PhysRevE.100.012104} {\bibfield
  {journal} {\bibinfo  {journal} {Phys. Rev. E}\ }\textbf {\bibinfo {volume}
  {100}},\ \bibinfo {pages} {012104} (\bibinfo {year} {2019})}\BibitemShut
  {NoStop}%
\bibitem [{\citenamefont {Martynec}\ \emph {et~al.}(2020)\citenamefont
  {Martynec}, \citenamefont {Klapp},\ and\ \citenamefont
  {Loos}}]{martynec_entropy_2020}%
  \BibitemOpen
  \bibfield  {author} {\bibinfo {author} {\bibfnamefont {T.}~\bibnamefont
  {Martynec}}, \bibinfo {author} {\bibfnamefont {S.~H.~L.}\ \bibnamefont
  {Klapp}},\ and\ \bibinfo {author} {\bibfnamefont {S.~A.~M.}\ \bibnamefont
  {Loos}},\ }\href {https://doi.org/10.1088/1367-2630/abb5f0} {\bibfield
  {journal} {\bibinfo  {journal} {New J. Phys.}\ }\textbf {\bibinfo {volume}
  {22}},\ \bibinfo {pages} {093069} (\bibinfo {year} {2020})}\BibitemShut
  {NoStop}%
\bibitem [{\citenamefont {Seara}\ \emph {et~al.}(2021)\citenamefont {Seara},
  \citenamefont {Machta},\ and\ \citenamefont
  {Murrell}}]{seara_irreversibility_2021}%
  \BibitemOpen
  \bibfield  {author} {\bibinfo {author} {\bibfnamefont {D.~S.}\ \bibnamefont
  {Seara}}, \bibinfo {author} {\bibfnamefont {B.~B.}\ \bibnamefont {Machta}},\
  and\ \bibinfo {author} {\bibfnamefont {M.~P.}\ \bibnamefont {Murrell}},\
  }\href {https://doi.org/10.1038/s41467-020-20281-2} {\bibfield  {journal}
  {\bibinfo  {journal} {Nat. Commun.}\ }\textbf {\bibinfo {volume} {12}},\
  \bibinfo {pages} {392} (\bibinfo {year} {2021})}\BibitemShut {NoStop}%
\bibitem [{\citenamefont {Le~Bellac}(1992)}]{LeBellac}%
  \BibitemOpen
  \bibfield  {author} {\bibinfo {author} {\bibfnamefont {M.}~\bibnamefont
  {Le~Bellac}},\ }\href@noop {} {\emph {\bibinfo {title} {Quantum and
  Statistical Field Theory}}}\ (\bibinfo  {publisher} {Oxford Science
  Publications},\ \bibinfo {address} {Oxford},\ \bibinfo {year}
  {1992})\BibitemShut {NoStop}%
\bibitem [{\citenamefont {Meibohm}\ and\ \citenamefont
  {Esposito}(2022)}]{Meibohm_2022}%
  \BibitemOpen
  \bibfield  {author} {\bibinfo {author} {\bibfnamefont {J.}~\bibnamefont
  {Meibohm}}\ and\ \bibinfo {author} {\bibfnamefont {M.}~\bibnamefont
  {Esposito}},\ }\href@noop {} {\bibfield  {journal} {\bibinfo  {journal}
  {Phys. Rev. Lett.}\ }\textbf {\bibinfo {volume} {128}},\ \bibinfo {pages}
  {110603} (\bibinfo {year} {2022})}\BibitemShut {NoStop}%
\bibitem [{\citenamefont {Holtzman}\ and\ \citenamefont
  {Raz}(2022)}]{Holtzman_2022}%
  \BibitemOpen
  \bibfield  {author} {\bibinfo {author} {\bibfnamefont {R.}~\bibnamefont
  {Holtzman}}\ and\ \bibinfo {author} {\bibfnamefont {O.}~\bibnamefont {Raz}},\
  }\href@noop {} {\bibfield  {journal} {\bibinfo  {journal} {Commun. Phys.}\
  }\textbf {\bibinfo {volume} {5}},\ \bibinfo {pages} {280} (\bibinfo {year}
  {2022})}\BibitemShut {NoStop}%
\bibitem [{\citenamefont {Aron}\ and\ \citenamefont
  {Chamon}(2020)}]{Aron_2020}%
  \BibitemOpen
  \bibfield  {author} {\bibinfo {author} {\bibfnamefont {C.}~\bibnamefont
  {Aron}}\ and\ \bibinfo {author} {\bibfnamefont {C.}~\bibnamefont {Chamon}},\
  }\href@noop {} {\bibfield  {journal} {\bibinfo  {journal} {SciPost Phys.}\
  }\textbf {\bibinfo {volume} {8}},\ \bibinfo {pages} {074} (\bibinfo {year}
  {2020})}\BibitemShut {NoStop}%
\bibitem [{\citenamefont {Guislain}\ and\ \citenamefont
  {Bertin}(2023)}]{guislain_nonequil2023}%
  \BibitemOpen
  \bibfield  {author} {\bibinfo {author} {\bibfnamefont {L.}~\bibnamefont
  {Guislain}}\ and\ \bibinfo {author} {\bibfnamefont {E.}~\bibnamefont
  {Bertin}},\ }\href@noop {} {\bibfield  {journal} {\bibinfo  {journal} {Phys.
  Rev. Lett.}\ }\textbf {\bibinfo {volume} {130}},\ \bibinfo {pages} {207102}
  (\bibinfo {year} {2023})}\BibitemShut {NoStop}%
\bibitem [{\citenamefont {Fruchart}\ \emph {et~al.}(2021)\citenamefont
  {Fruchart}, \citenamefont {Hanai}, \citenamefont {Littlewood},\ and\
  \citenamefont {Vitelli}}]{Fruchart_non-reciprocal_2021}%
  \BibitemOpen
  \bibfield  {author} {\bibinfo {author} {\bibfnamefont {M.}~\bibnamefont
  {Fruchart}}, \bibinfo {author} {\bibfnamefont {R.}~\bibnamefont {Hanai}},
  \bibinfo {author} {\bibfnamefont {P.}~\bibnamefont {Littlewood}},\ and\
  \bibinfo {author} {\bibfnamefont {V.}~\bibnamefont {Vitelli}},\ }\href@noop
  {} {\bibfield  {journal} {\bibinfo  {journal} {Nature}\ }\textbf {\bibinfo
  {volume} {592}},\ \bibinfo {pages} {363} (\bibinfo {year}
  {2021})}\BibitemShut {NoStop}%
\bibitem [{\citenamefont {Martin}\ \emph {et~al.}(2023)\citenamefont {Martin},
  \citenamefont {Daniel~Seara}, \citenamefont {Avni}, \citenamefont
  {Fruchart},\ and\ \citenamefont {Vitelli}}]{Martin_exact_2023}%
  \BibitemOpen
  \bibfield  {author} {\bibinfo {author} {\bibfnamefont {D.}~\bibnamefont
  {Martin}}, \bibinfo {author} {\bibfnamefont {D.}~\bibnamefont
  {Daniel~Seara}}, \bibinfo {author} {\bibfnamefont {Y.}~\bibnamefont {Avni}},
  \bibinfo {author} {\bibfnamefont {M.}~\bibnamefont {Fruchart}},\ and\
  \bibinfo {author} {\bibfnamefont {V.}~\bibnamefont {Vitelli}},\ }\href@noop
  {} {\  (\bibinfo {year} {2023})},\ \Eprint
  {https://arxiv.org/abs/arXiv:2307.08251} {arXiv:2307.08251} \BibitemShut
  {NoStop}%
\bibitem [{\citenamefont {Collet}(2014)}]{collet_macroscopic_2014}%
  \BibitemOpen
  \bibfield  {author} {\bibinfo {author} {\bibfnamefont {F.}~\bibnamefont
  {Collet}},\ }\href {https://doi.org/10.1007/s10955-014-1105-9} {\bibfield
  {journal} {\bibinfo  {journal} {J. Stat. Phys.}\ }\textbf {\bibinfo {volume}
  {157}},\ \bibinfo {pages} {1301} (\bibinfo {year} {2014})}\BibitemShut
  {NoStop}%
\bibitem [{\citenamefont {Sinelschikov}\ \emph {et~al.}(2023)\citenamefont
  {Sinelschikov}, \citenamefont {Poggialini}, \citenamefont {Abbate},\ and\
  \citenamefont {De~Martino}}]{Sinelschikov_emergence_2023}%
  \BibitemOpen
  \bibfield  {author} {\bibinfo {author} {\bibfnamefont {D.}~\bibnamefont
  {Sinelschikov}}, \bibinfo {author} {\bibfnamefont {A.}~\bibnamefont
  {Poggialini}}, \bibinfo {author} {\bibfnamefont {M.~F.}\ \bibnamefont
  {Abbate}},\ and\ \bibinfo {author} {\bibfnamefont {D.}~\bibnamefont
  {De~Martino}},\ }\href@noop {} {\bibinfo {title} {Emergence of collective
  self-oscillations in minimal lattice models with feedback}} (\bibinfo {year}
  {2023}),\ \Eprint {https://arxiv.org/abs/arXiv:2306.01823} {arXiv:2306.01823}
  \BibitemShut {NoStop}%
\bibitem [{\citenamefont {Touchette}(2009)}]{touchette_2009}%
  \BibitemOpen
  \bibfield  {author} {\bibinfo {author} {\bibfnamefont {H.}~\bibnamefont
  {Touchette}},\ }\href@noop {} {\bibfield  {journal} {\bibinfo  {journal}
  {Physics Reports}\ }\textbf {\bibinfo {volume} {478}} (\bibinfo {year}
  {2009})}\BibitemShut {NoStop}%
\bibitem [{\citenamefont {Knessl}\ \emph {et~al.}(1985)\citenamefont {Knessl},
  \citenamefont {Matkowsky}, \citenamefont {Schuss},\ and\ \citenamefont
  {Tier}}]{knessl_1985}%
  \BibitemOpen
  \bibfield  {author} {\bibinfo {author} {\bibfnamefont {C.}~\bibnamefont
  {Knessl}}, \bibinfo {author} {\bibfnamefont {B.~J.}\ \bibnamefont
  {Matkowsky}}, \bibinfo {author} {\bibfnamefont {Z.}~\bibnamefont {Schuss}},\
  and\ \bibinfo {author} {\bibfnamefont {C.}~\bibnamefont {Tier}},\ }\href@noop
  {} {\bibfield  {journal} {\bibinfo  {journal} {SIAM J. Appl. Math.}\ }\textbf
  {\bibinfo {volume} {46}},\ \bibinfo {pages} {1006} (\bibinfo {year}
  {1985})}\BibitemShut {NoStop}%
\bibitem [{\citenamefont {Graham}\ and\ \citenamefont
  {T\'el}(1987)}]{Graham_nonequilibrium1987}%
  \BibitemOpen
  \bibfield  {author} {\bibinfo {author} {\bibfnamefont {R.}~\bibnamefont
  {Graham}}\ and\ \bibinfo {author} {\bibfnamefont {T.}~\bibnamefont {T\'el}},\
  }\href@noop {} {\bibfield  {journal} {\bibinfo  {journal} {Phys. Rev. A}\
  }\textbf {\bibinfo {volume} {35}},\ \bibinfo {pages} {1328} (\bibinfo {year}
  {1987})}\BibitemShut {NoStop}%
\bibitem [{\citenamefont {Graham}\ and\ \citenamefont
  {T{\'e}l}(1984)}]{graham.tel.1984a}%
  \BibitemOpen
  \bibfield  {author} {\bibinfo {author} {\bibfnamefont {R.}~\bibnamefont
  {Graham}}\ and\ \bibinfo {author} {\bibfnamefont {T.}~\bibnamefont
  {T{\'e}l}},\ }\href@noop {} {\bibfield  {journal} {\bibinfo  {journal}
  {Journal of Statistical Physics}\ }\textbf {\bibinfo {volume} {35}},\
  \bibinfo {pages} {729} (\bibinfo {year} {1984})}\BibitemShut {NoStop}%
\bibitem [{\citenamefont {Graham}(1989)}]{graham_1989}%
  \BibitemOpen
  \bibfield  {author} {\bibinfo {author} {\bibfnamefont {R.}~\bibnamefont
  {Graham}},\ }\href@noop {} {\bibfield  {journal} {\bibinfo  {journal}
  {Journal of Statistical Physics}\ }\textbf {\bibinfo {volume} {54}},\
  \bibinfo {pages} {1207} (\bibinfo {year} {1989})}\BibitemShut {NoStop}%
\bibitem [{\citenamefont {Gillespie}(2007)}]{gillespie_stochastic_2007}%
  \BibitemOpen
  \bibfield  {author} {\bibinfo {author} {\bibfnamefont {D.~T.}\ \bibnamefont
  {Gillespie}},\ }\href
  {https://doi.org/10.1146/annurev.physchem.58.032806.104637} {\bibfield
  {journal} {\bibinfo  {journal} {Annu. Rev. Phys. Chem.}\ }\textbf {\bibinfo
  {volume} {58}},\ \bibinfo {pages} {35} (\bibinfo {year} {2007})}\BibitemShut
  {NoStop}%
\bibitem [{SM()}]{SM}%
  \BibitemOpen
  \bibinfo {note} {See Supplemental Material at XXX.}\BibitemShut {Stop}%
\bibitem [{\citenamefont {Schnackenberg}(1976)}]{schnakenberg_1976}%
  \BibitemOpen
  \bibfield  {author} {\bibinfo {author} {\bibfnamefont {J.}~\bibnamefont
  {Schnackenberg}},\ }\href@noop {} {\bibfield  {journal} {\bibinfo  {journal}
  {Rev. Mod. Phys.}\ }\textbf {\bibinfo {volume} {48}},\ \bibinfo {pages} {571}
  (\bibinfo {year} {1976})}\BibitemShut {NoStop}%
\bibitem [{\citenamefont {Gaspard}(2004)}]{gaspard_time-reversed_2004}%
  \BibitemOpen
  \bibfield  {author} {\bibinfo {author} {\bibfnamefont {P.}~\bibnamefont
  {Gaspard}},\ }\href {https://doi.org/10.1007/s10955-004-3455-1} {\bibfield
  {journal} {\bibinfo  {journal} {J. Stat. Phys.}\ }\textbf {\bibinfo {volume}
  {117}},\ \bibinfo {pages} {599} (\bibinfo {year} {2004})}\BibitemShut
  {NoStop}%
\end{thebibliography}
\end{document}